\documentclass[]{article}

\title{A Rigorous Study of Hawking Radiation \\on Collapsing Charged Spherically Symmetric Spacetimes}
\author{Fred Alford,\\ fta17@ic.ac.uk}
\date{Department of Mathematics,\\
	Imperial College London\\[2ex]
	\today}
\newcommand{\p}{\partial}
\newcommand{\IE}{I.E.[Y_{l,m}\psi_+,v_c,u_1,u_0]}
\newcommand{\IEE}[1]{I.E.[Y_{l,m}\psi_+,v_c,#1,u_0]}
\newcommand{\IT}{I.T.[Y_{l,m}\psi_+]}
\newcommand{\lur}{\langle u\rangle}
\newcommand{\R}{\mathbb{R}}
\newcommand{\RN}{Reissner--Nordstr\"om}
\newcommand{\RNS}{Reissner--Nordstr\"om }
\usepackage{amsmath,amssymb,amsthm}
\usepackage{geometry}
\usepackage{slashed}
\usepackage{array}
\usepackage{tikz}
\usepackage{wrapfig}
\usepackage{graphicx}
\usepackage{cutwin}
\usepackage{paracol}
\usepackage[numbers]{natbib}
\usepackage{bbm}
\usetikzlibrary{decorations.pathmorphing}
\newtheorem{Theorem}{Theorem}[subsection]

\newtheorem{Lemma}[Theorem]{Lemma}
\newtheorem{Corollary}[Theorem]{Corollary}
\newtheorem{Proposition}[Theorem]{Proposition}
\newtheorem{Remark}[Theorem]{Remark}
\newtheorem{Theorem1}{Theorem}[section]

\newtheorem{Remark1}[Theorem1]{Remark}
\newtheorem{Proposition1}[Theorem1]{Proposition}
\newtheorem{definition}{Definition}[section]

\usetikzlibrary{decorations.pathmorphing}
\tikzset{snake it/.style={decorate, decoration=snake}}
\usepackage{xpatch}
\makeatletter 
\xpatchcmd{\@thm}{\thm@headpunct{.}}{\thm@headpunct{}}{}{}
\makeatother
\numberwithin{equation}{section}
\usepackage{thmtools}
\usepackage{thm-restate}
\renewcommand{\theequation}{\arabic{section}.\arabic{subsection}.\arabic{equation}}
\begin{document}

\maketitle

\begin{abstract}
	In this paper, we give a rigorous mathematical treatment of the late time Hawking radiation of massless, uncharged bosons emitted by a class of collapsing, spherically symmetric, charged models of black hole formation, including both extremal and sub-extremal black holes. We will also prove a bound on the rate at which the radiation emitted approaches this late time limit. This includes an integrable decay rate of radiation emitted by extremal black holes, for which the late time limit vanishes. Thus, we show that the total expected quantity of any massless, uncharged boson emitted by an extremal \RNS black hole is finite.
\end{abstract}

%\tableofcontents
%\pagebreak

\begin{wrapfigure}{r}{5cm}\label{Fig:Penrose}
	\begin{tikzpicture}[scale =1.2]
	\node (I)    at ( 0,0) {};
	
	\path 
	(I) +(90:2)  coordinate[label=90:$i^+$]  (Itop)
	+(-90:2) coordinate (Imid)
	+(0:2)   coordinate[label=0:$i^0$] (Iright)
	+(-1,1) coordinate (Ileft)
	+(-0.6,1.4) coordinate[label=0:\tiny ($t^*_c$\text{, }$r_+$)] (BHH)
	+(-1,-3) coordinate[label=0:$i^-$] (Ibot)
	;
	\draw (Ileft) -- 
	node[midway, above left]    {$\mathcal{H}^+$}
	(Itop) --
	node[midway, above, sloped] {$\mathcal{I}^+$}
	(Iright) -- 
	node[midway, below, sloped] {$\mathcal{I}^-$}
	(Ibot) --
	node[midway, above, sloped]    {\small }    
	(Ileft) -- cycle;
	\draw[fill=gray!80] (Ibot) to[out=60, in=-60]
	node[midway, below, sloped] {\tiny $r=r_b$} (BHH)--(Ileft)--cycle;
	\draw (1.14,-0.86) to node[midway, above, sloped] {\tiny $\Sigma_{v}$} (-0.23,0.5);
	\end{tikzpicture}
\vspace{-3mm}
	\caption{Penrose Diagram of RNOS Model, with null hyper surface $\Sigma_{v}$.}\label{fig:PenRef}
\end{wrapfigure}
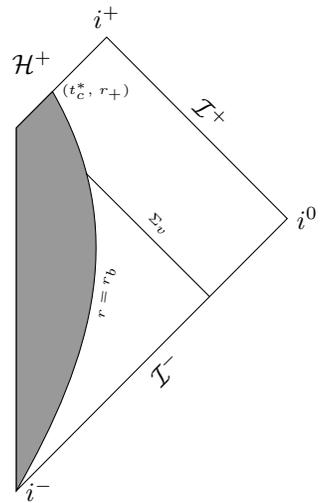

\section{Introduction}
\renewcommand{\theequation}{\arabic{section}.\arabic{equation}}

Classically, black holes, once formed, are permanent. The discovery of Hawking Radiation was therefore a major breakthrough in understanding how, in the context of quantum theories, black holes can in fact disintegrate over time, as it describes a mechanism to decrease the mass of black holes and potentially cause them to evaporate entirely. Since Hawking proposed this phenomenon in $1974$ \cite{hawking1974,hawking1975}, there have been hundreds of papers on the topic within the physics literature. For an overview of the physical aspects of Hawking radiation, we refer the reader to \cite{WaldQFT}. Concerning a mathematically rigorous treatment of Hawking radiation, however, there are substantially fewer works (see already \cite{BachelotHawking} and Section \ref{Sec:PrevWork} for a discussion of further references), and the mathematical status of Hawking radiation still leaves much to be desired. As a result, it has not been possible yet to ask more quantitative questions about Hawking radiation, which are necessary if one wants to eventually understand this phenomenon in the non-perturbative setting. This paper contributes towards solving this problem by giving a new physical space approach to Hawking's calculation allowing one to also obtain a rigorous bound on the rate at which emission approaches black body radiation.

The mathematical problem of Hawking radiation for massless, chargeless zero-spin bosons can be formulated as follows:

We first consider the exteriors of collapsing spherically symmetric spacetimes. In these exteriors, the spacetimes are a subset of \RNS spacetime \cite{Reissner}. Therefore we have coordinates $t^*$, $r$, $\theta$, $\varphi$ for which the metric takes the form
\begin{align}\label{eq:overviewmetric}
g=-\left(1-\frac{2M}{r}+\frac{q^2 M^2}{r^2}\right){dt^*}^2+2\left(\frac{2M}{r}-\frac{q^2 M^2}{r^2}\right)&dt^*dr+\left(1+\frac{2M}{r}-\frac{q^2 M^2}{r^2}\right)dr^2+r^2g_{S^2}\\\nonumber
t^*\in\R\qquad& r\in[\tilde{r}_b(t^*),\infty),
\end{align}
where $g_{S^2}$ is the metric on the unit $2$-sphere. Here $\tilde{r}_b(t^*)$ is initially the area radius of the boundary of the collapsing matter cloud as a function of $t^*$, and then becomes $r_+$ (defined in \eqref{eq:r+}) for sufficiently late time - see Section \ref{Sec:RNOS} for further details. Note that this problem requires a collapsing setting, as the problem is trivial (and physically incorrect) in the non-collapsing case. We refer to $M$ as the mass of the underlying \RNS spacetime, and $q\in[-1,1]$ as the charge to mass ratio. In particular, we are allowing the extremal case, $\vert q\vert=1$.

These collapsing charged model exteriors will include the exterior of the Oppenheimer--Snyder Model \cite{S-O} (for which $q=0$), and we will refer to these more general models as \RNS Oppenheimer--Snyder (RNOS) models \cite{Mine2}.

We now consider a function $\psi_+$ on future null infinity, $\mathcal{I}^+$. Let $\phi$ be the solution to the linear wave equation
\begin{equation}\label{eq:wave}
	\Box_g\phi:=\frac{1}{\sqrt{-g}}\p_a\left(\sqrt{-g}g^{ab}\p_b\phi\right)=0
\end{equation}
which vanishes on $\mathcal{H}^+$, and has future radiation field equal to $\psi_+(u-u_0,\theta,\varphi)$. We will be imposing Dirichlet conditions on the boundary of the matter cloud, i.e.~$\phi=0$ on $\{r=r_b(t^*)\}$. Note that we have existence of such a solution thanks to our companion paper \cite{Mine2}. Let us denote the past radiation field of $\phi$ by $\psi_{-,u_0}$. This is a function on past null infinity, $\mathcal{I}^-$. %By considering the quantum calculation done in \cite{hawking1974,hawking1975} (or see \cite{ReallBH} for a more in depth explanation), we know that the expected number of these particles emitted by the formation of the black hole is given by 
The Hawking radiation calculation is to determine
%\begin{equation}
%	\int_{\omega=-\infty}^0\int_{\varphi=0}^{2\pi}\int_{\theta=0}^\pi\vert\omega\vert\vert\hat{\psi}_-(\omega,\theta,\varphi)\vert^2\sin\theta d\theta d\varphi d\omega,
%\end{equation}
%
%
\begin{equation}\label{eq:LateTimeRadiation}
	\lim_{u_0\to\infty}\left(\int_{\omega=-\infty}^0\int_{\varphi=0}^{2\pi}\int_{\theta=0}^\pi\vert\omega\vert\vert\hat{\psi}_{-,u_0}(\omega,\theta,\varphi)\vert^2\sin\theta d\theta d\varphi d\omega\right),
\end{equation}
where $\hat{\psi}_{-,u_0}$ corresponds to the Fourier transform of $\psi_{-,u_0}$ with respect to the advanced time $u$ coordinate. 

Note that here, $\psi_{-,u_0}$ will depend on $u_0$. While there is no immediately obvious classical reason that this limit should exist, in \cite{hawking1974}, Hawking made a heuristic argument for both the existence and value of this limit.

The main Theorem of this paper verifies Hawking's prediction for the value of \eqref{eq:LateTimeRadiation}, and is stated below:
\begin{restatable}[Late Time Emission of Hawking Radiation]{thm}{HawkingVague}\label{Thm:HawkingVague}
	Let $\alpha:[-1,1]\times\R\to\R$ be the smooth function given by:
	\begin{equation}
		\alpha(q,\omega)=\begin{cases}
			\left(e^{\frac{2\pi\vert\omega\vert}{\kappa(q)}}-1\right)^{-1}&q\neq \pm 1\\
			0& q=\pm 1			
		\end{cases},
	\end{equation}
	where $\kappa$ is the surface gravity of the \RNS black hole, i.e.
	\begin{equation}\label{eq:kappa}
		\kappa=\frac{\sqrt{1-q^2}}{\left(2+2\sqrt{1-q^2}-q^2\right)M}.
	\end{equation}

	Let $\psi_+(u, \theta, \varphi)$ be a Schwartz function on the $\R\times S^2$, with $\hat{\psi}_+$ only supported on positive frequencies ($[0,\infty)\times S^2$). Fix $\mathcal{M}$ an RNOS spacetime (see Section \ref{Sec:RNOS}) with related $M>0$, $q\in[-1,1]$ and $r_b$. Let $\phi$ be the solution of \eqref{eq:wave}, as given by Theorem $7.2$ of \cite{Mine2}, such that
\begin{align}
\lim_{v\to\infty}r(u,v)\phi(u,v,\theta,\varphi)&=\psi_+(u-u_0,\theta,\varphi)\\
\lim_{u\to\infty}r(u,v)\phi(u,v,\theta,\varphi)&=0\quad \forall v\geq v_c,
\end{align}
Define the function $\psi_{-,u_0}$ by
\begin{equation}
\lim_{u\to-\infty}r(u,v)\phi(u,v,\theta,\varphi)=\psi_{-,u_0}(v,\theta,\varphi).
\end{equation}

Then there exists constants $A(M, r_b, \psi_+)$, $U(M, r_b, \psi_+)$, independent of $q$, such that
\begin{equation}\label{eq:RoughHawking}
	\left\vert\int_{\omega=-\infty}^0\int_{\varphi=0}^{2\pi}\int_{\theta=0}^\pi\vert\omega\vert\vert\hat{\psi}_{-,u_0}(\omega, \theta,\varphi)\vert^2 \sin\theta d\omega d\theta d\varphi -\int_{\omega=-\infty}^\infty\int_{\varphi=0}^{2\pi}\int_{\theta=0}^\pi\alpha(q,\omega)\vert\omega\vert\vert\hat{\psi}_{\mathcal{H}^-}(\omega,\theta,\varphi)\vert^2\sin\theta d\omega d\theta d\varphi\right\vert
	\leq \frac{A\log{u_0}}{u_0^2},
\end{equation}
for all $u_0\geq U$.

Further, in the case $\vert q\vert<1$, for each $n\in\mathbb{N}$, there exist constants $A_n(M, r_b, q,\psi_+)$, $U(M, r_b, q,\psi_+)$ such that
\begin{equation}\label{eq:RoughSubextremalHawking}
\left\vert\int_{\omega=-\infty}^0\int_{\varphi=0}^{2\pi}\int_{\theta=0}^\pi\vert\omega\vert\vert\hat{\psi}_{-,u_0}(\omega, \theta,\varphi)\vert^2 \sin\theta d\omega d\theta d\varphi -\int_{\omega=-\infty}^\infty\int_{\varphi=0}^{2\pi}\int_{\theta=0}^\pi\alpha(q,\omega)\vert\omega\vert\vert\hat{\psi}_{\mathcal{H}^-}(\omega,\theta,\varphi)\vert^2\sin\theta d\omega d\theta d\varphi\right\vert
\leq A_nu_0^{-n},
\end{equation}
for all $u_0>U$.

Here, $\hat{f}$ is the Fourier transform of $f$ with respect to its non-angular coordinate, $\psi_{\mathcal{H}^-}$ is the transmission of $\psi_+$ in pure \RNS spacetime (see Theorem \ref{Thm:RNExist}).
\end{restatable}

\begin{Remark}[The Reflective Boundary Condition]
	In this paper, we have chosen to impose reflective boundary conditions on the surface of the dust cloud. This is primarily done for mathematical simplicity - there are no Oppenheimer--Snyder type charged dust models collapsing to form Reissner--Nordstr\"om black holes that can be written down explicitly. If one were to include the interior of the dust cloud (with permeating boundary conditions) there are two additional calculations the must be considered:
	\begin{itemize}
		\item Firstly, one would need to determine energy boundedness of the scattering map from the surface ${v=v_c}$ backwards to $\mathcal{I}^-$. This is a non-trivial result, which the author has proven in the Oppenheimer--Snyder case \cite{Mine}, but is difficult to prove in a more general setting.
		\item Secondly, one would need to replace Section \ref{Sec:Reflection}. This would be less challenging, as this is the statement of a high frequency solution permeating through a matter cloud along geodesics, and is therefore an application of geometric optics approximation.
	\end{itemize}
	If one is able to complete both these steps, then proving the result for any given matter model should follow through identically to the rest of this paper.
\end{Remark}

The proof of Theorem \ref{Thm:HawkingVague}  will rely on certain scattering results for solutions of \eqref{eq:wave}, which are proven in our companion paper \cite{Mine2}. It will also make use of scattering results in pure \RNS spacetime, some rigorous high frequency approximations, and finally will use an $r$ weighted energy estimate, based very closely on the estimates given in \cite{ERNScat}.

The physical interpretation of this is the following: Given a single particle on future null infinity, $\mathcal{I}^+$, this will have a quantum state given by applying a creation operator to the vacuum state. This creation operator will have a corresponding \emph{positive frequency} function on $\mathcal{I}^+$, which will be the $\psi_+$ in the statement of the theorem. By considering the quantum calculation done in \cite{hawking1974,hawking1975} (or see \cite{ReallBH} for a more in depth explanation), we know that the expected number of these particles emitted by the formation of the black hole is given by
\begin{equation}
	\int_{\omega=-\infty}^0\int_{\varphi=0}^{2\pi}\int_{\theta=0}^\pi\vert\omega\vert\vert\hat{\psi}_-(\omega,\theta,\varphi)\vert^2\sin\theta d\theta d\varphi d\omega,
\end{equation}
where again $\hat{\psi}_-$ corresponds to the Fourier transform of $\psi_-$ with respect to the advanced time $u$ coordinate.

In this paper, we are concerned with the particles emitted at late times in the formation of the black hole. Thus instead of a fixed future radiation field, we will consider the family of future radiation fields given by $\psi_{+,u_0}(u,\theta,\varphi):=\psi_+(u-u_0,\theta,\varphi)$, parametrised by $u_0$. The integral \eqref{eq:LateTimeRadiation} therefore represents the late time limit of the expected number of particles (associated with the function $\psi_+$) emitted by the collapsing black hole.

The limit \eqref{eq:RoughHawking} has the physical interpretation that a \RNS black hole forming in the collapse of a matter cloud gives off radiation approaching that of a black body with temperature $\frac{\kappa}{2\pi}$ (in units where $\hbar=G=c=1$). Thus sub-extremal \RNS black holes will emit infinitely many particles in the future. In the extremal case however, this limit suggests that the amount of radiation emitted by a forming, extremal \RNS black hole tends towards $0$. This is therefore a rigorous result confirming Hawking's original calculation in both extremal and sub-extremal settings.

Equation \eqref{eq:RoughHawking} and \eqref{eq:RoughSubextremalHawking} also give estimates for the rate at which the limit predicted by Hawking is approached. In the case of \eqref{eq:RoughSubextremalHawking}, this rate is very fast. In the extremal case, though the bound decays more slowly, it manages to be integrable. This has the physical interpretation that, as the `final' temperature is zero, the total quantity of any given particle of radiation emitted by an extremal \RNS black hole that forms from collapse is finite. It remains an important open problem to find the mathematical representation of the total radiation (all such particles) emitted by an extremal \RNS black hole, and determine if this is indeed finite, and thus, whether extremal black holes will indeed evaporate due to Hawking radiation.

\subsection{Overview}

Before proving Theorem \ref{Thm:HawkingVague}, we will first discuss previous mathematical works on Hawking radiation in Section \ref{Sec:PrevWork}. We will then discuss the background ``RNOS" spacetime models in Section \ref{Sec:RNOS}, and the notation used in the rest of this paper in Section \ref{Sec:Notation}. Lastly before the proof, we will discuss previous scattering results on \RNS spacetimes (Section \ref{Sec:Coeff}) and on RNOS models (Section \ref{Sec:RNOSScat}).

The proof of Theorem \ref{Thm:HawkingVague} will be split up into three main parts. Firstly, in Section \ref{Sec:H^1/2 Norm} we calculate a very similar limit to \eqref{eq:LateTimeRadiation}, but integrating over all frequencies (rather than only negative frequencies). The bounds on the rate at which this limit is approached will be in terms of weighted energies of $\psi_+$. Secondly, in Section \ref{Sec:Integrated Error Terms}, we bound the decay rates of these weighted energies. Finally, in Section \ref{Sec:Proof}, we bring the results of Sections \ref{Sec:H^1/2 Norm} and \ref{Sec:Integrated Error Terms} together to obtain the final result.

\section{Previous Work}\label{Sec:PrevWork}
Hawking radiation on collapsing spacetimes has been mathematically studied in several other settings, for example \cite{Bachelot,HafnerD2001Acft,DrouotAlexis2017AQVo,fredenhagen1990}. Each of these papers primarily work in frequency space, and work in different contexts to this paper. Let us discuss some of these differences in more detail.

\subsection{Mathematical Works}

The original mathematical study of Hawking radiation by Bachelot, \cite{BachelotHawking}, considered Hawking radiation of massive or massless (but still uncharged) non-interacting bosons for a spherically symmetric uncharged collapsing model, and performs this calculation almost entirely in frequency space. This paper obtains what can be viewed as a partial result towards Theorem \ref{Thm:HawkingVague}, where the surface of the collapsing star is assumed to remain at a fixed radius for all sufficiently far back times, and no rate at which the limit is approached is calculated.

In \cite{HafnerD2001Acft}, H\"afner studies Hawking radiation of fermions for sub-extremal charged, rotating black holes, described by the Dirac equation rather than the wave equation. This was the first work on black holes outside spherical symmetry. The Dirac equation itself has a $0^{\text{th}}$ order conserved current, which avoids many of the difficulties of considering the linear wave equation, for which no such current exists.

In the paper \cite{DrouotAlexis2017AQVo}, Drouot considers the full Klein--Gordon equation, but on the sub-extremal Schwarzschild de-Sitter metric. This paper is also the first paper in this setting to obtain a rate at which the limit is obtained, independent of the angular mode. This is easier than the asymptotically flat case considered in Hawking's original work (due to the lack of a cosmological horizon at a finite radius in the asymptotically flat case).

The paper \cite{eskin2019hawking} looks at calculating the Hawking radiation of extremal and subextremal \RNS black holes in one fewer dimension, with no rate obtained. This paper also considers Hawking radiation in the context of the Unruh-type vacuum rather than Hawking radiation generated from a collapsing spacetime.

There are two ways to obtain an analogue of Hawking radiation by considering quantum states on a (non-collapsing) \RNS background. (Note that an understanding of these is not required to understand of the rest of this paper). Firstly, one can construct the Unruh state \cite{UnruhW.G1976Nobe}. If one considers quantum states on the \RNS black hole, one can show that there is a unique state that coincides with the vacuum state on $\mathcal{I}^-$ and is well behaved at $\mathcal{H}^+$ (i.e.~is a Hadamard state). This state evaluated on $\mathcal{I}^+$ is a thermal state of temperature $\kappa/2\pi$ (see \cite{DimockKayHartle}, for example).

The second analogue can be obtained by constructing the Hartle--Hawking--Israel state. If one again considers quantum states on the permanent \RNS black hole, one can show that there is a unique state that is well behaved (Hadamard) at $\mathcal{I}^+$, $\mathcal{I}^-$, $\mathcal{H}^+$, $\mathcal{H}^-$ \cite{KayWaldHartle}. This state is that of a thermal black body, again of temperature $\kappa/2\pi$. The interpretation of this is that the black hole is in equilibrium with this level of thermal radiation, and is therefore emitting the radiation of a black body of this temperature. This result has been considered in a mathematically rigorous manner on Schwarzschild \cite{KayHartle,DimockKayHartle}, and more recently in a more general setting \cite{SandersKo2015OtCo,GerardChristian2021THso}. The present paper, however, will be focused on the collapsing setting, as it is believed that the collapsing spacetime method will generalise more readily, as the Hartle--Hawking--Israel state has been shown not to exist in Kerr spacetimes \cite{KayWaldHartle}.

\subsection{Selected Works Within the Physics Literature}
The physics literature on this topic is vast, so we will only mention some select results here.

Hawking radiation on a charged background has been considered in several other papers in the physics literature, the most relevant being \cite{HawkingExtremal,Balbinot_2007}. The second of these, \cite{Balbinot_2007}, considers Hawking radiation emitted by extremal black holes in the style of Hawking's original paper. Many papers also make use of the surprising fact that the extremal \RNS Hawking radiation calculation is very similar to an accelerated mirror in Minkowski space \cite{MirrorExtremalHawking}. 

A more thorough discussion of the physical derivation of Hawking radiation in general, along with a full explanation of Hawking's original method for the calculation, can be found in chapter 14.4 of General Relativity by Wald, \cite{WaldRobertM1984Gr}.

\subsection{Further Remarks}

In contrast to many of the above works, the considerations of this paper are largely in physical space, despite the fact that the final statement involves the Fourier transform. We will be using the Friedlander radiation formalism, \cite{friedlander_1980}, for the radiation field, and we hope this will make the proof more transparent to the reader.

While we will make use of results on the classical scattering map on Reisner--Nordstr\"om spacetimes, we will not here discuss previous works on this classical scattering map. For an in depth discussion of scattering, we refer the reader to our previous works \cite{Mine,Mine2} and references therein.

\section{RNOS Models}\label{Sec:RNOS}
\renewcommand{\theequation}{\arabic{section}.\arabic{subsection}.\arabic{equation}}
In this section, we introduce our background models of spherically symmetric charged matter cloud collapse. Note these models only cover the exterior of the matter clouds, as we will be imposing reflective boundary conditions on the surface of these clouds. (We are thus not modelling the cloud per se, but only its boundary.) For a more in depth discussion of these models, we refer the reader to our companion paper \cite{Mine2}. Here, we will just state the properties of these \RNS Oppenheimer--Snyder type models (RNOS models) that we will be using.

\begin{wrapfigure}{r}{5cm}
	\begin{tikzpicture}[scale =0.8]
		\node (I)    at ( 0,0) {};
		
		\path 
		(I) +(0,2)  coordinate[label=90:$i^+$]  (Itop)
		+(0,-2) coordinate (Imid)
		+(2,0)   coordinate[label=0:$i^0$] (Iright)
		+(-2,0) coordinate[label=180 :$\mathcal{B}$] (Ileft)
		+(0,-2) coordinate[label=0:$i^-$] (Ibot)
		;
		\draw (Ileft) -- 
		node[midway, above, sloped]    {$\mathcal{H}^+$, $r=r_+$}
		(Itop) --
		node[midway, above, sloped] {$\mathcal{I}^+$}
		(Iright) -- node[midway, below, sloped]    {$\mathcal{I}^-$}
		(Ibot) --
		node[midway, below, sloped]    {$\mathcal{H}^-$, $r=r_+$} 
		(Ileft) -- cycle;
		\draw[fill=gray!80] (Ibot) to[out=70, in=-60]
		node[midway, above, sloped] {\tiny $r=r_b$} (-0.5,1.5) --(Ileft)--cycle;
	\end{tikzpicture}
	\caption{Penrose diagram of pure \RNS spacetimes}\label{fig:RNPenrose}
\end{wrapfigure}

The RNOS models are a class of collapsing spacetime exteriors, which can individually be viewed as a submanifold of a member of the $2$ parameter family of exterior \RNS spacetimes, or ``pure \RNS spacetimes". 

\subsection{\RNS Spacetimes}
For mass parameter $M>0$ and charge to mass ratio parameter $q\in[-1,1]$, the pure \RNS spacetime (Figure \ref{fig:RNPenrose}) takes the form:
\begin{equation}
	\mathcal{M}_{RN}=\R \times [r_+,\infty)\times S^2
\end{equation}
\begin{equation}
	g=-\left(1-\frac{2M}{r}+\frac{q^2 M^2}{r^2}\right){dt^*}^2+2\left(\frac{2M}{r}-\frac{q^2 M^2}{r^2}\right)dt^*dr+\left(1+\frac{2M}{r}-\frac{q^2 M^2}{r^2}\right)dr^2+r^2g_{S^2}
\end{equation}
where $g_{S^2}$ is the metric on the unit $2$-sphere, and
\begin{equation}\label{eq:r+}
	r_+:=M(1+\sqrt{1-q^2}).
\end{equation} 

The $t^*$ coordinate can be writen in terms of the ``usual" $t$ coordinate by
\begin{equation}
	t^*=t+\int \left(1-\frac{2M}{r}+\frac{q^2 M^2}{r^2}\right)^{-1} dr.
\end{equation}
We use $t^*$ in this paper as it extends to the future event horizon $\mathcal{H}^+:=\{(t^*,r_+,\theta,\varphi)\in\mathcal{M}_{RN}\}$, where the usual $(t,r)$ coordinate system degenerates.

\begin{Remark}[Inclusion of $\mathcal{H}^-$]
	As written, our manifold $\mathcal{M}_{RN}$ does not include the past event horizon, $\mathcal{H}^-$. If we wished to, we could attach $\mathcal{H}^-$ as a limit as $t^*+r\to -\infty$. In the sub-extremal case, we would then also want to attach the bifurcation sphere, $\mathcal{B}$, as the ``corner" between the future directed limit of $\mathcal{H}^-$ and the past directed limit of $\mathcal{H}^+$. However, in practice, we will only be considering $\mathcal{H}^-$ as a limit, so it is not strictly necessary to include this in our manifold.
\end{Remark}

\subsection{RNOS Definition}
\setcounter{equation}{0}
As mentioned previously, the RNOS models will be a submanifold of $\mathcal{M}_{RN}$, given by restricting $\mathcal{M}_{RN}$ to a region $\{r\geq \tilde{r}_b(t^*)\}$. Thus, the RNOS models have the same mass parameter $M> 0$ and charge to mass ratio $q\in[-1,1]$ and also an additional dependence on the $H^2_{loc}$ function $r_b:(-\infty,t^*_c]\to[0,\infty)$. We impose the following conditions on the function $r_b$:
\begin{align}\label{eq:rbnonincreasing}
	\dot{r_b}(t^*):=\frac{dr}{dt^*}&\in(-1,0]\\\label{eq:t^*_cdefinition}
	\exists t^*_c \text{ s.t. }r_b(t^*_c&)=r_+, r_b(t^*)>r_+\quad\forall t^*<t^*_c\\\label{eq:rbtimelike}
	(1,\dot{r}_b(t^*),0,0)&\in T(\mathcal{M})\text{ is timelike for all } t^*\in (-\infty,t^*_c],
\end{align}
where $r_+$ is the black hole horizon for the Reissner--Nordstr\"om spacetime given by \eqref{eq:r+}. 

We allow 2 possible past asymptotic behaviours for $r_b$. Firstly,
\begin{equation}\label{eq:FixedBoundary}
	\int_{-\infty}^{t^*_c} \vert\dot{r}_b(t^*)\vert+\vert\ddot{r}_b(t^*)\vert dt^*<\infty.
\end{equation}
This is known as the `fixed boundary' case, as it required $r_b(t^*)\to r_-$ as $t^*\to -\infty$ for some $r_-$.

The second allowed past asymptotic behaviour will be referred to as the `expanding boundary' case, and requires:
\begin{align}\label{eq:ExpandingBoundary}
	\int_{-\infty}^{t^*_c}\frac{1}{r_b(t^*)^2}dt^*&<\infty\\\nonumber
	\dot{r}_b(t^*)\in[-1+\epsilon,0]& \text{ for some }\epsilon>0.
\end{align}
This model includes any past boundary condition for which $\dot{r}_b\to a\in(-1,0)$, and also includes the Oppenheimer--Snyder model, as this has $r_b(t^*)\sim(-t^*)^{2/3}$. Note this case requires $r_b\to\infty$ as $t^*\to-\infty$.

\begin{definition}[RNOS Manifold]
	An RNOS manifold is given by:
	\begin{align}\label{eq:Manifold}
		\mathcal{M}&=\bigcup_{t^*\in\R}\{t^*\}\times[\tilde{r}_b(t^*),\infty)\times S^2\subset \mathcal{M}_{RN}\\\label{eq:metric}
		g=-\left(1-\frac{2M}{r}+\frac{q^2 M^2}{r^2}\right){dt^*}^2&+2\left(\frac{2M}{r}-\frac{q^2 M^2}{r^2}\right)dt^*dr+\left(1+\frac{2M}{r}-\frac{q^2 M^2}{r^2}\right)dr^2+r^2g_{S^2},
	\end{align}
	where we define
	\begin{equation}
		\tilde{r}_b(t^*):=\begin{cases}r_b(t^*)&t^*\leq t^*_c\\
			r_+& t^*>t^*_c\end{cases},
	\end{equation}
	with the function $r_b$ satisfying \eqref{eq:rbnonincreasing}-\eqref{eq:rbtimelike}, and either \eqref{eq:FixedBoundary} or \eqref{eq:ExpandingBoundary}.
	
	Here, the range of the second coordinate, $r$, depends on the first coordinate, $t^*$, and $g_{S^2}$ is the flat metric on the unit sphere.
\end{definition}

The RNOS models have the same exterior Penrose diagram as the original Oppenheimer-Snyder model (see Figure \ref{fig:PenRef}, derived in \cite{Mine}, for example).

\subsection{Double Null Coordinates}
\setcounter{equation}{0}

For both RNOS and pure Resinner--Nordstr\"om, we will use double null coordinates given by:
\begin{align}\label{eq:udefinition}
u&=t^*-\int_{s=3M}^r\frac{1+\frac{2M}{s}-\frac{q^2M^2}{s^2}}{1-\frac{2M}{s}+\frac{q^2M^2}{s^2}}ds+C_u\\\label{eq:vdefinition}
v&=t^*+r+C_v\\
\p_u&=\frac{1}{2}\left(1-\frac{2M}{r}+\frac{q^2M^2}{r^2}\right)\left(\p_{t^*}-\p_r\right)\\
\p_v&=\frac{1}{2}\left(\left(1+\frac{2M}{r}-\frac{q^2M^2}{r^2}\right)\p_{t^*}+\left(1-\frac{2M}{r}+\frac{q^2M^2}{r^2}\right)\p_r\right)\\
g&=-\left(1-\frac{2M}{r}+\frac{q^2M^2}{r^2}\right)dudv+r(u,v)^2g_{S^2}\\\label{eq:r^*definition}
r^*&=\int_{s=3M}^{r}\left(1-\frac{2M}{s}+\frac{q^2M^2}{s^2}\right)^{-1}ds.
\end{align}

In the definition of $u$, $v$ and $r^*$, there is an arbitrary choice of constant. Here, we have chosen $r^*$ to vanish at $r=3M$. In the extremal case, we will fix $C_u$ when determining the behaviour of the boundary of the matter cloud below, but will otherwise leave these constants arbitrary.

Much of the discussion will be concerning $u$ and $v$ coordinates. Therefore, we will find it useful to parameterise the surface of the cloud by $u$ and $v$. That is, given any $u$, define $v_b(u)$ to be the unique solution to
\begin{equation}\label{eq:vbDef}
r(u,v_b(u))=r_b(t^*(u,v_b(u))),
\end{equation}
We will also define $u_b$ in the domain $v\leq v_c$ as the inverse of $v_b$, \textit{i.e.~}
\begin{equation}
u_b(v):=v_b^{-1}(v), \quad i.e.\quad r(u_b(v),v)=r_b(t^*(u_b(v),v))
\end{equation}

We will be making use of the following properties of $v_b$:
\begin{align}
v_b(u)\to v_c:=v(t^*_c,r_+) \qquad \text{as } u\to\infty\\
v_c-v_b(u) = \begin{cases}
Ae^{-\kappa u}+ O(e^{-2\kappa u})&\vert q\vert<1\\
\frac{A}{u}+O(u^{-2}\log u)&\vert q\vert=1.
\end{cases}\\
v_b'(u)=\begin{cases}
Ae^{-\kappa u}+ O(e^{-2\kappa u})&\vert q\vert<1\\
\frac{A}{u^2}+O(u^{-3}\log u)&\vert q\vert=1
\end{cases},
\end{align}
for constants $A=A(\mathcal{M})>0$ depending on the choice of RNOS spacetime.

These are straightforward calculations, once we note in the extremal case we can choose the constant $C_u$ in \eqref{eq:udefinition} to remove the $u^{-2}$ term in the expansion of $v_c-v_b$. Here, $\kappa$ is the surface gravity of the \RNS black hole that our cloud is forming, as given by \eqref{eq:kappa}.

\subsection{Killing Fields}
\setcounter{equation}{0}

For both pure \RNS and RNOS spacetimes, we will make use of the existence of Killing vector fields.

In the interior of pure \RN, there exists a timelike (becoming null on $\mathcal{H}^+$) Killing vector field, given in our coordinates by $\p_{t^*}$. This is tangent to the event horizon $\mathcal{H}^+$, and $\kappa$ defined above obeys the usual equation for surface gravity of a black hole:
\begin{equation}
\p_{t^*}^a\nabla_a\p_{t^*}^b=\kappa \p_{t^*}^b.
\end{equation}
While this is a Killing field in the interior of RNOS spacetimes, it is not tangent to the boundary of the matter cloud $\{r=r_b\}$ and thus does not provide an energy conservation law. However, it will still be very useful when studying RNOS models.

Finally, we have three independent angular Killing vector fields in both Reisnner--Nordstr\"om and RNOS spacetimes, labelled $\{\Omega_i\}_{i=1}^3$, which span all angular derivatives and are tangent to $\{r=r_b\}$. When given in $\theta$, $\varphi$ coordinates, these take the form:
\begin{align}\nonumber
\Omega_1&=\p_{\varphi}\\\label{eq:AngularKilling}
\Omega_2&=\cos\varphi\p_{\theta}-\sin\varphi\cot\theta\p_{\varphi}\\\nonumber
\Omega_3&=-\sin\varphi\p_{\theta}-\cos\varphi\cot\theta\p_{\varphi}.
\end{align}

\section{Notation}\label{Sec:Notation}
\renewcommand{\theequation}{\arabic{section}.\arabic{equation}}
We will be considering the following hypersurfaces (see Figure \ref{fig:PenHyperSurfaces}) in the manifolds $\mathcal{M}_{RN}$ and $\mathcal{M}$, equipped with normals and volume forms induced by these normals. \textbf{Note these normals will not necessarily be unit normals} (and hence the volume forms are non-standard), but have been chosen such that divergence theorem can still be applied without introducing extra factors.
\begin{align}
&\Sigma_{u_0}:=\{(t^*,r,\theta,\varphi):u(t^*,r)=u_0\}&dV=\left(1-\frac{2M}{r}+\frac{q^2M^2}{r^2}\right)r^2\sin\theta d\theta d\varphi dv\qquad&dn=-du\\
&\Sigma_{v_0}:=\{(t^*,r,\theta,\varphi):v(t^*,r)=v_0\}&dV=\left(1-\frac{2M}{r}+\frac{q^2M^2}{r^2}\right)r^2\sin\theta d\theta d\varphi du\qquad&dn=-dv\\
&\Sigma_{t_0^*}:=\{(t_0^*,r,\theta,\varphi)\}&dV=r^2\sin\theta d\theta d\varphi dr\qquad&dn=-dt
\end{align}
\begin{equation}\label{eq:SigmaBarDefinition}
\bar{\Sigma}_{t_0^*,R}:=\left(\Sigma_{u=t_0^*+R}\cap\{r^*\leq-R\}\right)\cup\left(\Sigma_{t_0^*}\cap\{r^*\in[-R,R]\}\right)\cup\left(\Sigma_{v=t_0^*+R}\cap\{r^*\geq R\}\right).
\end{equation}

The volume form of $\bar{\Sigma}_{t_0^*,R}$ matches that of $\Sigma_{u_0}$, $\Sigma_{v_0}$ and $\Sigma_{t_0^*}$ in each section.

When considering these surfaces in the RNOS model, we will also impose that $r\geq r_b(t^*)$.

We define future/past null infinity as abstract sets by:
\begin{equation}
	\mathcal{I}^+:=\R\times S^2\qquad dV=\sin\theta d\theta d\varphi du\qquad\mathcal{I}^-:=\R\times S^2\qquad dV=\sin\theta d\theta d\varphi dv.
\end{equation}

\begin{wrapfigure}{r}{8cm}
	\vspace{-1cm}
	\begin{tikzpicture}[scale =2]
		\node (I)    at ( 0,0) {};
		
		\path 
		(I) +(90:2)  coordinate[label=90:$i^+$]  (Itop)
		+(-90:2) coordinate (Imid)
		+(0:2)   coordinate[label=0:$i^0$] (Iright)
		+(-1,1) coordinate (Ileft)
		+(-0.6,1.4) coordinate[label=0:\tiny ($t^*_c$\text{, }$r_+$)] (BHH)
		+(-1,-3) coordinate[label=0:$i^-$] (Ibot)
		;
		\draw (Ileft) -- 
		node[midway, above left]    {$\mathcal{H}^+$}
		(Itop) --
		node[midway, above, sloped] {$\mathcal{I}^+$}
		(Iright) -- 
		node[midway, below, sloped] {$\mathcal{I}^-$}
		(Ibot) --
		node[midway, above, sloped]    {\small }    
		(Ileft) -- cycle;
		\draw[fill=gray!80] (Ibot) to[out=60, in=-60]
		node[midway, below, sloped] {\tiny $r=r_b$} (BHH)--(Ileft)--cycle;
		\draw (1.14,-0.86) to node[midway, above, sloped] {\tiny $\Sigma_{v_0}$} (-0.23,0.5);
		\draw (0.93,1.07) to node[midway, above, sloped] {\tiny $\Sigma_{u_0}$} (-0.14,0);
		\draw (-0.12,-0.5) to[out=0, in=180] node[midway, above, sloped] {\tiny $\Sigma_{t^*_0}$} (2,0);
		\draw (-0.23,-1.2) to (0.17,-0.8) to node[midway, above, sloped] {\tiny $\bar{\Sigma}_{t^*_0,R}$} (0.7,-0.8) to (0.95,-1.05);
	\end{tikzpicture}
	\vspace{-6mm}
	\caption{Penrose Diagram of RNOS Model, with various hyper surfaces labelled.}\label{fig:PenHyperSurfaces}
\end{wrapfigure}

Past null infinity can be viewed as the limit of $\Sigma_{u_0}$ as $u_0\to\infty$. For an appropriate function $f(u,v,\theta,\varphi)$, we will define the function ``evaluated on $\mathcal{I}^+$" to be
\begin{equation}
	f(v,\theta,\varphi)\vert_{\mathcal{I}^-}:=\lim_{u\to-\infty}f(u,v,\theta,\varphi).
\end{equation}

Similarly, $\mathcal{I}^+$ can be viewed as the limit of $\Sigma_{v_0}$ as $v_0\to\infty$. For an appropriate function $f(u,v,\theta,\varphi)$, we will define the function ``evaluated on $\mathcal{I}^+$" to be
\begin{equation}
	f(u,\theta,\varphi)\vert_{\mathcal{I}^+}:=\lim_{v\to\infty}f(u,v,\theta,\varphi).
\end{equation}

Given a vector field $X$ and a scalar $w$, we will be considering their associated energy currents, given by:
\begin{align}
T_{\mu\nu}(\phi)&=\nabla_\mu\phi\nabla_\nu\bar{\phi}-\frac{1}{2}g_{\mu\nu}\nabla^\rho\phi\nabla_\rho\bar{\phi}\\
J^X_{\mu}&=X^\nu T_{\mu\nu}\\
K^X&=\nabla^\mu J^X_\mu\\
\label{eq:modcurrent}
J^{X,w}_\mu&=X^\nu T_{\mu\nu}+w\nabla_\mu(\vert\phi\vert^2)-\vert\phi\vert^2\nabla_\mu w\\
K^{X,w}&=\nabla^\nu J^{X,w}_\nu=K^X+2w\nabla_\mu\bar{\phi}\nabla^\mu\phi-\vert\phi\vert^2\Box_gw\\
\label{eq:energydef}X\text{-energy}(\phi,S)&=\int_Sdn(J^X),
\end{align}
where $\bar{\phi}$ is the complex conjugate of $\phi$.

For $\psi$ a Schwartz function on $\mathcal{I}^{\pm}$, we define
\begin{align}
\Vert\psi\Vert^2_{L^2(\mathcal{I}^+)}&:=\int_{\mathcal{I}^+}\vert\psi\vert^2\sin\theta d\theta d\varphi dv\\
\Vert\psi\Vert^2_{L^2(\mathcal{I}^-)}&:=\int_{\mathcal{I}^-}\vert\psi\vert^2\sin\theta d\theta d\varphi du\\\Vert\psi_0\Vert^2_{\dot{H}^1(\mathcal{I}^+)}&:=\int_{\mathcal{I}^+}\vert\p_{v}\psi\vert^2\sin\theta d\theta d\varphi dv\\
\Vert\psi\Vert^2_{\dot{H}^1(\mathcal{I}^-)}&:=\int_{\mathcal{I}^-}\vert\p_{u}\psi\vert^2\sin\theta d\theta d\varphi du.
\end{align}

We will be using Fourier transforms, for which we will use the following convention: Let $f:\R\times S^2\to \mathbb{C}$. Then
\begin{equation}
\hat{f}(\omega):=\int_{x=-\infty}^\infty e^{-ix\omega}f(x)dx.
\end{equation}

\section{Previous Results on Scalar Waves in Pure \RNS Spacetimes}\label{Sec:Coeff}
\renewcommand{\theequation}{\arabic{section}.\arabic{subsection}.\arabic{equation}}

As large portions of the Hawking radiation calculation take place in subsets of pure \RN, we will be making use of several key results already proven on this family of spacetimes.

\subsection{Scattering Results on Pure \RN}
\setcounter{equation}{0}

We will first consider some basic scattering results in pure \RN:

\begin{Theorem}[Existence of Scattering Solutions in pure \RN]\label{Thm:RNExist}
	Let $\psi_+(u,\theta,\varphi)$ be a smooth function, compactly supported in $[u_-,u_+]\times S^2$ on the 3-cylinder. Then there exists a unique finite $\p_{t^*}$-energy smooth solution, $\phi(u,v,\theta,\varphi)$ to \eqref{eq:wave} in pure Reisner--Nordstr\"om spacetime $\mathcal{M}_{RN}$ such that 
	\begin{align}
	\lim_{v\to\infty}r(u,v)\phi(u,v,\theta,\varphi)&=\psi_+(u,\theta,\varphi)\\
	\lim_{u\to\infty}r(u,v)\phi(u,v,\theta,\varphi)&=0.
	\end{align}
	
	There exist functions $\psi_{RN}$ and $\psi_{\mathcal{H}^-}$ such that
	\begin{align}
	\lim_{v\to-\infty}r(u,v)\phi(u,v,\theta,\varphi)&=\psi_{\mathcal{H}^-}(u,\theta,\varphi)\\
	\lim_{u\to-\infty}r(u,v)\phi(u,v,\theta,\varphi)&=\psi_{RN}(v,\theta,\varphi).
	\end{align}
	
	Furthermore, let us consider separating $\psi_+$ into spherical harmonics $Y_{l,m}$:
	\begin{equation}
		\psi_+(u,\theta,\varphi)=\sum_{l\geq 0}\sum_{m\in\mathbb{Z}}\psi_{+l,m}(u)Y_{l,m}(\theta,\varphi).
	\end{equation}
	Then $\psi_{\mathcal{H}^+}$ and $\psi_{RN}$ can be expressed in Fourier space as
	\begin{align}\label{eq:psiH-}
		\psi_{\mathcal{H}^-}(v)&=\frac{1}{2\pi}\sum_{l,m}\int_{\omega=-\infty}^{\infty}\hat{\psi}_{+l,m}(\omega)\tilde{T}_{\omega,l,m}e^{i\omega v}d\omega\\\label{eq:psiRN}
		\psi_{RN}(u)&=\frac{1}{2\pi}\sum_{l,m}\int_{\omega=-\infty}^{\infty}\hat{\psi}_{+l,m}(\omega)\tilde{R}_{\omega,l,m}e^{i\omega u}d\omega,
	\end{align}
	where $\tilde{T}_{\omega,l,m}$ and $\tilde{R}_{\omega,l,m}$ are transmission and reflection coefficients defined below by \eqref{eq:TandRdefinition}, and $\hat{\psi}_{+l,m}$ are the Fourier transform of $\psi_{+l,m}$.
	
	Finally, $\phi(u,v,\theta,\varphi)=0$ for all $u\geq u_+$.
\end{Theorem}

\begin{proof}[Sketch of Proof]
	For the existence of the radiation fields, the sub-extremal case can be deduced from the harder sub-extremal Kerr case \cite{KerrScatter} and the extremal case can be deduced from the scattering theory in \cite{ERNScat}.
	
	We will now outline the proof of \eqref{eq:psiH-},\eqref{eq:psiRN}, the existence of reflection and transmission coefficients. For a more in depth discussion of these, we refer the reader to \cite{KerrScatter,ChandrasekharS}.
	
	We will define the transmission and reflection coefficients in the same way as \cite{KerrScatter}. We first change coordinates to the tortoise radial function, $r^*$, and then consider fixed frequency solutions of the wave equation, $\psi=e^{i\omega t}u_{\omega,m,l}(r^*)Y_{l,m}(\theta,\varphi)$. The equation obeyed by this $u_{\omega,l,m}(r^*)$ is
	\begin{equation}\label{eq:RadialODE}
		u''+(\omega^2-V_l)u=0,
	\end{equation}
	where 
	\begin{equation}
		V_l(r)=\frac{1}{r^2}\left(l(l+1)+\frac{2M}{r}\left(1-\frac{q^2M}{r}\right)\right)\left(1-\frac{2M}{r}+\frac{q^2M^2}{r^2}\right).
	\end{equation}
	
	Considering asymptotic behaviour of possible solutions, there exist unique solutions $U_{hor}$ and $U_{inf}$ \cite{ChandrasekharS}, characterised by
	\begin{align}
		U_{hor}&\sim e^{-i\omega r^*}\textrm{ as }r^*\to -\infty\\
		U_{inf}&\sim e^{i\omega r^*}\textrm{ as }r^*\to \infty.
	\end{align}
	
	We can also see that $\bar{U}_{hor}$ and $\bar{U}_{inf}$ are solutions to \eqref{eq:RadialODE}. Moreover $U_{inf}$ and $\bar{U}_{inf}$ are linearly independent, so we can write $U_{hor}$ in terms of $U_{inf}$ and $\bar{U}_{inf}$:
	\begin{equation}\label{eq:TandRdefinition}
		\tilde{T}_{\omega,l,m}U_{hor}=\tilde{R}_{\omega,l,m}U_{inf}+\bar{U}_{inf}.
	\end{equation}
	
	Here $\tilde{R}$ and $\tilde{T}$ are what we refer to as the reflection and transmission coefficients, respectively. 
	
	Now we return to physical space. For a Schwartz function $\psi_+(u)$, we can impose the future radiation field $\psi_+ (u) Y_{l,m}(\theta,\varphi)$ on $\mathcal{I}^+$, and $0$ on $\mathcal{H}^+$. Therefore the solution of the wave equation is of the form $\phi=\frac{Y_{l,m}(\theta,\varphi)}{r}\psi$. We then rewrite the wave equation in terms of $\psi$.% and take a Fourier transform with respect to the timelike coordinate $t$, where $\p_t$ is our timelike Killing vector field. We obtain that $\hat{\phi}(\omega,r^*)$ obeys \eqref{eq:RadialODE} for each fixed value of $\omega$. By considering $r^*\to-\infty$, we know that
%	\begin{equation}
%		\psi(r^*,t)=\frac{1}{2\pi}\int_{\omega=-\infty}^{\infty}\hat{\psi}(\omega,r^*)e^{i\omega t}d\omega\sim\frac{1}{2\pi}\int_{\omega=-\infty}^{\infty}\hat{\psi}_{\mathcal{H}^+}(\omega)e^{i\omega(t+r^*)}+\hat{\psi}_{\mathcal{H}^-}(\omega)e^{i\omega(t-r^*)}d\omega\textrm{ as }r^*\to-\infty.
%	\end{equation}
%	Similarly we can consider $r^*\to\infty$:
%	\begin{equation}
%		\psi(r^*,t)=\frac{1}{2\pi}\int_{\omega=-\infty}^{\infty}\hat{\psi}(\omega,r^*)e^{i\omega t}d\omega\sim\frac{1}{2\pi}\int_{\omega=-\infty}^{\infty}\hat{\psi}_{\mathcal{I}^-}(\omega)e^{i\omega(t+r^*)}+\hat{\psi}_{\mathcal{I}^+}(\omega)e^{i\omega(t-r^*)}d\omega\textrm{ as }r^*\to\infty.
%	\end{equation}
	
	Using that $\psi_{\mathcal{H}^+}=0$ and $\psi_{\mathcal{I}^+}=\psi_+ (u)$, we can formally write
	\begin{equation}
		\hat{\psi}(r^*, \omega)=\hat{\psi}_+(\omega)\tilde{T}_{\omega,l,m}U_{hor}(r^*)=\hat{\psi}_+(\omega)\left(\tilde{R}_{\omega,l,m}U_{inf}(r^*)+\bar{U}_{inf}(r^*)\right).
	\end{equation}
	
	Therefore, given appropriate convergence of $\psi_+$, we can obtain an expression of $\psi$ on $\mathcal{H}^-$ and $\mathcal{I}^-$:
	\begin{align}
		\psi_{\mathcal{H}^-}(v)&=\frac{1}{2\pi}\int_{\omega=-\infty}^{\infty}\hat{\psi}_+(\omega)\tilde{T}_{\omega,l,m}e^{i\omega v}d\omega\\
		\psi_{RN}(u)&=\frac{1}{2\pi}\int_{\omega=-\infty}^{\infty}\hat{\psi}_+(\omega)\tilde{R}_{\omega,l,m}e^{i\omega u}d\omega.
	\end{align}

	The final result follows from orthogonality of $Y_{l,m}$.
\end{proof}

We will use two properties of $\tilde{R}$ and $\tilde{T}$, given by the following proposition

\begin{Proposition}[Boundedness and Decay of Reflection and Transmission Coefficients]\label{Prop:Bounded and Decay of R, T}
	Let $\tilde{R}_{\omega,l,m}$ and $\tilde{T}_{\omega,l,m}$ be defined as in Theorem \ref{Thm:RNExist}. Then
	\begin{equation}\label{eq:T, R energy}
		\vert\tilde{R}_{\omega,l,m}\vert^2+\vert\tilde{T}_{\omega,l,m}\vert^2=1.
	\end{equation}
	\begin{equation}\label{eq:ReflectionCoeffBound}
		\vert\tilde{R}_{\omega,l,m}\vert^2\leq\frac{C(l+1)^2}{1+M^2\omega^2}.
	\end{equation}
\end{Proposition}

\begin{proof}
	The first result \eqref{eq:T, R energy} can be deduced easily from $T$ energy conservation. The second result is proven in Appendix A in \cite{AlfordFrederickThesis}.
\end{proof}

\subsection{Further Properties of \RN}
\setcounter{equation}{0}

This section introduces known results concerning solutions of the wave equation on \RNS which we will make use of.

\begin{Proposition}[$T$-energy Conservation]\label{Prop:T-energyConservation}
	Let $T=\p_{t^*}$ be the vector field in $\mathcal{M}_{RN}$. Let $\Omega\subset\mathcal{M}_{RN}$ be a compact region with a regular boundary $\p\Omega$. Let $\phi$ be a solution to \eqref{eq:wave} on $\mathcal{M}_{RN}$. Then
	\begin{equation}
		T\text{-energy}(\phi,\p\Omega)=0.
	\end{equation}
where all the normals used in the definition of $T$-energy \eqref{eq:energydef} are outward pointing.
\end{Proposition}

\begin{proof}
	This is an immediate application of divergence theorem (or Generalised Stokes' Theorem).
\end{proof}

This result also holds for sufficiently well behaved non-compact regions, if one then includes $\mathcal{I}^{\pm}$ in the boundary.

\begin{Proposition}[Scattering $T$-energy Conservation]\label{Prop:T-energyConservationScattering}
	Let $T=\p_{t^*}$ be the vector field in $\mathcal{M}_{RN}$. Let $\Omega\subset\mathcal{M}_{RN}$ be a non-compact region with a regular boundary $\p\Omega$, equal to a finite union of regions of $\mathcal{I}^\pm$, $\mathcal{H}^\pm$, $\Sigma_{t^*}$, $\Sigma_u$ and $\Sigma_v$. Let $\phi$ be a solution to \eqref{eq:wave} on $\mathcal{M}_{RN}$. Then
	\begin{equation}
		T\text{-energy}(\phi,\p\Omega)=0.
	\end{equation}
	where all the normals used in the definition of $T$-energy \eqref{eq:energydef} are outward pointing.
\end{Proposition}

\begin{proof}
	This is a consequence of the scattering map \cite{KerrScatter}. $T$-energy conservation allows us to obtain this result, provided that there is some decay on the energy of our solutions towards $i^\pm$.
\end{proof}

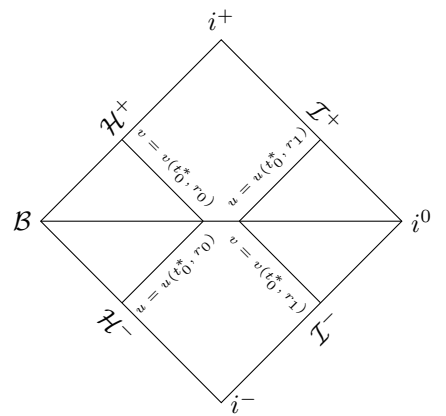
\begin{wrapfigure}{r}{5cm}\vspace{-5mm}
	\begin{tikzpicture}[scale =1.2]
		\node (I)    at ( 0,0) {};
		
		\path 
		(I) +(0,2)  coordinate[label=90:$i^+$]  (Itop)
		+(0,-2) coordinate (Imid)
		+(2,0)   coordinate[label=0:$i^0$] (Iright)
		+(-2,0) coordinate[label=180 :$\mathcal{B}$] (Ileft)
		+(0,-2) coordinate[label=0:$i^-$] (Ibot)
		;
		\draw (Ileft) -- 
		node[midway, above, sloped]    {$\mathcal{H}^+$}
		(Itop) --
		node[midway, above, sloped] {$\mathcal{I}^+$}
		(Iright) -- node[midway, below, sloped]    {$\mathcal{I}^-$}
		(Ibot) --
		node[midway, below, sloped]    {$\mathcal{H}^-$} 
		(Ileft) -- cycle;
		\draw (-2,0) to node[midway, above, sloped] {} (2,0);
		\draw (-0.2,0) to node[midway, above, sloped] {\tiny $v=v(t^*_0,r_0)$ } (-1.1,0.9);
		\draw (-0.2,0) to node[midway, below, sloped] {\tiny $u=u(t^*_0,r_0)$ } (-1.1,-0.9);
		\draw (0.2,0) to node[midway, above, sloped] {\tiny $u=u(t^*_0,r_1)$ } (1.1,0.9);
		\draw (0.2,0) to node[midway, below, sloped] {\tiny $v=v(t^*_0,r_1)$ } (1.1,-0.9);
	\end{tikzpicture}
	\caption{The Domain of Dependence}\label{fig:Domain}
\end{wrapfigure}

\begin{Theorem}[Domain of Dependence of the wave equation]\label{Thm:Domain of Dependence}
	Let $\phi(t_0,r,\theta,\varphi)$ be a smooth solution of \eqref{eq:wave} on pure Reisner--Nordstr\"om spacetime $\mathcal{M}_{RN}$, such that on surface $\Sigma_{t_0}$, $\phi(t_0, r, \theta, \varphi)$ and $\nabla\phi(t_0,r,\theta,\varphi)$ is supported on $r\in[r_0,r_1]$.
	
	Then $\phi$ vanishes in the $4$ regions $\{t>t_0\}\cap\{v\leq v(t_0,r_0)\}$, $\{t>t_0\}\cap\{u\leq u(t_0,r_1)\}$, $\{t<t_0\}\cap\{v\geq v(t_0,r_1)\}$ and $\{t<t_0\}\cap\{u\geq u(t_0,r_0)\}$. (See Figure \ref{fig:Domain}.)
\end{Theorem}

In this case, this result is a trivial consequence of $T$-energy conservation. However, it can also be derived from more general domain of dependence results concerning hyperbolic PDEs.

The final result we will be using is (part of) Proposition $7.4$ in \cite{ERNScat}, which we will restate here:

\begin{Proposition}\label{Prop:Non-DegenerateScatering}
	Let $\phi$ be a sufficiently well behaved solution to \eqref{eq:wave} in an extremal \RNS spacetime $\mathcal{M}_{RN}$, with $\psi=r\phi$. Let $M<r_0<2M$. Then there exists a constant, $C=C(M, r_0)>0$ such that
	\begin{align}\nonumber
	\int_{\Sigma_{u_1}\cap\{r\leq r_0\}}&\left(1-\frac{r}{M}\right)^{-2}\vert\p_v\psi\vert^2\sin\theta d\theta d\varphi dv+\int_{\Sigma_{v=u_1+r^*_0}\cap\{r^*>0\}}\vert\p_u\psi\vert^2\sin\theta d\theta d\varphi du\\
	&\leq C\int_{\mathcal{H}^-\cap\{u\leq u_1\}}\left(M^2+(u-u_1)^2\right)\vert\p_u\psi\vert^2+\left\vert\mathring{\slashed\nabla}\psi\right\vert^2\sin\theta d\theta d\varphi du\\\nonumber
	&\qquad+C\int_{\mathcal{I}^-\cap\{v\leq u_1\}}(M^2+(v-u_1-r^*_0)^2)\vert\p_v\psi\vert^2+\left\vert\mathring{\slashed\nabla}\psi\right\vert^2\sin\theta d\theta d\varphi dv
	\end{align}
	for all $u_1\in\R$.
\end{Proposition}

\begin{proof}
	To prove this, we have taken the $r_{\mathcal{I}^+}$ in the original statement of the proposition in \cite{ERNScat} to be where $r^*=0$.
\end{proof}

Section \ref{Sec:Integrated Error Terms} will be concerned with a sub-extremal equivalent of this result, proven in a very similar manner to \cite{ERNScat}.

\section{Classical Scattering in RNOS spacetimes}\label{Sec:RNOSScat}
\renewcommand{\theequation}{\arabic{section}.\arabic{equation}}
We now return to our collapsing exterior spacetime, $\mathcal{M}$. In order to discuss properties of solutions to the linear wave equation \eqref{eq:wave}, we first need a result on the existence of such solutions, for which we will use a result from \cite{Mine2}:

\begin{Theorem1}[Existence of Smooth Solutions to the Linear Wave Equation on RNOS]\label{Thm:Existence}
	Let $\psi_+(u,\theta,\varphi)$ be a smooth, compactly supported function on $\mathcal{I}^+$. Fix an RNOS spacetime, $(\mathcal{M},g)$. Then there exists a unique finite $\p_{t^*}$-energy $\phi\in C^{\infty}(\mathcal{M})$ such that $\phi$ is a solution to \eqref{eq:wave} on $\mathcal{M}$, $\phi$ vanishes on $r=r_b(t^*)$, and
	\begin{align}
	\lim_{v\to\infty}r(u,v)\phi(u,v,\theta,\varphi)&=\psi_+(u,\theta,\varphi)\\
	\lim_{u\to\infty}r(u,v)\phi(u,v,\theta,\varphi)&=0\quad \forall v\geq v_c.
	\end{align}
	
	Moreover, if we define
	\begin{equation}
	\psi_-(v,\theta,\varphi):=\lim_{u\to -\infty}r(u,v)\phi(u,v,\theta,\varphi),
	\end{equation}
	then 
	\begin{equation}\label{eq:FiniteI-Energy}
	\int_{\mathcal{I}^-}\vert\p_v\psi_-\vert^2\sin\theta d\theta d\varphi dv<\infty.
	\end{equation}
	
	Finally, we have the following global conservation law:
	\begin{equation}\label{eq:ConservedCurrent}
	\int_{\mathcal{I}^+}\psi_+\p_u\bar{\psi}_+-\bar{\psi}_+\p_u\psi_+=\int_{\mathcal{I}^-}\psi_-\p_u\bar{\psi}_--\bar{\psi}_-\p_u\psi_-.
	\end{equation}
\end{Theorem1}
\begin{proof}
	Here, existence of a unique, smooth solution to \eqref{eq:wave} is given by Theorem $7.1$ and $7.2$ in \cite{Mine2}. To obtain \eqref{eq:FiniteI-Energy}, we note that the solution $\phi$ must remain supported away from $r=r_+$, by a domain of dependence argument. As $\phi$ is supported away from $r=r_+$, we have that $\phi$ has finite non-degenerate energy on $\Sigma_{t^*_c}$, and thus by Theorem $7.1$ in \cite{Mine2}, the integral in \eqref{eq:FiniteI-Energy} converges.
	
	The current
	\begin{equation}
	\phi\nabla_{\mu}\bar\phi-\bar{\phi}\nabla_{\mu}\phi
	\end{equation}
	is divergence free by \eqref{eq:wave}, and therefore applying divergence theorem in the region $\{u\geq u_-, v\leq v_+\}$ and taking limits as $u_-\to-\infty, v_+\to\infty$ gives us \eqref{eq:ConservedCurrent}. Taking this limit makes use of decay rates given in \cite{Mine2}.
\end{proof}

\section{Convergence Rates of the $\dot{H}^{1/2}$ Norm Controlled by Integrated Error Terms ($I.E.$)}\label{Sec:H^1/2 Norm}
\renewcommand{\theequation}{\arabic{section}.\arabic{equation}}

To motivate this section, we consider the following Lemma:

\begin{Lemma}[Reduction to $\dot{H}^{1/2}$ Norm]\label{Lem:Reduction}
	For $\alpha$, $\psi_+$, $\psi_-$ as in the statement of Theorem \ref{Thm:HawkingVague}, we have that
	\begin{equation}
		2\left\vert\int_{\omega=-\infty}^0\vert\omega\vert\vert\hat{\psi}_-\vert^2 -\int_{\omega=-\infty}^\infty\alpha\vert\omega\vert\vert\hat{\psi}_{\mathcal{H}^-}\vert^2\right\vert=\left\vert\int_{\omega=-\infty}^\infty\vert\omega\vert\vert\hat{\psi}_-\vert^2 -\int_{\omega=-\infty}^\infty\vert\omega\vert\left(\coth\left(\frac{\pi}{\kappa}\vert\omega\vert\right)\vert\hat{\psi}_{\mathcal{H}^-}\vert^2+\vert\hat{\psi}_{RN}\vert^2\right)\right\vert.
	\end{equation}
	Here, when $\kappa=0$, we interpret $\coth\left(\frac{\pi}{\kappa}\vert\omega\vert\right)$ as identically $1$.
\end{Lemma}

\begin{proof}
	We first note that
	\begin{equation}\label{eq:ParticleCurrent}
		\int_{\omega=-\infty}^\infty\omega\vert\hat{\psi}_+\vert^2d\omega=\frac{i}{2}\int_{\mathcal{I}^+}\bar{\psi}\nabla\psi-\psi\nabla\bar{\psi}du=\frac{i}{2}\int_{\mathcal{I}^-}\bar{\psi}\nabla\psi-\psi\nabla\bar{\psi}dv=\int_{\omega=-\infty}^\infty\omega\vert\hat{\psi}_-\vert^2d\omega,
	\end{equation}
	by \eqref{eq:ConservedCurrent} in Theorem \ref{Thm:Existence}.
	
	Thus, we have
	\begin{align}
		2\int_{-\infty}^0 \vert \omega \vert \vert\hat{\psi}_-\vert^2 &= \int_{-\infty}^\infty \left(\vert \omega \vert-\omega\right) \vert\hat{\psi}_-\vert^2=\int_{-\infty}^\infty\vert \omega \vert \vert\hat{\psi}_-\vert^2-\int_{-\infty}^\infty \omega \vert\hat{\psi}_+\vert^2\\\nonumber
		&=\int_{-\infty}^\infty\vert \omega \vert \vert\hat{\psi}_-\vert^2-\int_{-\infty}^\infty\vert \omega \vert \vert\hat{\psi}_+\vert^2,
	\end{align}
	using \eqref{eq:ParticleCurrent} and that $\hat{\psi}_+$ is only supported on $\omega\geq 0$.
	
	Combining Theorem \ref{Thm:Existence} and Proposition \ref{Prop:Bounded and Decay of R, T}, we can see that
	
	\begin{equation}
		\int_{-\infty}^\infty\vert \omega \vert \vert\hat{\psi}_+\vert^2=\int_{-\infty}^\infty\vert \omega \vert\left( \vert\hat{\psi}_{\mathcal{H}^-}\vert^2+\vert\hat{\psi}_{RN}\vert^2\right)
	\end{equation}
	
	Finally, we note that $2\alpha=\coth \left(\frac{\pi\vert\omega\vert}{\kappa}\right) -1$ $\forall \omega, q$, which gives us the result.
\end{proof}

This Lemma reduces Theorem \ref{Thm:HawkingVague} to a result on the limit of the $\dot{H}^{1/2}$ norm, given by
\begin{equation}\label{eq:H1/2norm}
\int_{\omega=-\infty}^\infty\int_{S^2}\vert\omega\vert\vert\hat{\psi}_{-,u_0}\vert^2\sin\theta d\theta d\varphi d\omega,
\end{equation}
where $\hat{f}$ is the Fourier transform of $f$ with respect to its non-angular coordinate. This will be the hardest part of the proof of Theorem \ref{Thm:HawkingVague}.

We define the $I.T.$ (Integrated Terms) and $I.E.$ (Integrated Errors due to the tails of $\psi_+$, $\psi_{\mathcal{H}^-}$ and $\psi_{RN}$), as mentioned in the title. These are given by
\begin{align}\nonumber
	I.T.[\psi_+]&=\int_{-\infty}^{\infty}\int_{S^2}(M^2+u^2)\left(1+\left\vert\mathring{\slashed\nabla}\right\vert^4\right)\left(\vert\p_u\psi_+\vert^2+\vert\psi_+\vert\right)\sin\theta d\theta d\varphi du\\\nonumber
	I.E.[\psi_+,v_c,u_1,u_0]&=\int_{-\infty}^{u_1}\int_{S^2}\left[(M^2+(u_0-u)^2)\left(1+\left\vert\mathring{\slashed\nabla}\right\vert^4\right)(\vert\p_u\psi_{\mathcal{H}^-}\vert^2+\vert\psi_{\mathcal{H}^-}\vert^2)\right]\sin\theta d\theta d\varphi du\\
	&\qquad +\int_{-\infty}^{v_c}\int_{S^2}\left[(M^2+(v_c-v)^2)\left(1+\left\vert\mathring{\slashed\nabla}\right\vert^4\right)(\vert\p_v\psi_{RN}\vert^2+\vert\psi_{RN}\vert^2)\right]\sin\theta d\theta d\varphi dv \\\nonumber
	&\qquad +\int_{u=2u_0-u_1}^{\infty}\int_{S^2}(M^2+(u-u_0+u_1)^2)\left(1+\left\vert\mathring{\slashed\nabla}\right\vert^4\right)(\vert\p_u\psi_+\vert^2+\vert\psi_+\vert^2)\sin\theta d\theta d\varphi du
\end{align}
Here, $\mathring{\slashed\nabla}$ is the derivative on the unit sphere, and we write $\left\vert\mathring{\slashed\nabla}\right\vert^4\vert f\vert^2$ to mean $\left\vert\mathring{\slashed\nabla}^2f\right\vert^2$. Note that $I.T.[\psi_+]$ controls similarly weighted norms of $\psi_{\mathcal{H}^-}$ and $\psi_{RN}$, thanks to reflection and transmission coefficients being bounded above by $1$ (see Section \ref{Sec:Coeff}).

Finally, throughout this section, we will use the following notation:
\begin{align}\label{eq:deltar}
	\delta r &= r(u_1, v_c)-r_+\\\label{eq:langleu}
	\lur &= (M^2 +(u_0-u_1)^2)^{1/2}
\end{align}

This allows us to state the main Theorem of this section:
\begin{Theorem1}[Limit of the $\dot{H}^{1/2}$ Norm]\label{Thm:Hawking}
	Let $\psi_+(u, \theta, \varphi)$ be a Schwartz function on the $\R\times S^2$, with $\hat{\psi}_+$ only supported on positive frequencies. Let $\phi$ be the solution of \eqref{eq:wave} on an RNOS background $\mathcal{M}$, with associated $M,q,r_b$, such that
	\begin{align}
	\lim_{v\to\infty}r(u,v)\phi(u,v,\theta,\varphi)&=\psi_{+,u_0}:=\psi_+(u-u_0,\theta,\varphi)\\
	\lim_{u\to\infty}r(u,v)\phi(u,v,\theta,\varphi)&=0\quad \forall v\geq v_c,
	\end{align}
	as given by Theorem \ref{Thm:Existence}.
	
	Define the function $\psi_{-,u_0}$ by
	\begin{equation}
	\lim_{u\to-\infty}r(u,v)\phi(u,v,\theta,\varphi)=\psi_{-,u_0}(v,\theta,\varphi).
	\end{equation}
	Then there exists a constant $A(M, r_b)$ independent of $q$ such that
	\begin{align}\label{eq:Hawking}
	\Bigg\vert\int_{\omega=-\infty}^\infty\int_{ S^2}\vert\omega\vert\vert\hat{\psi}_{-,u_0}&(\omega, \theta,\varphi)\vert^2 \sin\theta d\omega d\theta d\varphi\\\nonumber &-\int_{\omega=-\infty}^\infty\int_{ S^2}\vert\omega\vert\left(\coth\left(\frac{\pi}{\kappa}\vert\omega\vert\right)\vert\hat{\psi}_{\mathcal{H}^-}(\omega,\theta,\varphi)\vert^2+\vert\hat{\psi}_{RN}(\omega,\theta,\varphi)\vert^2\right) d\omega \sin\theta d\omega d\theta d\varphi\Bigg\vert\\\nonumber
	&\qquad\leq A\left(\frac{\log{u_0}}{u_0^2}I.T.[\psi_+]+u_0^{2}I.E.[\psi_+,v_c,u_1,u_0]\right),
	\end{align}
	for sufficiently large $u_0$ and for $u_1$ such that $1-\frac{2M}{r}+\frac{q^2M^2}{r^2}\vert_{u_1,v_c} = B u_0^{-2}$ for fixed $B$. Here, when $\kappa=0$, we interpret $\coth\left(\frac{\pi}{\kappa}\vert\omega\vert\right)$ as identically $1$.

	Here $\psi_{\mathcal{H}^-}$ and $\psi_{RN}$ are the transmission and reflection of $\psi_+$ in pure \RNS spacetime, as defined by Theorem \ref{Thm:RNExist}, $\kappa$ is the surface gravity defined in \eqref{eq:kappa}.

	In the case $\vert q\vert <1$, there exists a constant $A(\mathcal{M})$ such that
	\begin{align}\label{eq:ExHawking}
	\Bigg\vert\int_{\omega=-\infty}^\infty\int_{ S^2}\vert\omega\vert\vert\hat{\psi}_{-,u_0}&(\omega, \theta,\varphi)\vert^2 \sin\theta d\omega d\theta d\varphi\\\nonumber &-\int_{\omega=-\infty}^\infty\int_{ S^2}\vert\omega\vert\left(\coth\left(\frac{\pi}{\kappa}\vert\omega\vert\right)\vert\hat{\psi}_{\mathcal{H}^-}(\omega,\theta,\varphi)\vert^2+\vert\hat{\psi}_{RN}(\omega,\theta,\varphi)\vert^2\right)2 d\omega \sin\theta d\omega d\theta d\varphi\Bigg\vert\\\nonumber
	&\qquad\leq A\left(e^{-\kappa u_1}I.T.[\psi_+]+e^{\kappa u_1}I.E.[\psi_+,v_c,u_1,u_0]\right),
	\end{align}
	for sufficiently large $u_0$ and $u_1$.
\end{Theorem1}

When using Theorem \ref{Thm:Hawking} to prove Theorem \ref{Thm:HawkingVague}, we have to choose $u_1$ large enough to obtain decay of the bulk term $I.T.$, but small enough to ensure decay of the $I.E.$ terms, despite the factor $e^{2\kappa u_1}$ factor in front of $I.E.$

\subsection{The Set-up and the Reduction to Fixed Spherical Harmonic $l$}
\renewcommand{\theequation}{\arabic{section}.\arabic{subsection}.\arabic{equation}}
\setcounter{equation}{0}

\begin{wrapfigure}{r}{5cm}
	\vspace{-2cm}
	\begin{tikzpicture}[scale =1.2]
	\node (I)    at ( 0,0) {};
	
	\path 
	(I) +(90:2)  coordinate[label=90:$i^+$]  (Itop)
	+(-90:2) coordinate (Imid)
	+(0:2)   coordinate[label=0:$i^0$] (Iright)
	+(-1,1) coordinate (Ileft)
	+(-0.6,1.4) coordinate[label=0:\tiny ($t^*_c$\text{, }$r_+$)] (BHH)
	+(-1,-3) coordinate[label=0:$i^-$] (Ibot)
	;
	\draw (Ileft) -- 
	node[midway, above, sloped]    {$\mathcal{H}^+, \phi=0$}
	(Itop) --
	node[midway, above, sloped] {$\mathcal{I}^+, r\phi\to Y_{l,m}\psi_+$}
	(Iright) -- 
	node[midway, below, sloped] {$\mathcal{I}^-, r\phi\to\psi_-Y_{l,m}$}
	(Ibot) --
	node[midway, above, sloped]    {\small }    
	(Ileft) -- cycle;
	\draw[fill=gray!80] (Ibot) to[out=60, in=-60]
	node[midway, below, sloped] {\tiny $r=r_b, \phi=0$} (BHH)--(Ileft)--cycle;
	\draw[->] (0.7,1) to[out=-130,in=130] (0.7,-1);
	\end{tikzpicture}
	\caption{The set-up for the Hawking radiation calculation}\label{fig:HawkingSet-up}
\end{wrapfigure}
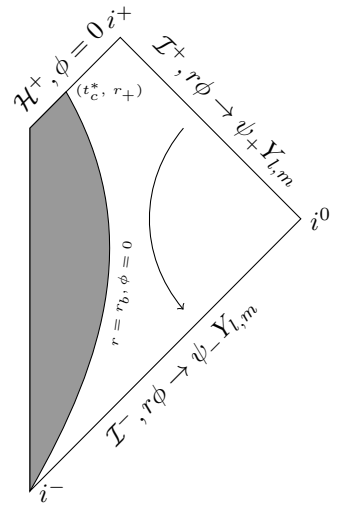

We will prove Theorem \ref{Thm:Hawking} by first restricting to the case $\psi_+(u-u_0,\theta,\varphi)=\psi_+(u-u_0)Y_{l,m}(\theta,\varphi)$. Here $Y_{l,m}$ is spherical harmonic, see for example \cite{SphericalHarmonics}. The full result will then follow from orthogonality of $Y_{l,m}$ and the fact that we will track dependence of constants on $l,m$. Let $\phi$ be the solution to $\eqref{eq:wave}$, subject to $\psi=0$ on $r=r_b(t^*)$, with future radiation field $Y_{l,m}(\theta, \varphi)\psi_+(u-u_0)$, and $\phi=0$ on $\mathcal{H}^+$, as given by Theorem \ref{Thm:Existence}.

We will generally be considering $\psi(u,v)$, given by
\begin{equation}
\psi(u,v)Y_{l,m}(\theta,\varphi):=r(u,v)\phi(u,v,\theta,\varphi)
\end{equation}
rather than $\phi$ itself. Note $\psi(u,v)$ is independent of $\theta, \varphi$, as spherical symmetry of our system implies that if we restrict scattering data in Theorems \ref{Thm:RNExist} and \ref{Thm:Existence} to one spherical harmonic, the solution will also be restricted to that harmonic.

Re-writing the wave equation for fixed $l,m$ in terms of $\psi$, we obtain:
\begin{equation}\label{eq:RadWave}
4\p_u\p_v\psi=-\frac{1}{r^2}\left(1-\frac{2M}{r^2}+\frac{q^2M^2}{r^2}\right)\left(l(l+1)+\frac{2M}{r}\left(1-\frac{q^2M}{r}\right)\right)\psi=:-4V(r)\psi.
\end{equation}
\begin{equation}\label{eq:RadBoundary}
\psi(u,v_b(u))=0,
\end{equation}
where $v_b$ is as given in \eqref{eq:vbDef}. The past radiation field will be denoted by $\psi_-$, and the backwards evolution we will be studying is demonstrated in Figure \ref{fig:HawkingSet-up}.

Note that $\psi$, $\phi$, $\psi_-$ all depend on $u_0$.

\begin{Remark}[Keeping Track of the $l$s]
	There will be several proofs in this section which are standard calculations in the literature. We have included these calculations as we will be required to keep count of factors of $l$ throughout - this will allow the final bound to be given in terms of an explicit energy.
\end{Remark}
\pagebreak
\subsection{The Proof of $\dot{H}^{1/2}$ Convergence}\label{Sec:SummaryOfProof}
\setcounter{equation}{0}

\columnratio{0.7}
\begin{paracol}{2}
	\begin{proof}[Proof of Theorem \ref{Thm:Hawking}]\renewcommand{\qedsymbol}{}
	We prove this Theorem by breaking the result up into $4$ parts:
	\begin{align}\nonumber
		\Bigg\vert\int\vert\omega\vert\vert\hat{\psi}_{-,u_0}&\vert^2-\int\vert\omega\vert\left(\coth\left(\frac{\pi}{\kappa}\vert\omega\vert\right)\vert\hat{\psi}_{\mathcal{H}^-}\vert^2+\vert\hat{\psi}_{RN}\vert^2\right)\Bigg\vert\\\label{eq:ApproxAns1}
		\leq&\left\vert\int\vert\omega\vert\coth\left(\frac{\pi}{\kappa}\vert\omega\vert\right)\left(\vert\hat{\psi}_0\vert^2-\vert\hat{\psi}_{\mathcal{H}^-}\vert^2\right)\right\vert\\\label{eq:ApproxAns3}	
		&+\left\vert\int\left(\vert\omega\vert\coth\left(\frac{\pi}{\kappa}\vert\omega\vert\right)\vert\hat{\psi}_0\vert^2 -\vert\omega\vert\vert\widehat{\psi_0\circ u_b}\vert^2\right)\right\vert\\\label{eq:ApproxAns2}
		&+\left\vert\int\vert\omega\vert\left(\vert\widehat{\psi_0\circ u_b}\vert^2-\vert\hat{\psi}_1\vert^2\right)\right\vert\\\label{eq:ApproxAns4}	&+\left\vert\int\vert\omega\vert\left(\vert\hat{\psi}_{-,u_0}\vert^2-\vert\hat{\psi}_{RN}\vert^2-\vert\hat{\psi}_1\vert^2\right)\right\vert,
	\end{align}
	where the functions $\psi_0$ and $\psi_1$ have yet to be chosen.
	
	\begin{enumerate}
	\item Section \ref{Sec:Evolution in Pure Reissner--Nordstrom} will define $\psi_0$ as the value of $\psi$ on the surface $\{v=v_c\}$, cut off for $u\leq u_1$. Corollary \ref{Cor:FinalPure} in this section will then bound \eqref{eq:ApproxAns1} by considering evolution in the region $R_1=\{v\geq v_c\}$ in Figure \ref{fig:HawkingRegions}. For this one must first note that
	\begin{equation*}
		\vert\omega\vert \coth\left(\frac{\pi}{\kappa}\vert\omega\vert\right)\leq \frac{\kappa}{\pi}+\vert\omega\vert.
	\end{equation*}
\end{enumerate}
\end{proof}
	\switchcolumn
	\begin{figure}\vspace{-5mm}
		\begin{tikzpicture}[scale =1.2]
		\node (I)    at ( 0,0) {};
		
		\path 
		(I) +(90:2)  coordinate[label=90:$i^+$]  (Itop)
		+(-90:2) coordinate (Imid)
		+(0:2)   coordinate[label=0:$i^0$] (Iright)
		+(-1,1) coordinate (Ileft)
		+(-0.6,1.4) coordinate[label=0: $R_2$] (BHH)
		+(-1,-3) coordinate[label=0:$i^-$] (Ibot)
		+(0.5,0.8) coordinate[label=0:$R_1$] (R1)
		+(0.4,-0.6) coordinate[label=0:$R_3$] (R3)
		;
		\draw (Ileft) -- 
		node[midway, above, sloped]    {$\mathcal{H}^+, \phi=0$}
		(Itop) --
		node[midway, above, sloped] {$\mathcal{I}^+, r\phi\to Y_{l,m}\psi_+$}
		(Iright) -- 
		node[midway, below, sloped] {$\mathcal{I}^-, r\phi\to\psi_-Y_{l,m}$}
		(Ibot) --
		node[midway, above, sloped]    {\small }    
		(Ileft) -- cycle;
		\draw[fill=gray!80] (Ibot) to[out=60, in=-60]
		node[midway, below, sloped] {\tiny $r=r_b$} (BHH)--(Ileft)--cycle;
		\draw (BHH) to node[midway, above, sloped] {\tiny $v=v_c$} (1.4,-0.6);
		\draw (-0.2,1) to node[midway, above, sloped] {} (-0.35,0.85);
		\draw[->] (-0.3,1.25) to (-0.3,1);
	\end{tikzpicture}
\caption{The regions we will consider in the Hawking radiation calculation}\label{fig:HawkingRegions}
\end{figure}
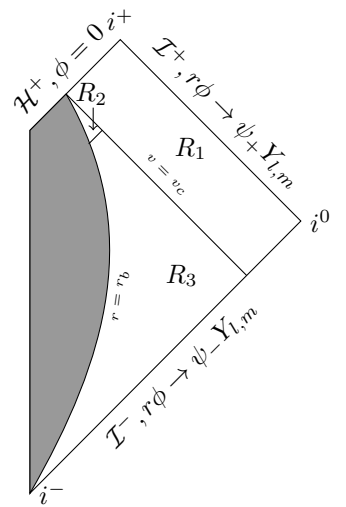
\end{paracol}
\begin{proof}[\unskip\nopunct]\vspace{-7mm}
	\begin{enumerate}
	\setcounter{enumi}{1}
		\item Section \ref{Sec:Bachelot} will generalise a useful Lemma by Bachelot (Lemma II.6 in \cite{BachelotLemma}, Lemmas \ref{Lemma:Bachelot} and \ref{Lemma:ExtremalBachelot} here), which allows Proposition \ref{Prop:Bachelot&Errors} to bound \eqref{eq:ApproxAns3}.
		\item Section \ref{Sec:Reflection} will define $\psi_1$ to be the value of $\psi$ on the surface $\{u=u_1\}$, cut off for $v>v_c$. Proposition \ref{Prop:FinalReflection} will then bound \eqref{eq:ApproxAns2} by considering the reflection of the solution off the surface of the matter cloud in region $R_2:=\{v\leq v_c, u\geq u_1\}\cap\{r\geq r_b\}$.
		\item Section \ref{Sec:HighFrequency} will consider the high frequency transmission of the solution from near the surface of the matter cloud to $\mathcal{I}^-$. This will occur in the region labelled $R_3:=\{v\leq v_c, u\leq u_1\}$ in Figure \ref{fig:HawkingRegions}. This will allow Proposition \ref{Prop:FinalHighFrequency} to bound \eqref{eq:ApproxAns4}.
	\end{enumerate}
\end{proof}

Although many aspects of this proof are firmly rooted in Fourier space, where possible we will use physical space calculations. We hope this will lead to a more transparent proof.

\subsection{Evolution in Pure \RN}\label{Sec:Evolution in Pure Reissner--Nordstrom}
\setcounter{equation}{0}

\begin{wrapfigure}{r}{5cm}
	\begin{tikzpicture}[scale =1.2]
		\node (I)    at ( 0,0) {};
		
		\path 
		(I) +(0,2)  coordinate[label=90:$i^+$]  (Itop)
		+(0,-2) coordinate (Imid)
		+(2,0)   coordinate[label=0:$i^0$] (Iright)
		+(-2,0) coordinate[label=180 :$\mathcal{B}$] (Ileft)
		+(0,-2) coordinate[label=0:$i^-$] (Ibot)
		+(0.1,0.3) coordinate[label=0:$R_1$] (R1)
		;
		\draw (Ileft) -- 
		node[midway, above, sloped]    {$\mathcal{H}^+, \phi=0$}
		(Itop) --
		node[midway, above, sloped] {$\mathcal{I}^+, r\phi\to Y_{l,m}\psi_+$}
		(Iright) -- node[midway, below, sloped]    {$\mathcal{I}^-$}
		(Ibot) --
		node[midway, below, sloped]    {$\mathcal{H}^-$} 
		(Ileft) -- cycle;
		\draw (-1.5,0.5) to node[midway, above, sloped] {\tiny $v=v_c, r\phi=Y_{l,m}\psi $} (0.5,-1.5);
	\end{tikzpicture}
	\caption{The first region we will consider in the Hawking radiation calculation}\label{fig:R1}
\end{wrapfigure}
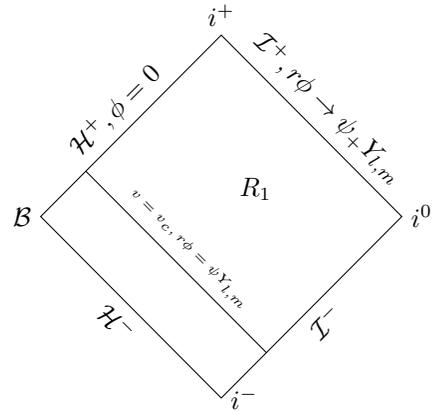

In this section we will be considering the following problem: In pure \RNS spacetime, if we impose radiation field $\psi_+(u)Y_{l,m}(\theta,\varphi)$ on $\mathcal{I}^+$ and that our solution vanishes on $\mathcal{H}^+$, what happens to the solution on a surface of constant $v$ as we let $v\to-\infty$? See Figure \ref{fig:R1}. The aim of this section will be to prove the following result:

\begin{Proposition}[$H^{1/2}$ Error from \RNS Transmission]\label{Prop:FinalPure}
	Let $\chi:\R\to\R$ be a smooth, cut off function such that
	\begin{equation}
		\chi(x)\begin{cases}
			=1 & x\geq0\\
			\in[0,1] & x\in[-1,0] \\
			=0 & x\leq -1
		\end{cases}.
	\end{equation}
	 Let $\psi_+:\mathbb{R}\to\mathbb{C}$ be a Schwartz function, with $\hat{\psi}_+$ supported on positive frequencies. Let $\psi$ be the solution of \eqref{eq:RadWave}, as given by Theorem \ref{Thm:RNExist}, with radiation field on $\mathcal{I}^+$ equal to $Y_{l,m}\psi_+$, and which vanishes on $\mathcal{H}^+$. Let $u_0$ be fixed. Let $v_c,u_1\in\R$, with $r(u_1,v_c)\leq 8M/3$ and $u_1<u_0$. Define
	\begin{equation}
		\psi_0(u,v):=\chi\left(\frac{u-u_1}{M}\right)\psi(u,v),
	\end{equation}
	that is $\psi_0$ is $\psi$, but smoothly cut-off when $u\leq u_1$.
	
	Then there exists a constant $A(M,\chi)>0$ such that
	\begin{equation}\label{eq:H1/2 near H-}
		\left\vert\int_{\omega=-\infty}^\infty\left(\kappa+\vert\omega\vert\right)\left(\vert\hat{\psi}_{\mathcal{H}^-}\vert^2-\vert\hat{\psi}_0\vert^2\right) d\omega\right\vert\leq A\delta r(\kappa+\delta r)\lur \IT+ A\left(1+\delta r\lur^3\right)\IE
	\end{equation}
	where $\kappa$, $\delta r$ and $\lur$ are given by \eqref{eq:kappa}, \eqref{eq:deltar} and \eqref{eq:langleu} respectively. Here $\IE$ and $I.T.[Y_{l,m}\psi_+,v_c,u_1,u_0]$ are both given in the statement of Theorem \ref{Thm:Hawking} and $\psi_{RN}, \psi_{\mathcal{H}^-}$ are as defined in Theorem \ref{Thm:RNExist}.
\end{Proposition}
\begin{Remark}
	We have chosen the above form of $\psi_0$ to be arbitrarily polynomially decaying for $\vert u\vert\to \infty$. This ensures that the integral in \eqref{eq:H1/2 near H-} converges.
\end{Remark}

The result actually used will be an immediate Corollary of Proposition \ref{Prop:FinalPure}. By propagating our solution along the Killing vector, $T$, by distance $u_0-v_c$, Proposition \ref{Prop:FinalPure} is equivalent to considering a solution with radiation field $\psi_+(u-u_0)Y_{l,m}(\theta,\varphi)$ on $\mathcal{I}^+$ on a surface of fixed $v$, and allowing $u_0\to\infty$:

\begin{Corollary}[$H^{1/2}$ Error from \RNS Transmission]\label{Cor:FinalPure}
	Let $\chi$ and $\psi_+$ be as in Proposition \ref{Prop:FinalPure}. Let $\psi$ be the solution of \eqref{eq:RadWave}, as given by Theorem \ref{Thm:RNExist}, with radiation field on $\mathcal{I}^+$ equal to $Y_{l,m}(\theta,\varphi)\psi_+(u-u_0)$, and which vanishes on $\mathcal{H}^+$. Let $v_c$ be fixed. Let $u_0,u_1\in\R$, with $r(u_1,v_c)\leq 8M/3$ and $u_1<u_0$. Define
	\begin{equation}
	\psi_0(u):=\chi\left(\frac{u-u_1}{M}\right)\psi(u,v_c)
	\end{equation}
	Then there exists a constant $A(M,\chi)>0$ such that
	\begin{align}
		\left\vert\int_{\omega=-\infty}^\infty\left(\kappa+\vert\omega\vert\right)\left(\vert\hat{\psi}_{\mathcal{H}^-}\vert^2-\vert\hat{\psi}_0\vert^2\right) d\omega\right\vert &\leq A\delta r(\kappa+\delta r)\lur\IT\\\nonumber
		&\qquad+A(1+\delta r\lur^3)\IE.
	\end{align}
	Here $\IE$ and $I.T.[Y_{l,m}\psi_+,v_c,u_1,u_0]$ are both given in the statement of Theorem \ref{Thm:Hawking}.
\end{Corollary}

\begin{proof}
	We apply Proposition \ref{Prop:FinalPure} to a time translation of $\psi$.
\end{proof}

We will start the proof of Proposition \ref{Prop:FinalPure} by considering the following bounds:

\begin{Proposition}[Reissner-Nordstr\"om Transmission]\label{Prop:PureReissner-Nordstrom}
	Define $\psi_+$, $\psi$, $\psi_{\mathcal{H}^-}$, $\chi$ as in the statement of Proposition \ref{Prop:FinalPure}. Define
	\begin{equation}
		\psi_0(u,v):=\chi\left(\frac{u-u_1}{M}\right)\psi(u,v)
	\end{equation}
	Note that $\psi_0(u,v_c)$ is equal to $\psi_0(u)$ given in Proposition \ref{Prop:FinalPure}.
	
	Then there exists constants $A(M), B(M)$ such that
	\begin{align}\label{eq:v2error}
		\int_{v=-\infty}^{v_c}(v_c-v)^2\left(\vert\p_v\psi(u_2,v)\vert^2+V\vert\psi(u_2,v)\vert^2\right)dv&\leq \frac{\IE}{(l+1)^4}
		\\\label{eq:ERNL2}
		\sup_{v\leq v_c}\left(\int_{u_1}^{\infty}\vert\psi_0(u,v)\vert^2 du\right)&\leq \frac{A \IT +A\langle u \rangle^2\IE}{(l+1)^4}\\
		\label{eq:RNH1Transmission}
		\int_{u=-\infty}^{\infty}\vert\p_u\psi_0(u,v_c)-\p_u\psi_{\mathcal{H}^-}(u)\vert^2du&\leq B\left(\delta r^2\IT +(1+\delta r^2\lur^2)\IE\right)\\
		\label{eq:RNTransmission}
		\int_{u=\infty}^{\infty}\vert\psi_0(u,v_c)-\psi_{\mathcal{H}^-}(u)\vert^2du&\leq B\left(\delta r^2\lur^2 \IT +(1+\delta r^2 \lur^4)\IE\right).
	\end{align}

	(The factor of $(l+1)^4$ here is only due to $I.T.$ having more factors of $(l+1)$ that necessary.)
	
	Here, $I.T.$ and $I.E.$ are as defined in Theorem \ref{Thm:Hawking}.
\end{Proposition}

\begin{proof}
We will here rewrite in $u,v$ coordinates the conserved $T$-energy on $\Sigma_v$ (Proposition \ref{Prop:T-energyConservationScattering}):
\begin{equation}\label{eq:T-energy}
	\text{T-energy}(\phi,\Sigma_{v})=\int_{-\infty}^{\infty}\vert\p_u\psi(u,v)\vert^2+V(r)\vert\psi(u,v)\vert^2du.
\end{equation}	

We begin with a slightly unconventional	one-sided weighted energy estimate on $\Sigma_u$. For any fixed $r^*$ with $r(r^*)<8M/3$ (this is less than the maximum value of $V$ for all $q$ and $l$), define
\begin{equation}
	F_n(x,r^*):=\int_{v=-\infty}^{x}(x-v)^n\left(\vert\p_v\psi(2r^*-x,v)\vert^2+V\vert\psi(2r^*-x,v)\vert^2\right)dv.
\end{equation}
We know that $F_0(x,r^*)$ is the T-energy of $\Sigma_{u=2r^*-v}\cap \{v\leq x\}$, and $\lim_{x\to-\infty}F_n(x,r^*)=0$ for $n\leq 2$.

\begin{equation}
	\frac{d}{dx}F_n(x,r^*)\leq n F_{n-1}(x, r^*).
\end{equation}

By integrating by parts and setting $2r^*=u_2+v_c$ for any fixed $u_2\leq u_1$, we obtain
\begin{equation}
	\int_{v=-\infty}^{v_c}(v_c-v)^n\left(\vert\p_v\psi(u_2,v)\vert^2+V\vert\psi(u_2,v)\vert^2\right)\leq \int_{-\infty}^{u_2}(u_2-u)^2\vert\p_u\psi_{\mathcal{H}^-}\vert^2du+\int_{-\infty}^{v_c}(v_c-v)^2\vert\p_v\psi_{RN}\vert^2\leq \frac{\IE}{(l+1)^4},
\end{equation}
giving us \eqref{eq:v2error}.

We can further bound, for all $u\leq u_1$,
\begin{equation}
	\vert\psi(u,v_c)\vert^2\leq 2\vert\psi_{\mathcal{H}^-}(u)\vert^2+2\left(\int_{-\infty}^{v_c}\vert\p_v\psi(u,v)\vert dv\right)^2\leq A\frac{\IE}{(l+1)^4}.
\end{equation}
	
We state Hardy's inequality (applied to a cut-off function): on any smooth function $\phi$ such that $\lim_{u\to\infty}(\phi(u,v))=0$, we have
\begin{equation}\label{eq:r2weightGivesL2}
	\int_{u=u_-}^{\infty}\vert\phi(u,v_c)\vert^2du\leq A\int_{u=u_-}^{\infty}(M^2+(u-u_-)^2)\left\vert\p_u\phi(u,v)\right\vert^2du.
\end{equation}

The remaining proof is using fairly standard weighted energy methods, which we will write out here in order to keep track of constants involving $l+1$.

By considering a weighted version of the $T$-energy in the region $u\geq u_0, v\geq v_c$, we obtain
\begin{align}\label{eq:u2bounds1}
	\int_{u_0}^{\infty} \left(M^2+(u-u_0)^2\right)\left(\vert\p_u\psi(u,v)\vert^2+V\vert\psi(u,v)\vert^2\right)du&=\int_{u_0}^{\infty}\left(M^2+(u-u_0)^2\right)\vert\p_u\psi_+\vert^2du\\\nonumber
	&\qquad-\int_{u\geq u_0, v'\geq v}2(u-u_0)\left(\vert\p_u\psi(u,v')\vert^2+V\vert\psi(u,v')\vert^2\right)dv'du\\\nonumber
	&\leq \int_{u_0}^{\infty}\left(M^2+(u-u_0)^2\right)\vert\p_u\psi_+\vert^2du\leq \frac{\IT }{(l+1)^4}.
\end{align}

We obtain a similar estimate in the region $u\in [u_1,u_0], v\leq v_c$:
\begin{align}\label{eq:u2bounds2}
	\int_{u_1}^{u_0} \left(M^2+(u-u_0)^2\right)\left(\vert\p_u\psi(u,v)\vert^2+V\vert\psi(u,v)\vert^2\right)du&=\int_{u_1}^{u_0}\left(M^2+(u-u_0)^2\right)\vert\p_u\psi_{\mathcal{H}^-}\vert^2du\\\nonumber
	&\qquad+\int_{-\infty}^{v_c}\lur^2\left(\vert\p_u\psi(u_1,v)\vert^2+V\vert\psi(u_1,v)\vert^2\right)dv\\\nonumber
	&\qquad-\int_{u\in[u_1,u_0], v'\geq v}2(u_0-u)\left(\vert\p_u\psi(u,v')\vert^2+V\vert\psi(u,v')\vert^2\right)dv'du\\\nonumber
	&\leq \int_{-\infty}^{u_0}\left(M^2+(u-u_0)^2\right)\vert\p_u\psi_{\mathcal{H}^-}\vert^2du+\int_{-\infty}^{v_c}\lur^2\vert\p_v\psi_{RN}\vert^2dv\\\nonumber
	&\leq \frac{\IT +\lur^2\IE}{(l+1)^4}.
\end{align}

This allows us to obtain \eqref{eq:ERNL2}:

\begin{align}
	\int_{u=-\infty}^{\infty}\vert \psi_0(u,v)\vert^2du&\leq \int_{u=u_1-M}^{\infty}\vert \psi(u,v)\vert^2du\leq \int_{-\infty}^{\infty}(M^2+(u_0-u)^2)\vert\p_u\psi(u,v)\vert^2du\\\nonumber
	&\leq \int_{u_1}^{\infty}(M^2+(u_0-u)^2)\vert\p_u\psi(u,v)\vert^2du+A\int_{u_1-M}^{u_1}(M^2+(u_0-u)^2)\vert\chi'\psi(u,v)\vert^2du\\\nonumber
	&\leq A\frac{\IT +\lur^2\IE}{(l+1)^4}
\end{align}

We then prove \eqref{eq:RNH1Transmission}:
\begin{align}\nonumber
\int_{u=-\infty}^{\infty}\vert\p_u\psi_0(u,v_c)-\p_u\psi_{\mathcal{H}^-}(u)\vert^2du&\leq A\int_{u=-\infty}^{\infty}\left\vert\int_{-\infty}^{v_c}\chi\p_v\p_u\psi dv\right\vert^2du+A\int_{u=-\infty}^{u_1}\vert\p_u\psi_0(u,v_c)-\p_u\psi_{\mathcal{H}^-}(u)\vert^2du\\\nonumber
&\leq A\int_{u=-\infty}^{\infty}\left\vert\int_{-\infty}^{v_c}V\psi_0 dv\right\vert^2du+A\IE\\
&\leq A\left(\int_{-\infty}^{v_c}\left(\int_{-\infty}^{\infty}V^2\vert\psi_0\vert^2du\right)^{1/2}dv\right)^2+A\IE\\\nonumber
&\leq A\sup_{v\leq v_c}\left(\int_{u_1}^{\infty}\vert\psi_0(u,v)\vert^2du\right)\left(\int_{-\infty}^{v_c}V(u_1,v)dv\right)^2+A\IE\\\nonumber
&\leq B\left(\delta r^2\IT +(1+\delta r^2\lur^2)\IE\right),
\end{align}
Which is sufficient to prove \eqref{eq:RNTransmission}, by Poincar\'e's inequality.
\end{proof}

This brings us on to the main proof of this section.

\begin{proof}[Proof of Proposition \ref{Prop:FinalPure}]
	We will consider the $\kappa$ term first (and therefore $\vert q \vert<1$). This bound is an easy consequence of Proposition \ref{Prop:PureReissner-Nordstrom}:
	\begin{align}
	\left\vert\int_{\omega=-\infty}^\infty\kappa\left(\vert\hat{\psi}_{\mathcal{H}^-}\vert^2-\vert\hat{\psi}_0\vert^2\right) d\omega\right\vert&\leq A\kappa\left(\int_{\omega=-\infty}^\infty\left\vert\hat{\psi}_{\mathcal{H}^-}-\hat{\psi}_0\right\vert^2d\omega\right)^{1/2}\left(\int_{\omega=-\infty}^\infty\left(\vert\hat{\psi}_{\mathcal{H}^-}\vert^2+\vert\hat{\psi}_0\vert^2\right)d\omega\right)^{1/2}\\\nonumber
	&\leq A\kappa\left(\delta r\lur\IT+(1+\delta r\lur^2)\IE\right)
	\end{align}
	as required.
	
	To bound the $\vert \omega\vert \vert \hat{\psi}\vert^2$ term, we consider
	\begin{equation}\label{eq:FourierWave}
		-i\omega \p_v\hat{\psi_0}=\widehat{V\psi_0}+\widehat{\chi'\p_v\psi},
	\end{equation}
	where $\chi=\chi\left(\frac{u-u_1}{M}\right)$, so $\chi'$ is only supported on $u\in [u_1-M,u_1]$.

	Here, $\hat\psi$ is the Fourier transform of $\psi$ \textbf{with respect to $u$}. This transform may not exist in an $L^2$ sense. However, as $V\psi$ is an $L^2$ function on $\Sigma_u$, this implies that $\p_v\hat{\psi}$ exists in a distributional sense. We denote $\delta\psi_0 := \psi_0-\psi_{\mathcal{H}^-}$. We will make use of the fact that $\hat{\psi_{\mathcal{H}^-}}$ is only supported on $\omega>0$. We then have
	\begin{align}
		\left\vert\int_{\omega=-\infty}^\infty\vert\omega\vert\left(\vert\hat{\psi}_{\mathcal{H}^-}\vert^2-\vert\hat{\psi}_0\vert^2\right) d\omega\right\vert&\leq \left\vert\int_{\omega=-\infty}^\infty\left(\vert\omega\vert-\omega\right)\left(\vert\hat{\psi}_{\mathcal{H}^-}\vert^2-\vert\hat{\psi}_0\vert^2\right) d\omega\right\vert +\left\vert\int_{\omega=-\infty}^\infty\omega\left(\vert\hat{\psi}_{\mathcal{H}^-}\vert^2-\vert\hat{\psi}_0\vert^2\right) d\omega\right\vert\\\nonumber
		&\leq 2\left\vert\int_{\omega=-\infty}^\infty\left(\frac{\vert\omega\vert}{\omega}-1\right)\int_{v=-\infty}^{v_c}\R\left(i\bar{\hat{\psi_0}}\left(\widehat{V \psi_0}+\widehat{\chi'\p_v\psi}\right)\right)dvdw\right\vert \\\nonumber
		&\qquad+\left\vert\int_{\omega=-\infty}^\infty\omega\left(\vert\hat{\psi}_{\mathcal{H}^-}\vert^2-\vert\hat{\psi}_0\vert^2\right) d\omega\right\vert.
	\end{align}
	We note that $\int_{-\infty}^{\infty}\omega\vert\psi\vert^2d\omega$ is given by by the particle current (\eqref{eq:ConservedCurrent} in Theorem \ref{Thm:Existence}). This allows us to easily bound the final term by $I.E.$ terms, leaving us to bound
	\begin{align}
		\left\vert\int_{\omega=-\infty}^\infty\left(\frac{\vert\omega\vert}{\omega}-1\right)\int_{v=-\infty}^{v_c}\R\left(i\bar{\hat{\psi_0}}\left(\widehat{V \psi_0}+\widehat{\chi'\p_v\psi}\right)\right)dvdw\right\vert &\leq \left\vert\int_{\omega=-\infty}^\infty\left(\frac{\vert\omega\vert}{\omega}-1\right)\int_{v=-\infty}^{v_c}\R\left(i\bar{\widehat{\delta\psi}_0}\left(\widehat{V \psi_0}+\widehat{\chi'\p_v\psi}\right)\right)dvdw\right\vert \\\nonumber
		&\leq 2\left\vert\int_{v=-\infty}^{v_c}\Vert\delta\psi_0\Vert_{\Sigma_v}\left(V(u_1,v)\Vert\psi_0\Vert_{\Sigma_v}+\Vert\chi'\p_v\psi\Vert_{\Sigma_v}\right) dv\right\vert \\\nonumber
		&\leq 2\sup_{v\leq v_c}(\Vert\delta\psi_0\Vert_{\Sigma_v}\Vert\psi_0\Vert_{\Sigma_v})\int_{v=-\infty}^{v_c}V(u_1,v)dv\\\nonumber
		&\qquad+2\sup_{v\leq v_c}\Vert\delta\psi_0\Vert_{\Sigma_v}\left(\int_{v=-\infty}^{v_c}(M^2+(v-v_c)^2)\Vert\chi'\p_v\psi\Vert_{\Sigma_v}^2dv\right)^{1/2}.
	\end{align}
These are bounded by Proposition \ref{Prop:PureReissner-Nordstrom}.

\end{proof}

\subsection{Generalisation of Bachelot}\label{Sec:Bachelot}
\setcounter{equation}{0}
In this section we will prove the following result:

\begin{Proposition}\label{Prop:Bachelot&Errors}
	Let $\psi_+$, $\psi$, $\psi_0$, $u_0$, $u_1$ and $v_c$ be as in Corollary \ref{Cor:FinalPure}. Then there exists a constant $A(M,\chi,r_b)$ independent of $q$ such that
	\begin{equation}
		\left\vert\int_{\omega=-\infty}^\infty\vert\omega\vert\vert\widehat{\psi_0\circ u_b}\vert^2-\vert\omega\vert\coth\left(\frac{\pi}{\kappa}\vert\omega\vert\right)\vert\hat{\psi_0}\vert^2 d\omega\right\vert\leq \frac{A\log(u_1)}{(l+1)^4u_1^2}\left(\IT +\langle u \rangle^2\IE\right).
	\end{equation}
	
	Further, in the case $\vert q\vert<1$, there exists a constant $A(M,\chi,r_b,q)$ such that
	\begin{equation}
		\left\vert\int_{\omega=-\infty}^\infty\vert\omega\vert\vert\widehat{\psi_0\circ u_b}\vert^2-\vert\omega\vert\coth\left(\frac{\pi}{\kappa}\vert\omega\vert\right)\vert\hat{\psi_0}\vert^2 d\omega\right\vert\leq \frac{Ae^{-\kappa u_1}}{(l+1)^4}\left(\IT +\langle u \rangle^2\IE\right).
	\end{equation}
\end{Proposition}
\begin{proof}
	For this calculation, we will be using Lemma $II.6$ from \cite{BachelotLemma}:
	\begin{Lemma}\label{Lemma:Bachelot}
		For $\beta>0$, $u\in C^{\infty}_0(\R)$, we define
		\begin{equation}\label{eq:FBachelot}
			F(\xi)=\int_{x=-\infty}^\infty e^{i\xi e^{\beta x}}u'(x)dx.
		\end{equation}
		Then we have
		\begin{equation}\label{eq:LemmaBachelot}
			\int_{\xi=-\infty}^\infty\vert\xi\vert^{-1}\vert F(\xi)\vert^2d\xi=\int_{\xi=-\infty}^\infty \vert\xi\vert \coth\left(\frac{\pi}{\beta}\vert\xi\vert\right)\vert\hat{u}(\xi)\vert^2d\xi.
		\end{equation}
	\end{Lemma}
	
	We will need to generalise this lemma to include the extremal case:
	\begin{Lemma}\label{Lemma:ExtremalBachelot}
		Let $A,\kappa\in\R_{>0}$, $v_c\in\R$ be constants. Define
		\begin{align}
			p_0(v)&=\frac{1}{\kappa}\log\left(\frac{A}{v_c-v}\right) \\
			p_1(v)&=\frac{A}{v_c-v}.
		\end{align}
		Then for all $f\in C^{\infty}_0(\R)$, we have
		\begin{align}
			\int_{\omega=-\infty}^\infty\vert\omega\vert\vert\widehat{f\circ p_0}\vert^2d\omega&=\int_{\omega=-\infty}^\infty\vert\omega\vert\coth\left(\frac{\pi}{\kappa}\vert\omega\vert\right)\vert\hat{f}\vert^2d\omega\\
			\int_{\omega=-\infty}^\infty\vert\omega\vert\vert\widehat{f\circ p_1}\vert^2d\omega&=\int_{\omega=-\infty}^\infty\vert\omega\vert\vert\hat{f}\vert^2d\omega.
		\end{align}
	\end{Lemma}
	\begin{proof}
		The $p_0$ case is a straightforward application of Lemma \ref{Lemma:Bachelot}, taking $\beta=\kappa$ and noting that $F(\xi)=-i\xi\widehat{u\circ p_0}$. For the $p_1$ case, the proof is a little different, but still proceeds in similar way to the proof of Lemma \ref{Lemma:Bachelot} (see \cite{BachelotLemma}).
		\begin{align}
			\int_{\omega=-\infty}^\infty\vert\omega\vert\vert\widehat{u\circ p}\vert^2d\omega&=\lim_{\epsilon\to 0}\left(\int_{\omega=-\infty}^\infty\vert\omega\vert e^{-\epsilon\vert\omega\vert}\vert\widehat{u\circ p}\vert^2d\omega\right)=\lim_{\epsilon\to 0}\left(\iiint_{x,x',\omega=-\infty}^\infty\vert\omega\vert e^{-\epsilon\vert\omega\vert}e^{i\omega(x'-x)}u\circ p(x)\overline{u\circ p}(x')dxdx'd\omega\right)\\\nonumber
			&=\lim_{\epsilon\to 0}\iint_{u,u'\in\R}\left(\int_{\omega=-\infty}^\infty\vert\omega\vert e^{-\epsilon\vert\omega\vert}e^{i\omega(\frac{A}{y}-\frac{A}{y'})}\right)u(y)\bar{u}(y')\frac{A^2}{y^2y'^2}dydy'\\\nonumber
			&=\lim_{\epsilon\to 0}\iint_{y,y'\in\R}\left(\frac{2\left(\epsilon^2-\left(\frac{A}{y}-\frac{A}{y'}\right)^2\right)}{\left(\epsilon^2+\left(\frac{A}{y}-\frac{A}{y'}\right)^2\right)^2}\right)u(y)\bar{u}(y')\frac{A^2}{y^2y'^2}dydy'\\\nonumber
			&=\lim_{\epsilon\to 0}\iint_{y,w\in\R}\left(\frac{2\left(\frac{\epsilon^2y^2(y-w)^2}{A^2}-w^2\right)}{\left(\frac{\epsilon^2y^2(y-w)^2}{A^2}+w^2\right)^2}\right)u(y)\bar{u}(y-w)dydw\\\nonumber
			&=\iint_{y,w\in\R}\lim_{\alpha\to 0}\left(\frac{2\left(\alpha^2-w^2\right)}{\left(\alpha^2+w^2\right)^2}\right)u(y)\bar{u}(y-w)dydw=\int_{w\in\R}\lim_{\alpha\to 0}(\widehat{\vert\omega\vert e^{-\alpha\vert\omega\vert}})\widehat{\left(\vert\hat{u}\vert^2\right)}dw\\\nonumber
			&=\int_{\omega=-\infty}^\infty\vert\omega\vert\vert\hat{u}\vert^2d\omega,
		\end{align}
		as required.
	\end{proof}
	
	We define $p_0$ and $p_1$ as above, noting that these are the leading order approximation to $u_b$ in the sub-extremal and extremal cases respectively.
	
	Then we apply Lemma \ref{Lemma:Bachelot} and \ref{Lemma:ExtremalBachelot} to obtain:
	\begin{align}\label{eq:UseBachelot}
		\int_{\omega=-\infty}^\infty\vert\omega\vert\vert\widehat{\psi_0\circ u_b}\vert^2d\omega=\begin{cases}
			\int_{\R}\vert\omega\vert \coth\left(\frac{\pi}{\kappa}\vert\omega\vert\right)\vert\widehat{\psi_0\circ u_b\circ p_i^{-1}}\vert^2d\omega & i=0\\
			\int_{\R}\vert\omega\vert\vert\widehat{\psi_0\circ u_b\circ p_i^{-1}}\vert^2d\omega & i=1
		\end{cases}.
	\end{align}
	Note that $\vert\omega\vert \coth\left(\frac{\pi}{\kappa}\vert\omega\vert\right)\leq \frac{\kappa}{\pi}+\vert\omega\vert$. 
	
	We now consider $u_b\circ p_i^{-1}$. In the sub-extremal case, we have that
	\begin{equation}
		u_b\circ p_1^{-1}= u + O(e^{-\kappa u}).
	\end{equation}
	Note that the coefficient of $e^{-\kappa u}$ includes some factors of $\kappa$ here.
	
	In the extremal case,
	\begin{equation}
		u_b\circ p_0^{-1}= u + O\left(\frac{\log(u)}{u^2}\right).
	\end{equation}
	
	We now note that in both the extremal and sub-extremal cases, we have:
	\begin{align}\nonumber
		\int_{\omega=-\infty}^\infty\left(\kappa+\vert\omega\vert\right)\Bigg\vert\vert\widehat{\psi_0\circ u_b\circ p_i^{-1}}&\vert^2-\vert\hat\psi_0\vert^2\Bigg\vert d\omega\leq A\left(\int_{u_1}^{\infty}\vert\p_u\psi_0\vert^2+\kappa^2\vert\psi_0\vert^2du\right)^{1/2}\left(\int_{u_1}^{\infty}\vert\psi_0\circ u_b\circ p_i-\psi_0\vert^2du\right)^{1/2}\\\label{eq:GammaError}
		&\leq A\left(\Vert\p_u\psi_0\Vert_{L^2}^2+\kappa^2 \Vert\psi_0\Vert_{L^2}^2\right)^{1/2}\left(\int_{u_1}^{\infty}\left\vert\int_{u}^{u_b(p_i(u))}\p_u\psi_0du\right\vert^2du\right)^{1/2}\\\nonumber
		&\leq A\left(\Vert\p_u\psi_0\Vert_{L^2}^2+\kappa^2 \Vert\psi_0\Vert_{L^2}^2\right)^{1/2}\left(\int_{u_1}^{\infty}\left\vert\int_{-\sup_{u'\geq u_1}\{\vert u_b(p_i(u'))-u'\vert\}}^{\sup_{u'\geq u_1}\{\vert u_b(p_i(u'))-u'\vert\}}\vert\p_u\psi_0(u+x)\vert dx\right\vert^2du\right)^{1/2}\\\nonumber
		&\leq A\left(\Vert\p_u\psi_0\Vert_{L^2}^2+\kappa^2 \Vert\psi_0\Vert_{L^2}^2\right)^{1/2}\int_{-\sup_{u'\geq u_1}\{\vert u_b(p_i(u'))-u'\vert\}}^{\sup_{u'\geq u_1}\{\vert u_b(p_i(u'))-u'\vert\}}\left(\int_{u_1}^{\infty}\vert\p_u\psi_0\vert^2 du\right)^{1/2}dx\\\nonumber
		&\leq \begin{cases}\frac{A e^{-\kappa u}}{(l+1)^4} \left(\IT +\langle u \rangle^2\IE\right)&\vert q\vert<1\\
			\frac{A\log u_1}{(l+1)^4u_1^2} \left(\IT +\langle u \rangle^2\IE\right)& \vert q\vert\leq 1\end{cases},
	\end{align}
	as required. We have used Minkowski's integral inequality to reach the penultimate line.
\end{proof}

\subsection{Reflection off the Matter Cloud}\label{Sec:Reflection}
\setcounter{equation}{0}

In this section we will consider evolving our solution in the small compact (in $r$, $t^*$ coordinates) region, given by $R_2:=\{v\leq v_c, u\leq u_1\}\cap\{r\geq r_b\}$.

We will consider the surface $r=r_b(t^*)$ to be instead parametrised by $v=v_b(u)$, or equivalently by $u=u_b(v)=v_b^{-1}(v)$, as in \eqref{eq:vbDef}. The final aim of this section will be to prove the following result:

\begin{Proposition}[$H^{1/2}$ Error from the Reflection]\label{Prop:FinalReflection}
	Let $\psi_+$, $\psi$, $\chi$ and $\psi_0$ be as in Corollary \ref{Cor:FinalPure}, $\delta r$ as in \eqref{eq:deltar}, and define
	\begin{equation}
		\psi_1(v)=\chi\left(\frac{v_c-v}{M}\right)\psi(u_1,v).
	\end{equation}
	Then there exists a constant $A(M,\chi,r_b)$ independent of $q$ such that
	\begin{equation}
		\left\vert\int_{\omega=-\infty}^\infty\vert\omega\vert\left(\vert\hat{\psi}_1\vert^2-\vert\widehat{\psi_0\circ u_b}\vert^2\right) d\omega\right\vert\leq \delta r^2 v_b'(u_1)\IT+(1+\delta r^2 v_b'(u_1)\lur^2)\IE.
	\end{equation}
\end{Proposition}

\begin{Proposition}[Reflection Energy Bounds]\label{Prop:Reflection}
	Let $\psi_+$, $\chi$, $\psi$, $\psi_0$, $u_0$, $u_1$ and $v_c$ be as in Corollary \ref{Cor:FinalPure}. Let $\psi_1$ be as in Proposition \ref{Prop:FinalReflection}. 
	
	Then there exists a constant $A$ such that
	\begin{align}\label{eq:psi_1L2}
	\int_{v=v_b(u_1)}^{v_c+M}\vert\psi_1\vert^2dv&\leq A\frac{\delta r^2\IT+(1+\delta r^2\lur^2)\IE)}{(l+1)^4}
	\\
	\int_{v_b(u_1)}^{v_c}\vert\p_v\psi_1+\p_v(\psi_0\circ u_b)\vert^2 dv& \leq A\left(\delta r^4\IT+(1+\lur^2 \delta r^4)\IE\right).
	\end{align}
\end{Proposition}

\begin{Remark}
	Note that if we define $\psi_{refl}$ as the solution to the equation
	\begin{equation}
	\p_u\p_v\psi_{refl}=0,
	\end{equation}
	with initial conditions
	\begin{equation}
	\psi_{refl}(u,v_c)=\psi(u,v_c),
	\end{equation}
	then this $\psi_{refl}$ takes the form:
	\begin{equation}
		\psi_{refl}(u,v)=\psi(u,v_c)-\psi(u_b(v),v_c).
	\end{equation}.
	Thus
	\begin{equation}
		\psi_{refl}(u_1,v)=-\psi(u_b(v),v_c).
	\end{equation}
	Therefore, $\psi_{refl}$ is reflected as if it were in $1+1$ dimensional Minkowski spacetime. Thus, Proposition \ref{Prop:Reflection} gives a bound on how much $\psi$ differs from a reflection in Minkowski.
\end{Remark}

\begin{proof}
	We firstly use \eqref{eq:RadWave} to show
	\begin{align}
		\int_{v=v_b(u_1)}^{v_c}\vert\p_v\psi(u_1,v)-\p_v\psi(u_b(v),v)\vert^2dv&=\int_{v=v_b(u_1)}^{v_c}\left\vert\int_{u=u_1}^{\infty}\p_u\p_v\psi dv\right\vert^2du=\int_{v=v_b(u_1)}^{v_c}\left\vert\int_{u=u_1}^{\infty} V\psi dv\right\vert^2du\\\nonumber
		&\leq A(l+1)^2\delta r \int_{v=v_b(u_1)}^{v_c}\int_{u=u_1}^{\infty}V\vert\psi\vert^2 dvdu,
	\end{align}
	where $\delta r$ is as defined in \eqref{eq:deltar}.

	As $\psi=0$ on $S_b$, we know that $\p_v\psi(u_b(v),v)=-u_b'(v)\p_u\psi(u_b(v),v)$. Therefore, 
	\begin{equation}
		\int_{S_b}\left\vert \p_v\psi(u_b(v),v)+(v_b'(u))^{-1}\p_u\psi(u_b(v),v)\frac{du}{dx}\right\vert^2dx=0,
	\end{equation} 
	for any parametrisation $x$.
	
	Thus, we then consider
	\begin{align}\nonumber
		\int_{u_1}^{\infty}(v_b'(u))^{-1}\vert\p_u\psi(u,v_c)-\p_u\psi(u,v_b(u))\vert^2du&\leq A \sup_{u\in [u_1,2\infty]}\left(\frac{(v_c-v_b(u))V(u,v_c)}{v_b'(u)}\right)\int_{u=u_1}^{\infty}V(u,v)\int_{v=v_b(u_1)}^{v_c}\vert\psi\vert^2dvdu\\
		&\leq A(l+1)^2 \delta r\int_{u=u_1}^{\infty}\int_{v=v_b(u_1)}^{v_c}V\vert\psi\vert^2dvdu.
	\end{align}
	
	Overall, we obtain
	\begin{equation}
		\int_{v=v_b(u_1)}^{v_c}\vert\p_v\psi(u_1,v)+\p_v(\psi(u_b(v),v_c))\vert^2dv\leq A (l+1)^2\delta r\int_{v\leq v_c, u=u_1}^{\infty}V\vert\psi\vert^2dvdu.
	\end{equation}
	
	In order to bound this, we require the following Lemma:
	
	\begin{Lemma}\label{Lem:CompactReflection}
		Let $\tilde{\psi}$ be a solution to \eqref{eq:RadWave} which, when restricted to $\Sigma_{v_c}$, is compactly supported between $u'$ and $u''$. Here, both $u'$ and $u''$ are greater than $u_1$, and $r(u_1,v_c)$ sufficiently close to $r_+$. Then there exists a constant $A(M, r_b)$ such that
		\begin{equation}\label{eq:Vpsiu1}
			\int_{\Sigma_{\tilde{u}}\cap \{v\leq v_c\}}\vert\tilde{\psi}\vert^2 dv \leq A \frac{(v_c-v_b(u'))^2\left(1-\frac{2M}{r}+\frac{q^2M^2}{r^2}\right)\vert_{u',v_b(u')}}{1-\frac{2M}{r}+\frac{q^2M^2}{r^2}\vert_{u'',v_b(u'')}}\int_{\Sigma_{v_c}}\vert\p_u\tilde{\psi}\vert^2+V \vert\tilde{\psi}\vert^2du,
		\end{equation}
		for all $\tilde{u}\geq u_1$.
	\end{Lemma}
	
	\begin{proof}[Proof of Lemma]
		This proof is mostly using standard results. We first note that, for any solution compactly supported on $\Sigma_{v_c}$, we have non-degenerate energy boundedness from \cite{Mine2}, that is
		\begin{equation}
			E(\Sigma_{v})=\int_{\Sigma_{v}}\frac{\vert \p_u\tilde{\psi}\vert^2}{1-\frac{2M}{r}+\frac{q^2M^2}{r^2}}+V \vert\tilde{\psi}\vert^2 du
		\end{equation}
		is bounded by its value on $\Sigma_{v_c}$, and similarly
		\begin{equation}
			E(\Sigma_u)=\int_{\Sigma_{u}\cap\{v\leq v_c\}}\frac{\vert \p_v\tilde{\psi}\vert^2}{1-\frac{2M}{r}+\frac{q^2M^2}{r^2}}+V \vert\tilde{\psi}\vert^2 dv\leq E(\Sigma_{v_c}).
		\end{equation}
		
		We first achieve the result for $\tilde{u}\geq u'$ by considering the non-degenerate energy above $u'$, and apply Poincar\'e's inequality to bound
		\begin{align}\label{eq:ReflLemmaProof}
			\int_{\Sigma_{\tilde{u}}\cap\{v\leq v_c\}}\vert\tilde{\psi}\vert dv&\leq (v_c-v_b(u'))^2\left(1-\frac{2M}{r}+\frac{q^2M^2}{r^2}\right)\vert_{u',v_c} E(\Sigma_{\tilde{u}})\\\nonumber
			&\leq \frac{(v_c-v_b(u'))^2\left(1-\frac{2M}{r}+\frac{q^2M^2}{r^2}\right)\vert_{u',v_c}}{\left(1-\frac{2M}{r}+\frac{q^2M^2}{r^2}\right)\vert_{u'',v_b(u'')}}\int_{\Sigma_{v_c}}\vert\p_u\tilde{\psi}\vert^2+V \vert\tilde{\psi}\vert^2du.
		\end{align}
		
		To prove the result for $\tilde{u}\leq u'$, a straightforward calculation gives us
		\begin{align}
			\int_{\Sigma_{u'}\cap [v_b(u'),v_c]} \vert\tilde{\psi}\vert^2dv +\int_{\Sigma_{v_c}\cap [\tilde{u},u']}\frac{1}{V}\vert\p_u\tilde{\psi}\vert^2du = \int_{\Sigma_{\tilde{u}}\cap [v_b(u'),v_c]} \vert\tilde{\psi}\vert^2dv +\int_{\Sigma_{v_b(u')}\cap [\tilde{u},u']}\frac{1}{V}\vert\p_u\tilde{\psi}\vert^2du\\\nonumber
			+\int_{S_b\cap[\tilde{u},u']}\frac{1}{V}\vert\p_u\tilde{\psi}\vert^2du - \int_{\{u\in[\tilde{u},u']\}} \frac{1}{V^2}\frac{dV}{dv}\vert\p_u\tilde{\psi}\vert^2dudv.
		\end{align}
		We can again use Gronwall's inequality on $f(u)=\int_{(S_b\cup\Sigma_u)\cap[u,u']}\frac{1}{V}\vert\p_u\tilde{\psi}\vert^2du$ to bound
		\begin{equation}\label{eq:HighFreqRefl}
			\int_{\Sigma_{\tilde{u}}\cap \{v\leq v_c\}}\vert\tilde{\psi}\vert^2dv\leq \frac{V(\tilde{u},v_c)}{V(\tilde{u},v_b(\tilde{u}))}\left(\int_{\Sigma_{u'}\cap \{v\leq v_c\}} \vert\tilde{\psi}\vert^2dv +\int_{\Sigma_{v_c}\cap [\tilde{u},u']}\frac{1}{V}\vert\p_u\tilde{\psi}\vert^2du\right).
		\end{equation}
		For this, we have used that $V^{-1}\frac{dV}{dv}$ is strictly increasing in $r$ for $r$ sufficiently close to $r_+$. (Here, sufficiently close can be chosen independent of $q,l$.)
		
		This gives the result, noting that $V(u,v_c)/V(u,v_b(u))$ is uniformly bounded for all $l$ and $q$, and for $r$ sufficiently close to $r_+$.
	\end{proof}

	To use Lemma \ref{Lem:CompactReflection}, we split $\psi$ up into $\sum_i \psi^i$. Here $\psi^i$ are solutions to \eqref{eq:RadWave} with initial conditions on $\Sigma_{v_c}$
	\begin{equation}
		\psi^i(u,v_c)=\chi\left(\frac{u_1+iM-u}{M}\right)\left(1-\chi\left(\frac{u_1+(i-1)M-u}{M}\right)\right)\psi(u,v_c),
	\end{equation}
	so $\psi^i(u,v_c)$ is supported on $u\in[u_1+(i-1)M,u_1+(i+1)M]$.
	
	As \eqref{eq:RadWave} is linear, $\Vert\psi\Vert_{L^2}\leq \sum_i\Vert\psi^i\Vert_{L^2}$. Therefore we can use Lemma \ref{Lem:CompactReflection} to show that, for $i\geq 1$
	\begin{equation}
		\int_{\Sigma_u\cap\{v\leq v_c\}}\vert\psi^i\vert^2dv\leq A(v_c-v_b(u))^2\int_{\Sigma_{v_c}}\vert\p_u\psi^i\vert^2+V\vert\psi^i\vert^2du,
	\end{equation}
	and therefore
	\begin{equation}
		\int_{\Sigma_u\cap\{v\leq v_c\}}\vert\psi\vert^2dv\leq \sum_i i^2\int_{\Sigma_u\cap\{v\leq v_c\}}\vert\psi^i\vert^2dv\leq A(v_c-v_b(u))^2\int_{\Sigma_{v_c}}(u-u_0)^2\vert\p_u\psi\vert^2+(M^{-2}+V)\vert\psi\vert^2du.
	\end{equation}
	We have here used that $\Vert\psi\Vert_{L^2(\Sigma_{v_c})}^2\leq\sum_i i^2\Vert\psi^i\Vert_{L^2(\Sigma_{v_c})}^2$ and $\sum_i i^2\Vert\psi^i\Vert_{L^2(\Sigma_{v_c})}^2\leq\Vert(M^2+(u-u_0)^2)^{1/2}\psi\Vert_{L^2(\Sigma_{v_c})}^2$, as each $\psi^i$ only overlaps with 1 other $\psi^i$ at any given value of $u$. This gives us \eqref{eq:psi_1L2}.

	Combining all the above, we obtain
	\begin{align}\nonumber
		\int_{v=v_b(u_1)}^{v_c}\vert\p_v\psi(u_1,v)+\p_v(\psi(u_b(v),v_c))\vert^2dv&\leq A \frac{\IT+\lur^2\IE}{(l+1)^2} \delta r\int_{u=u_1}^\infty V(u,v_c)(v_c-v_b(u))^2du\\
		&\leq A \left(\IT+\lur^2\IE\right) \delta r^4.
	\end{align}

	Finally we note that the error from initially writing $\psi$ rather than $\psi_0$ or $\psi_1$ can be removed by adding $\IE$ terms.
\end{proof}

This brings us to the proof of Proposition \ref{Prop:FinalReflection}.

\begin{proof}[Proof of Proposition \ref{Prop:FinalReflection}]
	This proof is a straightforward application of Poincar\'e's inequality.
	\begin{align}
	\int_{\omega=-\infty}^\infty\vert\omega\vert\left\vert\vert\hat{\psi}_1\vert^2-\vert\widehat{\psi_0\circ u_b}\vert^2\right\vert d\omega&\leq A\left(\int_{\omega=-\infty}^\infty\vert\omega\vert^2\left\vert\hat{\psi}_1+\widehat{\psi_0\circ u_b}\right\vert^2d\omega\right)^{1/2}\left(\int_{\omega=-\infty}^\infty\left(\vert\widehat{\psi_0\circ u_b}\vert^2+\vert\hat{\psi}_1\vert^2\right)d\omega\right)^{1/2}\\\nonumber
	&\leq A\left(\int_{v_b(u_1)}^{v_c}\left\vert\p_v\psi_1+\p_v\psi_0\circ u_b\right\vert^2dv\right)^{1/2}\left(\int_{v_b(u_1)}^{v_c}\vert\psi_1\vert^2+\vert\psi_0\circ u_b\vert^2dv\right)^{1/2}
	\end{align}
	Which gives us the required result, using Proposition \ref{Prop:Reflection}.
\end{proof}

\subsection{High Frequency Transmission}\label{Sec:HighFrequency}
\setcounter{equation}{0}

We now consider how our solution on $\Sigma_{u_1}$ is transmitted to $\mathcal{I}^-$, ultimately resulting in the following proposition: 
\begin{Proposition}[Hawking Radiation Error from High Frequency Transmission]\label{Prop:FinalHighFrequency}
	Let $\psi_+$, $\psi$, $u_0$, $u_1$ and $v_c$ be as in Corollary \ref{Cor:FinalPure}. Let $\psi_1$ be as in Proposition \ref{Prop:FinalReflection}, and let $\psi_-$ be the past radiation field of $\psi$. Then there exists a constant $A(\mathcal{M})$ such that
	\begin{equation}\label{eq:GeoOpt}
		\left\vert\int_{\omega=-\infty}^\infty\vert\omega\vert\left(\vert\hat{\psi}_-\vert^2-\vert\hat{\psi}_1\vert^2-\vert\hat{\psi}_{RN}\vert^2\right) d\omega\right\vert\leq A\left(\delta r \IT+\left(\delta r^2 \lur^2+\frac{1}{1-\frac{2M}{r}+\frac{q^2M^2}{r^2}\vert_{u_1,v_c}}\right)\IE\right)
	\end{equation}
	Here $\kappa$ is the surface gravity, as in \eqref{eq:kappa} and $I.T.[\psi_+]$, $I.E.[\psi_+,v_c,u_1,u_0]$ are as defined in the statement of Theorem \ref{Thm:Hawking}.
\end{Proposition}

We start this section by bounding the energy through the surface $\Sigma_{v_b(u_1)}$, as all other energy is transmitted to $\mathcal{I}^-$. However, the map taking solutions on space-like surfaces back to their past radiation fields is bounded with respect to the \emph{non-degenerate} energy (Theorem $7.1$ in \cite{Mine2}). Thus we need to look at non-degenerate energy through $\Sigma_{v_b(u_1)}$.  The non-degenerate energy on a surface $\Sigma_v$ takes the form
\begin{equation}
\int_{\Sigma_v}\frac{\vert\p_u\psi\vert^2}{1-\frac{2M}{r}+\frac{M^2q^2}{r^2}}+V\vert\psi\vert^2 du
\end{equation}

In order to bound this energy, we have the following proposition:
\begin{Proposition}[High Frequency Transmission in Pure \RN]\label{Prop:RNTransmission}
	Let $\psi$, $v_c$, $u_1$, $\psi_1$ be as in Corollary \ref{Cor:FinalPure}. Then there exists a constant $A(M)$ independent of $q$ such that
	\begin{align}\label{eq:HighFreqBounds}
	\int_{\Sigma_{v_b(u_1)}}\frac{r^2\vert\p_u\psi\vert^2}{1-\frac{2M}{r}+\frac{q^2M^2}{r^2}}du+&\sup_{u_2\leq u_1}\left(\int_{\Sigma_{u_2}\cap\{v\leq v_c\}}\frac{Vr^2\vert\psi\vert^2}{1-\frac{2M}{r}+\frac{q^2M^2}{r^2}}dv\right)\\\nonumber
	&\leq A\left(\int_{\Sigma_{v_c}\cap \{u\leq u_1\}}\frac{r^2\vert\p_u\psi\vert^2}{1-\frac{2M}{r}+\frac{q^2M^2}{r^2}}du+l(l+1)\int_{\Sigma_{u_1}\cap \{v\leq v_c\}}\vert\psi\vert^2 dv\right).
	\end{align}
	
%	In the case $l=0$ we similarly obtain a constant $A(M)$ such that
%	\begin{align}
%		\int_{\Sigma_{v_b(u_1)}}\frac{r^3\vert\p_u\psi\vert^2}{1-\frac{2M}{r}+\frac{q^2M^2}{r^2}}du+&\sup_{u_2\leq u_1}\left(\int_{\Sigma_{u_2}\cap\{v\leq v_c\}}\frac{Vr^3\vert\psi\vert^2}{1-\frac{2M}{r}+\frac{q^2M^2}{r^2}}dv\right)\\\nonumber
%		&\leq A\left(\int_{\Sigma_{v_c}\cap \{u\leq u_1\}}\frac{r^3\vert\p_u\psi\vert^2}{1-\frac{2M}{r}+\frac{q^2M^2}{r^2}}du+\int_{\Sigma_{u_1}\cap \{v\leq v_c\}}\vert\psi\vert^2 dv\right).
%	\end{align}
	
	Furthermore, let $\psi_-$ be the past radiation field for $\psi$. Then there exists a constant $A(M)$ (again independent of $q$) such that
	\begin{equation}\label{eq:HighFreqDifferenceBounds}
	\int_{v=v_b(u_1)}^{v_c}\left\vert\p_v\psi_1-\p_v\psi_-\right\vert^2dv\leq A\left((l+1)^4\int_{\Sigma_{u_1}\cap \{v\leq v_c\}}\vert\psi\vert^2du+\frac{\IEE{u_1}}{1-\frac{2M}{r}+\frac{q^2M^2}{r^2}\vert_{u_1,v_c}}\right).
	\end{equation}
\end{Proposition}

\begin{proof}
	For this, we will consider, for any $u_2\leq u_1$,
	\begin{align}\label{eq:HighTransmissionProof}
		\int_{\Sigma_{u_1}\cap\{v\leq v_c\}}&\frac{Vr^2}{1-\frac{2M}{r}+\frac{q^2M^2}{r^2}}dv+\int_{\Sigma_{v_c}\cap\{u\in [u_2,u_1]\}}\frac{r^2\vert\p_u\psi\vert^2}{1-\frac{2M}{r}+\frac{q^2M^2}{r^2}}du\\\nonumber
		&=\int_{\Sigma_{u_2}\cap\{v\in 	[v_b(u_1),v_c]\}}\frac{Vr^2}{1-\frac{2M}{r}+\frac{q^2M^2}{r^2}}dv+\int_{(\Sigma_{v_2}\cup S_b)\cap\{u\in [u_2,u_1]\}}\frac{r^2\vert\p_u\psi\vert^2}{1-\frac{2M}{r}+\frac{q^2M^2}{r^2}}du\\\nonumber
		&\qquad+\int_{\{u,v\in[u_2,u_1]\times[v_b(u_1),v_c]\}}\left(1-\frac{2M}{r}+\frac{q^2M^2}{r^2}\right)\frac{2M}{r^2}\left(1-\frac{2Mq^2}{r^2}\right)\vert\psi\vert^2+\frac{2r\left(1-\frac{6M}{r}+\frac{4q^2M^2}{r^2}\right)}{1-\frac{2M}{r}+\frac{q^2M^2}{r^2}}\vert\p_u\psi\vert^2dudv.
	\end{align}
	The bulk coefficients of $\vert\psi\vert^2$ and $\vert\p_u\psi\vert^2$ are both negative for $r$ sufficiently large, so we obtain a finite interval on which these have the incorrect sign. We can use Hardy's inequality to bound
	\begin{equation}
		\int_{\Sigma_v\cap \{u\in[u_2,u_1]\}}\left(1-\frac{2M}{r}+\frac{q^2M^2}{r^2}\right)\frac{\vert\psi\vert^2}{r^2}\leq A\vert\psi(u_1,v)\vert^2+A\int_{\Sigma_v\cap \{u\in[u_2,u_1]\}}\frac{\vert\p_u\psi\vert^2}{1-\frac{2M}{r}+\frac{q^2M^2}{r^2}}du.
	\end{equation}
	Thus \eqref{eq:HighTransmissionProof} allows us to apply Gronwall's inequality to $\int_{\Sigma_v\cap \{u\in[u_2,u_1]\}}\frac{\vert\p_u\psi\vert^2}{1-\frac{2M}{r}+\frac{q^2M^2}{r^2}}du$, which gives us \eqref{eq:HighFreqBounds}.

	In order to prove \eqref{eq:HighFreqDifferenceBounds}, we proceed by considering
	\begin{align}
		\int_{v=-\infty}^{v_c}\vert\p_v\psi_1-\p_v\psi_-\vert^2dv&=\int_{v=-\infty}^{v_c}\left\vert\int_{u=-\infty}^{u_1}V\psi du\right\vert^2dv\leq\left(\int_{u=-\infty}^{u_1}\left(\int_{\Sigma_u\cap\{v\leq v_c\}}V^2\vert\psi\vert^2dv\right)^{1/2}du\right)^2\\\nonumber
		&\leq A\sup_{u\leq u_1}\left(\int_{\Sigma_u\cap\{v\leq v_c\}}\frac{Vr^2\vert\psi\vert^2}{1-\frac{2M}{r}+\frac{q^2M^2}{r^2}}dv\right)\int_{u=-\infty}^{u_1}\sqrt{\frac{V(1-\frac{2M}{r}+\frac{q^2M^2}{r^2})}{r^2}}du\\\nonumber
		&\leq A(l+1)^2\sup_{u\leq u_1}\left(\int_{\Sigma_u\cap\{v\leq v_c\}}\frac{Vr^2\vert\psi\vert^2}{1-\frac{2M}{r}+\frac{q^2M^2}{r^2}}dv\right),
	\end{align} 
	giving the desired result.
\end{proof}
Before proving the final result for this section, we have one small result we will need.
\begin{Lemma}[Smallness of Past Error]\label{Lem:Smallpsi2}
Let $\psi_+$, $\psi$, $\psi_-$, $\psi_1$ be as in Corollary \ref{Cor:FinalPure}. Define
\begin{equation}
	\psi_2:=\psi_--\psi_1-\psi_{RN}.
\end{equation}

Then there exists a constant $A(M,r_b)$ such that
\begin{align}
	\int_{u=-\infty}^\infty\vert\p_u\psi_2\vert^2du\leq A\left(\delta r^2\IT+\lur^2\delta r^2\IE+\frac{\IE}{1-\frac{2M}{r}+\frac{q^2M^2}{r^2}\vert_{u_1,v_c}}\right)\\
	\int_{u=-\infty}^\infty\vert\psi_2\vert^2du\leq A \left(\delta r^2 \IT+\lur^2\delta r^2\IE+\frac{\IE}{1-\frac{2M}{r}+\frac{q^2M^2}{r^2}\vert_{u_1,v_c}}\right).
\end{align}
\begin{proof}
	Note that $\psi_2$ is only supported in $v\leq v_c+M$, as $v>v_c$ is out of the past light cone of the collapsing cloud. Thus, the solution in $v>v_c$ coincides with that of \RN, and $\psi_1$ is only supported in $v\in[v_b(u_1),v_c+M]$.
	
	We can obtain the first equation fairly easily thanks to Proposition \ref{Prop:RNTransmission} paired with non-degenerate energy boundedness:
	\begin{equation}\label{eq:psi2H1}
		\Vert\psi_2\Vert_{\dot{H}^1}^2\leq 4\Vert\psi_--\psi_1\Vert_{\dot{H}^1(\mathcal{I}^-)\cap\{v\in [v_b,v_c]\}}^2+4\Vert\psi_{RN}\Vert_{\dot{H}^1(\cap\{v\leq v_c\})}^2+4\Vert\psi_-\Vert_{\dot{H}^1(\{v\leq v_b(u_1)\})}^2+4\Vert\psi_1\Vert_{\dot{H}^1(\{v\geq v_c\})}^2.
	\end{equation}
	
	To bound the $L^2$ norm of $\psi_2$, we break the calculation up slightly differently
	\begin{equation}
		\Vert\psi_2\Vert_{L^2}^2\leq 3\Vert\psi_--\psi_1\Vert_{L^2(\mathcal{I}^-)\cap\{v\leq v_c\}}^2+3\Vert\psi_{RN}\Vert_{L^2(\cap\{v\leq v_c\})}^2+4\Vert\psi_1\Vert_{L^2(\{v\geq v_c\})}^2.
	\end{equation} 
We can bound $\Vert \psi_--\psi_1\Vert_{L^2(\{v\leq v_c\})}$ using Proposition \ref{Prop:RNTransmission}, and we can bound both $\Vert \psi_1\Vert_{L^2(\{v\geq v_c\})}$ and $\Vert\psi_{RN}\Vert_{L^2(\cap\{v\leq v_c\})}$ by $I.E.$ terms.
\end{proof}
\end{Lemma}
We can now prove the main result for this section.
\begin{proof}[Proof of Proposition \ref{Prop:FinalHighFrequency}]
	
	We can expand \eqref{eq:GeoOpt} to get:
	\begin{equation}\label{eq:Expansion}
	\int_{-\infty}^{\infty}\vert\omega\vert\left(\vert\hat{\psi}_-\vert^2-\vert\hat{\psi}_1\vert^2-\vert\hat{\psi}_{RN}\vert^2\right) d\omega=\int_{-\infty}^{\infty}\vert\omega\vert\left(\vert\hat{\psi}_2\vert^2+2\R\left(\left(\hat{\psi}_1+\hat{\psi}_2\right)\bar{\hat{\psi}}_{RN}+\hat{\psi}_1\bar{\hat{\psi}}_2\right)\right) d\omega.
	\end{equation}
	
	We can then bound
	\begin{align}
	\int_{\infty}^{\infty}\vert\omega\vert \R\left(\hat{\psi}_2\bar{\hat{\psi}}_{RN}\right)d\omega&\leq\Vert\psi_{RN}\Vert_{L^2(\mathcal{I}^-)}\Vert\psi_2\Vert_{\dot{H}^1(\mathcal{I}^-)}\\
	\int_{\infty}^{\infty}\vert\omega\vert \R\left(\hat{\psi}_1\bar{\hat{\psi}}_{RN}\right)d\omega&\leq\Vert\psi_1\Vert_{L^2(\mathcal{I}^-)}\Vert\psi_{RN}\Vert_{\dot{H}^1(\mathcal{I}^-)}\\
	\int_{\infty}^{\infty}\vert\omega\vert \R\left(\hat{\psi}_1\bar{\hat{\psi}}_2\right)d\omega&\leq\Vert\psi_1\Vert_{L^2(\mathcal{I}^-)}\Vert\psi_2\Vert_{\dot{H}^1(\mathcal{I}^-)}\\
	\int_{-\infty}^{\infty}\vert\omega\vert\vert\hat{\psi}_2\vert^2d\omega&\leq\Vert\psi_2\Vert_{L^2(\mathcal{I}^-)}\Vert\psi_2\Vert_{\dot{H}^1(\mathcal{I}^-)}.
	\end{align}
	
	We have already bounded $\Vert\psi_1\Vert_{L^2(\mathcal{I}^-)}$ in Proposition \ref{Prop:Reflection}, $\Vert \psi_2\Vert_{\dot{H}^1}$ and $\Vert \psi_2\Vert_{L^2}$ in Lemma \ref{Lem:Smallpsi2}, which give the required results.

\end{proof}

\section{Treatment of Error Terms}\label{Sec:Integrated Error Terms}
\renewcommand{\theequation}{\arabic{section}.\arabic{equation}}

In this section we show the arbitrary polynomial decay of the $I.E.$ terms, provided that $\hat{\psi}_+$ vanishes and has all derivatives vanishing at $\omega=0$. Most of the work to show this behaviour has been previously done in the extremal ($\vert q\vert=1$) case, in \cite{ERNScat}. This gives our first Theorem:

\begin{Theorem1}\label{Thm:ExtremalI.E.decay}[Decay of the I.E. terms in the $\vert q\vert=1$ case]
	Let $\psi_+$ be a Schwartz function on the cylinder, with $\hat{\psi}_+$ compactly supported on $\omega\geq 0$. Then for each $n$, there exists an $A_n(M,\psi_+)$ such that
	\begin{equation}
	I.E.[\psi_+,v_c,u_1,u_0]\leq A_n\left((u_0-u_1)^{-n}+(u_0-v_c)^{-n}\right).
	\end{equation}
	Here, $I.E.$ is as defined in Theorem \ref{Thm:Hawking}, in the case of an extremal ($\vert q\vert=1$) RNOS model.
\end{Theorem1}

\begin{proof}
	As $\hat\phi_+$ and all it's $\omega$ derivatives vanish at $\omega=0$, then $\hat\psi_{-n}:=\omega^{-n}\psi_+$ is also a Schwartz function. Instead of imposing $\psi_+$ as our radiation field on $\mathcal{I}^+$, we can use $\psi_{-n}$. The resulting solution has the property
	\begin{equation}
	\p_{t^*}^n\psi_{-n}=\psi,
	\end{equation}
	by uniqueness of solutions to the wave equation (Theorem \ref{Thm:RNExist}).
	
	We then apply Theorem $4.2$ (with $u_0$ as the origin) from \cite{ERNScat} to $\psi_{-n}$, to see
	\begin{align}
	\int_{\mathcal{H}^-}(1+(u-u_0)^2)^n\vert\p_u\psi_{\mathcal{H}^-}\vert^2&+(1+(u-u_0)^2)^n\vert\mathring{\slashed\nabla}\psi_{\mathcal{H}^-}\vert^2\sin\theta d\theta d\varphi du\\\nonumber
	&+\int_{\mathcal{I}^-}(1+(v-u_0-R)^2)^n\vert\p_u\psi_{RN}\vert^2+(1+(v-u_0-R)^2)^n\vert\mathring{\slashed\nabla}\psi_{RN}\vert^2\sin\theta d\theta d\varphi du\leq A_n[\psi_+].
	\end{align}
	
	Restricting the integrals to $u\leq u_1$ and $v\leq v_c$, we can see that
	\begin{align}
	I.E.[\psi_+,v_c,u_1,u_0]&\leq A\int_{u=-\infty}^{u_1}(1+(u-u_0)^2)^{3/2}\vert\p_u\psi_{\mathcal{H}^-}\vert^2+(1+(u-u_0)^2)^{3/2}\vert\mathring{\slashed\nabla}\psi_{\mathcal{H}^-}\vert^2\sin\theta d\theta d\varphi du \\\nonumber
	&\qquad+\int_{v=-\infty}^{v_c}(1+(v-u_0-R)^2)^{3/2}\vert\p_u\psi_{\mathcal{RN}}\vert^2+(1+(v-u_0-R)^2)^{3/2}\vert\mathring{\slashed\nabla}\psi_{RN}\vert^2\sin\theta d\theta d\varphi du\\\nonumber
	&\leq A_n[\psi_+]\left((1+(u_1-u_0)^2)^{-n+3/2}+(1+(u_0-v_c)^2)^{-n+3/2}\right),
	\end{align}
	giving our result.
\end{proof}

We now look to extend this result to the sub-extremal case. The following section will closely follow the equivalent proof of the extremal case \cite{ERNScat}. The next ingredient needed for the $r^{*p}$ method is integrated local energy decay, or $ILED$. This will be done in a manner similar to \cite{RedShift}. 

\begin{Proposition1}[ILED for sub-extremal \RN]\label{Prop:ILED}
	Let $\phi$ be a solution of \eqref{eq:wave} on a \RNS background $\mathcal{M}_{RN}$. Let $t_0$ be a fixed value of $t$, and let $R$ be a large fixed constant. Then there exists a constant $A=A(M,R,n)$ such that
	\begin{align}
	\int_{-\infty}^{t_0}\left(\int_{\Sigma_t\cap\{\vert r^*\vert\leq R\}}\vert\p_r\phi\vert^2\right)dt&\leq A\int_{\bar{\Sigma}_{t_0,R}}dn(J^{\p_t})\leq A\int_{\Sigma_{t_0}}-dt(J^{\p_t})\\
	\int_{-\infty}^{t_0}\left(\int_{\Sigma_t\cap\{\vert r^*\vert\leq R\}}-dt(J^{\p_t})\right)dt&\leq A\sum_{\vert\alpha\vert +j\leq 1}\int_{\bar{\Sigma}_{t_0,R}}dn(J^{\p_t}[\p_t^j\Omega^\alpha\phi])\leq A\sum_{\vert\alpha\vert +j\leq 1}\int_{\Sigma_{t_0}}-dt(J^{\p_t}[\p_t^j\Omega^\alpha\phi]).
	\end{align}
\end{Proposition1}
\begin{proof}
	Consider Reisnner--Nordstr\"om spacetime in $r^*,t,\theta,\varphi$ coordinates. For ease of writing, we will denote
	\begin{equation}
	D(r^*)=1-\frac{2M}{r}+\frac{M^2q^2}{r^2}.
	\end{equation}
	
	As done so far in this paper, we will restrict to a spherical harmonic. We will first consider the case $l\geq 1$. We choose $\omega=h'/4+h D/2r$, $h'(r^*)=(A^2+(r^*-R)^2)^{-1}$, and consider divergence theorem applied to 
	\begin{align}
	J^\textbf{X}&:=J^{X,\omega}+\frac{h'}{D}\beta\phi^2\p_{r^*}\\\nonumber
	\beta&=\frac{D}{r}-\frac{r^*-R}{A^2+(r^*-R)^2},
	\end{align}
	for $A$ and $R$ yet to be chosen.
	
	Note that the flux of this current through any $t=const$ surface is bounded by the $\p_{r^*}\phi$ and $\p_t\phi$ terms of the $T$ energy. The bulk term of this is given by
	\begin{align}\label{eq:K^X}
	K^\textbf{X}=\nabla^\nu J^\textbf{X}_\nu=\frac{h'}{D}\left(\p_r\phi+\beta\phi\right)^2+\left(\frac{(r^*-R)^2-A^2}{2D((r^*-R)^2+A^2)^3}+\left(\frac{l(l+1)}{r^2}\left(\frac{D}{r}-\frac{D'}{2D}\right)+\frac{D'}{2r^2}-\frac{D''}{2Dr}\right)h\right)\phi^2.
	\end{align}
	
	Calculating the coefficient of $h\phi^2$ gives us
	\begin{align}
	\frac{l(l+1)}{r^2}\left(\frac{D}{r}-\frac{D'}{2D}\right)&+\frac{D'}{2r^2}-\frac{D''}{2Dr}\\\nonumber
	&=\frac{M^4}{r^7}\left(l(l+1)\left(\frac{r}{M}\right)^4-3(l(l+1)-1)\left(\frac{r}{M}\right)^3+(2q^2l(l+1)-4q^2-8)\left(\frac{r}{M}\right)^2+15q^2\left(\frac{r}{M}\right)-6q^4\right)\\\nonumber
	&=\frac{M^4}{r^7}\left((x-1)(x-2)(l(l+1)x^2+3(x-1))-(1-q^2)((2l(l+1)-4)x^2+15x-6(1+q^2))\right)\\\nonumber
	&=l(l+1)(x-3)x^3+(3x-8)x^2+q^2((2l(l+1)-4)x^2+15x-6q^2),
	\end{align}
	where $x=r/M$. Searching for roots of this, we can see there is a root at $r=M$, but this is strictly less than $0$ for $r<2M$, and strictly greater than $0$ for $r>3M$. In this interval, we consider the function
	\begin{align}
	f(x)&=l(l+1)x-3(l(l+1)-1)+(2q^2l(l+1)-4q^2-8)x^{-1}+15q^2x^{-2}-6q^4x^{-3}\\
	f'(x)&=l(l+1)(1-2q^2x^{-2})+(8+4q^2)x^{-2}-30q^2x^{-3}+18q^4x^{-4}>0,
	\end{align}
	for $x>2$. Therefore the coefficient of $h\phi^2$ in \eqref{eq:K^X} has exactly one root, in a bounded region (independent of $q$) of $r^*$. We label this point $r^*_0$, and we let
	\begin{equation}
	h(r^*_0)=0.
	\end{equation}
	As $h$ has a positive gradient, this means that $f(x)h\geq 0$, with a single quadratic root at $r^*_0$. Provided $R>r^*_0$, we also know $h>\pi/2A$ for sufficiently large values of $r^*$. Thus to ensure $K^\textbf{X}$ is positive definite, it is sufficient to show that $R$ and $A$ can be chosen such that
	\begin{equation}
	\frac{(r^*-R)^2-A^2}{2D((r^*-R)^2+A^2)^3}+\frac{M}{r^4}f\left(\frac{r}{M}\right)h>0.
	\end{equation}
	We only need to consider the region $\vert r^*-R\vert<A$. By choosing $R-r^*_0-A>>M$, we can ensure that in this region, $D>1-\epsilon$, $\frac{M}{r}f\left(\frac{r}{M}\right)\geq l(l+1)(1-\epsilon)$, and $r\leq r^*(1-\epsilon)$. Thus it is sufficient to choose $R$ and $A$, with $R-r^*_0-A>>M$, such that
	\begin{equation}
	l(l+1)\pi(1-\epsilon)-\frac{A\left(A^2-(r^*-R)^2\right)r^{*3}}{((r^*-R)^2+A^2)^3}>0.
	\end{equation}
	Let $y=\frac{r^*-R}{A}$, then we are looking for the maximum of 
	\begin{equation}
	\frac{(1-y^2)(y+\frac{R}{A})^3}{(1+y^2)^3}.
	\end{equation}
	If we choose $R-r^*_0=1.001A$, and choose $0.001A>>M$, then 
	\begin{equation}
	\sup_{-1\leq y\leq 1}\frac{(1-y^2)(y+1.001)^3}{(1+y^2)^3}<\frac{\pi}{2},
	\end{equation}
	and we have $K^\textbf{X}$ is positive definite.
	\begin{align}
	K^\textbf{X}&\geq\frac{\epsilon\vert\p_r\phi+\beta\phi\vert^2}{D(M^2+r^{*2})}+\epsilon\left(\frac{l(l+1)D\tanh\left(\frac{r^*-r^*_0}{M}\right)^2}{r^3}+\frac{1}{D(M^2+r^{*2})^2}\right)\vert\phi\vert^2\\\nonumber
	&\geq\frac{\epsilon\vert\p_r\phi\vert^2}{D(M^2+r^{*2})}+\epsilon\left(\frac{l(l+1)D\tanh\left(\frac{r^*-r^*_0}{M}\right)^2}{r^3}+\frac{1}{D(M^2+r^{*2})^2}\right)\vert\phi\vert^2.
	\end{align}
	
	To bound the $T$-energy locally, we can thus consider
	\begin{align}
	A\sum_{\vert\alpha\vert +j\leq 1}K^\textbf{X}[\p_t^j\Omega^\alpha\phi]&\geq \frac{-dt(J^{\p_t})}{M^2+r^{*2}}\\
	A\sum_{\vert\alpha\vert +j\leq 1}K^\textbf{X}[\p_t^j\Omega^\alpha\phi]&\geq A (-dt(J^{\p_t}))\qquad \forall \vert r^*\vert\leq R,
	\end{align}
	where $\Omega$ are the angular Killing Fields, as given by \eqref{eq:AngularKilling}.
	
	For the $l=0$ case, we again follow the example of \cite{RedShift} and take $X=\p_{r^*}$. Given that all angular derivatives vanish, applying divergence theorem to $J^X$ in the interval $r^*\in (-\infty, r^*_0)$, we obtain
	\begin{equation}\label{eq:K l=0}
	\int_{-\infty}^{t_0}(\p_t\phi(r^*_0))^2+(\p_{r^*}\phi(r^*_0))^2r^2\sin\theta d\theta d\varphi dt+\int_{r^*=-\infty}^{r^*_0}\frac{2D}{r}\int_{-\infty}^{t_0}\left(-(\p_t\phi)^2+(\p_{r^*}\phi)^2\right)r^2\sin\theta d\theta d\varphi dt dr^*\leq 4T\text{-energy}(\Sigma_{t^*_0}).
	\end{equation}
	
	Let
	\begin{equation}
	F(r^*):=\int_{r^*=-\infty}^{r^*_0}\frac{2D}{r}\int_{-\infty}^{t_0}(\p_t\phi)^2r^2\sin\theta d\theta d\varphi dtdr^*.
	\end{equation}
	Then \eqref{eq:K l=0} implies
	\begin{equation}
	F'(r^*)\leq \frac{2D}{r}F(r^*)+\frac{8D}{r}T\text{-energy}(\Sigma_{t^*_0}).
	\end{equation}
	
	Noting that $\int_{r^*=-\infty}^{r^*_0}\frac{2D}{r}dr^*=2\log\left(\frac{r}{r^+}\right)$, an application of Gronwall's inequality yields
	\begin{equation}
	F(r^*)\leq A\left(\frac{r^2}{r_+^2}\right)T\text{-energy}(\Sigma_{t^*_0}).
	\end{equation}
	
	By applying this to \eqref{eq:K l=0}, we can obtain
	\begin{align}
	\left(\frac{r_+^2}{r_0^2}\right)\int_{t_0}^{\infty}\int_{r^*=-\infty}^{r^*_0}\frac{2D}{r}\left(\p_t\phi)^2+(\p_{r^*}\phi)^2\right)\sin\theta d\theta d\varphi dr^* dt&\leq A T\text{-energy}\\
	\int_{-\infty}^{t_0}\left(\int_{\Sigma_t\cap\{\vert r^*\vert\leq R\}}-dt(J^{\p_t})\right)dt&\leq AT\text{-energy}
	\end{align}
	We now have the result for all $l$ using $\Sigma_{t_0}$.
	
	Once we note that the region $\{t\leq t_0, \vert r^*\vert\leq R\}$ is entirely in the domain of dependence of $\bar{\Sigma}_{t_0,R}$, we can consider the alternative solution, $\tilde\phi$, given by the data of $\phi$ on $\bar{\Sigma}_{t_0,R}$, but vanishing on $\mathcal{H}^-$ and $\mathcal{I}^-$ to the future of $\bar{\Sigma}_{t_0,R}$. We evolve this forward to $\Sigma_{t_0}$, we can apply the above result. As $\tilde\phi=\phi$ to the past of $\bar{\Sigma}_{t_0,R}$, we have the result.
\end{proof}

\begin{Remark1}[Degeneracy at the Photon Sphere]
	For the $l\geq 1$ case, as $l\to\infty$, the root of the $h$ function chosen tends towards the root of 
	\begin{equation}
	1-\frac{3M}{r}+\frac{2M^2q^2}{r^2}=0,
	\end{equation}
	known as the photon sphere, $r=r_p$. If we do not require control of the $T$-energy at this particular value, then we do not need to include angular derivatives
	\begin{equation}
	\int_{t_0}^\infty\left(\int_{\Sigma_t\cap\{\epsilon\leq\vert r^*-r^*_p\vert\leq R\}}-dt(J^{\p_t})\right)dt\leq A\sum_{j=0}^1\int_{\Sigma_{t_0}}-dt(J^{\p_t}[\p_t^j\phi]).
	\end{equation}
\end{Remark1}

\begin{Remark1}[Forward and higher order ILED]
	By sending $t\to-t$, Proposition \ref{Prop:ILED} immediately gives us the result in the forward direction:
	\begin{equation}
	\int_{t_0}^\infty\left(\int_{\Sigma_t\cap\{\vert r^*\vert\leq R\}}-dt(J^{\p_t})\right)dt\leq A\sum_{j+\vert\alpha\vert\leq 1}\int_{\Sigma_{t_0}}-dt(J^{\p_t}[\p_t^j\Omega^\alpha\phi]).
	\end{equation}
	
	We can also apply the Proposition \ref{Prop:ILED} to $\p_t^j\Omega^\alpha\phi$ to obtain
	\begin{equation}
	\int_{t_0}^\infty\left(\int_{\Sigma_t\cap\{\vert r^*\vert\leq R\}}\vert\nabla^n\phi\vert^2\right)dt\leq A\sum_{j+\vert\alpha\vert\leq n}\int_{\Sigma_{t_0}}-dt(J^{\p_t}[\p_t^j\Omega^\alpha\phi]),
	\end{equation}
	where we have rewritten terms in $\nabla^n\phi$ involving more than one $r^*$ derivative using \eqref{eq:wave}.
\end{Remark1}

\begin{Proposition1}[Boundedness of $r^*$ Weighted Energy]\label{Prop:r^* Weighted Bound}
	Let $\psi_+$ be a Schwartz function. Let $\psi$ be the solution to \eqref{eq:RadWave} on a \RNS background $\mathcal{M}_{RN}$, as given by Theorem \ref{Thm:RNExist}, with radiation field on $\mathcal{I}^+$ equal to $\psi_+$, and which vanishes on $\mathcal{H}^+$. Let $R$ be a constant, and let $t_0$ be a fixed value of $t$. Then for each $n\in\mathbb{N}_0$, we have the following bounds:
	\begin{equation}
	\sum_{j+\vert\alpha\vert\leq n}\int_{\bar{\Sigma}_{t_0,R}}(M^{2+2j}+\vert r^*\vert^{2+2j})dn(J^{\p_t}[\Omega^\alpha\p_t^{j}\phi])\leq A_n\sum_{1\leq j+\vert m\vert\leq n+1}\int_{-\infty}^\infty\left(M^{2j}+u^{2j}\right)(l+1)^{2m}\left\vert\p_u^j\psi_+\right\vert^2du,
	\end{equation}
	where $A_n=a_n(M,t_0,R,n)$.
\end{Proposition1}

\begin{proof}
	We start by bounding an $r^p$ weighted norm on $\Sigma_{u_0}\cap{r^*\leq -R}$ for some $u_0\in\R$ and  $R$ large by induction.
	\begin{align}
	\int_{\Sigma_{u_0}\cap\{r^*\leq -R\}}(-R-r^*)^p\vert \p_v\psi\vert^2dv&=\int_{u\geq u_0, r^*\leq -R}-p(-r^*-R)^{p-1}\vert \p_v\psi\vert^2+(-R-r^*)^pV\p_v(\vert\psi\vert^2)dudv\\\nonumber
	&\leq-\int_{u\geq u_0, r^*\leq -R}\p_v((-R-r^*)^pV)\vert\psi\vert^2dudv\\\nonumber
	&\leq A\int_{u=u_0}^{\infty}\int_{\Sigma_v}V(-R-r^*)^{p-1}\vert\psi\vert^2dvdu\\\nonumber
	&\leq A\int_{u=u_0}^{\infty}\int_{u'=u}^{\infty}(u-u_0)^{p-1}\vert\p_u\psi_+\vert^2du'du=A\int_{u=u_0}^{\infty}(u-u_0)^p\vert\p_u\psi_+\vert^2du.
	\end{align}
	Here we have used Hardy's inequality and the inductive step to reach the last line, along with an explicit calculation to show that $-\p_v((-R-r^*)^pV)\leq A(-R-r^*)^{p-1} V$. $A$ is a constant which depends on $M$ and the choice of $R$. Note this calculation applies for all $p\in\mathbb{N}$ for sub-extremal \RN, but in the extremal case this only applies up to $p=2$. By applying this result to $\p_t^j\Omega^\alpha\phi$, we obtain the required bound for $\bar{\Sigma}_{t_0,R}\cap\{r^*\leq -R\}$.
	
	For $r^*\in[-R,R]$, we note that $T$-energy boundedness of $\p_t^j\Omega^\alpha\phi$ is sufficient for our result, as the constant $A_n$ may depend on our choice of $R$.
	
	For the equivalent result on $\Sigma_{v_0}\cap{r^*\geq R}$, a similar approach does not work, as the $T$ energy on $\Sigma_v$ does not approach $0$ as $v\to\infty$. Instead, we will make use of the vector field multiplier $u^2\p_u+v^2\p_v$. Let $u_0\leq v_0-R$. This will closely follow the proof of Proposition $8.1$ in \cite{ERNScat}.
	
	\begin{align}\label{eq:K energy}
	\int_{\Sigma_{v_0}\cap{u\leq v_0-R}}u^2\vert \p_u\psi\vert^2+v^2V\vert\psi\vert^2du+&\int_{\Sigma_{u_0}\cap\{v\geq v_0\}}v^2\vert \p_v\psi\vert^2+u^2V\vert\psi\vert^2du\\\nonumber
	&=\int_{\mathcal{I}^+\cap\{u\leq v_0-R\}}u^2\vert \p_u\psi\vert^2+v^2V\vert\psi\vert^2du+\int_{\Sigma_{u=v_0-R}\cap\{v\geq v_0\}}v^2\vert \p_v\psi\vert^2+u^2V\vert\psi\vert^2du\\\nonumber
	&\qquad+\int_{u\in[u_0,v_0-R], v\geq v_0}\left(\p_v(v^2V)+\p_u(u^2V)\right)\vert\psi\vert^2dudv.
	\end{align}
	
	We then note
	\begin{equation}
	\p_v(v^2V)+\p_u(u^2V)=2tV+tr^*V'=t(2V+r^*V')\leq\begin{cases}
	\frac{A\vert t\vert}{r^3}\leq Ar^{-2}&l=0\\
	\frac{A V \vert t\vert \log\left(\frac{r}{M}\right)}{r}\leq A V\log\left(\frac{r}{M}\right)&l\neq 0
	\end{cases},
	\end{equation}
	using that $\vert t\vert\leq r^*+\max\{v_0-R,-v_0\}$ in the region we are considering. Here $A$ depends on the choice of $v_0$ and $R$. We can then take a supremum of \eqref{eq:K energy} over $u_0\leq v_0-R$ and $v\geq v_0$ to obtain
	\begin{align}\label{eq:K energy sup}
	\sup_{v\geq v_0}\int_{\Sigma_{v}\cap{u\leq v_0-R}}u^2\vert \p_u\psi\vert^2+v^2V\vert\psi\vert^2du+&\sup_{u\leq v_0-R}\int_{\Sigma_{u}\cap\{v\geq v_0\}}v^2\vert \p_v\psi\vert^2+u^2V\vert\psi\vert^2dv\\\nonumber
	&\leq\int_{\mathcal{I}^+\cap\{u\leq v_0-R\}}u^2\vert \p_u\psi\vert^2+v^2V\vert\psi\vert^2du+\int_{\Sigma_{u=v_0-R}\cap\{v\geq v_0\}}v^2\vert \p_v\psi\vert^2+u^2V\vert\psi\vert^2dv\\\nonumber
	&\qquad+A\int_{u\leq v_0-R, v\geq v_0}\left(V\log\left(\frac{r}{M}\right)+r^{-2}\right)\vert\psi\vert^2dudv.
	\end{align}
	
	We can bound the final integral using the following:
	\begin{align}
	\int_{u=-\infty}^{v_0-R}\int_{v=v_0}^{\infty}\left(V \log\left(\frac{r}{M}\right)+r^{-2}\right)\vert\psi\vert^2dvdu&\leq 	\int_{u=-\infty}^{v_0-R}\int_{v=v_0}^{\infty}\left(V \log\left(\frac{-u}{M}\right)+V \log\left(\frac{v}{M}\right)+u^{-2}\right)\vert\psi\vert^2dvdu\\\nonumber
	&\leq A\int_{u=-\infty}^{v_0-R}u^{-2}\log\left(\frac{-u}{M}\right)\int_{v=v_0}^{\infty}u^2V\vert\psi\vert^2dvdu\\\nonumber
	&\qquad+A\int_{v=v_0}^{\infty}v^{-2}\log\left(\frac{v}{M}\right)\int_{u=-\infty}^{v_0-R}v^2V\vert\psi\vert^2dudv\\\nonumber
	&\qquad+A\int_{u=-\infty}^{v_0-R}u^{-2}\int_{v=v_0}^{\infty}\vert\psi\vert^2dudv\\\nonumber
	&\leq \epsilon\sup_{u\leq v_0-R}\int_{\Sigma_{u}\cap\{v\geq v_0\}}v^2\vert \p_v\psi\vert^2+u^2V\vert\psi\vert^2du\\\nonumber
	&\qquad+\epsilon\sup_{v\geq v_0}\int_{\Sigma_{v}\cap{u\leq v_0-R}}u^2\vert \p_u\psi\vert^2+v^2V\vert\psi\vert^2du\\\nonumber
	&\qquad+\epsilon\sup_{u\leq v_0-R}\int_{\Sigma_{u}\cap\{v\geq v_0\}}\vert\psi\vert^2dv,
	\end{align} 
	where $v_0$ and $R$ are sufficiently large.
	
	We can then apply Hardy's inequality to get
	\begin{equation}
	\sup_{u\leq v_0-R}\int_{\Sigma_{u}\cap\{v\geq v_0\}}\vert\psi\vert^2dv\leq A \sup_{u\leq v_0-R}\int_{\Sigma_{u}\cap\{v\geq v_0\}}V\vert\psi\vert^2dv+\sup_{u\leq v_0-R}\int_{\Sigma_{u}\cap\{v\geq v_0\}}v^2\vert \p_v\psi\vert^2dv.
	\end{equation}
	
	We can then rearrange \eqref{eq:K energy sup} to see
	\begin{align}
	\sup_{v\geq v_0}\int_{\Sigma_{v}\cap{u\leq v_0-R}}u^2\vert \p_u\psi\vert^2+v^2V\vert\psi\vert^2du+&\sup_{u\leq v_0-R}\int_{\Sigma_{u}\cap\{v\geq v_0\}}v^2\vert \p_v\psi\vert^2+u^2V\vert\psi\vert^2dv\\\nonumber
	&\leq A\int_{\mathcal{I}^+\cap\{u\leq v_0-R\}}u^2\vert \p_u\psi\vert^2+v^2V\vert\psi\vert^2du+A \int_{\Sigma_{u=v_0-R}\cap\{v\geq v_0\}}v^2\vert \p_v\psi\vert^2+u^2V\vert\psi\vert^2dv.
	\end{align}
	
	By taking an appropriate limit of this, we can see that
	\begin{align}\label{eq:Spacelike Infinity}
	\int_{\Sigma_{v_0}\cap{u\leq v_0-R}}u^2\vert \p_u\psi\vert^2+v^2V\vert\psi\vert^2du+&\int_{\mathcal{I}^-\cap\{v\geq v_0\}}v^2\vert \p_v\psi\vert^2+u^2V\vert\psi\vert^2dv\\\nonumber
	&\leq A\int_{\mathcal{I}^+\cap\{u\leq v_0-R\}}u^2\vert \p_u\psi\vert^2+v^2V\vert\psi\vert^2du+A \int_{\Sigma_{u=v_0-R}\cap\{v\geq v_0\}}v^2\vert \p_v\psi\vert^2+u^2V\vert\psi\vert^2dv.
	\end{align}
	
	We can also consider a time reversal of this statement to get
	\begin{align}
	\int_{\mathcal{I}^+\cap\{u\leq v_0-R\}}u^2\vert \p_u\psi\vert^2+v^2V\vert\psi\vert^2du+& \int_{\Sigma_{u=v_0-R}\cap\{v\geq v_0\}}v^2\vert \p_v\psi\vert^2+u^2V\vert\psi\vert^2dv
	\\\nonumber
	&\leq A\int_{\Sigma_{v_0}\cap{u\leq v_0-R}}u^2\vert \p_u\psi\vert^2+v^2V\vert\psi\vert^2du+A\int_{\mathcal{I}^-\cap\{v\geq v_0\}}v^2\vert \p_v\psi\vert^2+u^2V\vert\psi\vert^2dv.
	\end{align}
	 
	In order to add more $u$ and $v$ weighting to this, we commute with the vector field $S=u\p-U+v\p_v$.
	\begin{equation}
	(\p_u\p_v+V)S(f)=S[(\p_u\p_v+V)f]+(2V-r^*\p_{r^*}V)f+2(\p_u\p_v+V)f.
	\end{equation}
	
	Thus an easy induction argument gives
	\begin{equation}
	\vert (\p_u\p_v+V)S^n\psi\vert\leq A\vert V-r^*\p_{r^*}V\vert\sum_{k=0}^{n-1}S^k\psi,
	\end{equation}
	noting that
	\begin{equation}
	\left\vert\p_{r^*}^n(V-r^*\p_{r^*}V)\right\vert\leq A\vert V-r^*\p_{r^*}V\vert\leq\frac{A(l+1)^2\log\left(\frac{r}{M}\right)}{r^3}.
	\end{equation}
	
	Repeating \eqref{eq:K energy sup}, but applied to $S^n\psi$, we obtain
	
	\begin{align}
	F_n:=\sup_{v\geq v_0}\int_{\Sigma_{v}\cap{u\leq v_0-R}}u^2&\vert \p_uS^n\psi\vert^2+v^2V\vert S^n\psi\vert^2du+\sup_{u\leq v_0-R}\int_{\Sigma_{u}\cap\{v\geq v_0\}}v^2\vert \p_vS^n\psi\vert^2+u^2V\vert S^n\psi\vert^2du\\\nonumber
	&\leq\int_{\mathcal{I}^+\cap\{u\leq v_0-R\}}u^2\vert \p_uS^n\psi\vert^2+v^2V\vert S^n\psi\vert^2du+\int_{\Sigma_{u=v_0-R}\cap\{v\geq v_0\}}v^2\vert \p_vS^n\psi\vert^2+u^2V\vert S^n\psi\vert^2du\\\nonumber
	&\qquad+A\int_{u\leq v_0-R, v\geq v_0}\left(V\log\left(\frac{r}{M}\right)+r^{-2}\right)\vert S^n\psi\vert^2dudv\\\nonumber
	&\qquad+A\sum_{k=0}^{n-1}\int_{u\leq v_0-R, v\geq v_0}\frac{(l+1)^2\log\left(\frac{r}{M}\right)}{r^3}\vert S^k\psi\vert\left\vert u^2\p_uS^n\psi+v^2\p_v S^n\psi\right\vert dudv\\\nonumber
	&\leq\int_{\mathcal{I}^+\cap\{u\leq v_0-R\}}u^2\vert \p_uS^n\psi\vert^2+v^2V\vert S^n\psi\vert^2du+\int_{\Sigma_{u=v_0-R}\cap\{v\geq v_0\}}v^2\vert \p_vS^n\psi\vert^2+u^2V\vert S^n\psi\vert^2du\\\nonumber
	&\qquad+A\epsilon F_n+AF_n^{1/2}\sum_{k=0}^{n-1}\left(\int_{u=-\infty}^{v_0-R}\frac{(l+1)^4\log\left(\frac{r(u,v_0)}{M}\right)^2}{u^2V(u,v_0)r(u,v_0)^6}du\right)^{1/2}F_k^{1/2}\\\nonumber
	&\leq A\int_{\mathcal{I}^+\cap\{u\leq v_0-R\}}u^2\vert \p_uS^n\psi\vert^2+v^2V\vert S^n\psi\vert^2du+A\int_{\Sigma_{u=v_0-R}\cap\{v\geq v_0\}}v^2\vert \p_vS^n\psi\vert^2+u^2V\vert S^n\psi\vert^2du\\\nonumber
	&\qquad+A\epsilon(l+1)F_n^{1/2}\sum_{k=0}^{n-1}F_k^{1/2}\\\nonumber
	&\leq A\int_{\mathcal{I}^+\cap\{u\leq v_0-R\}}u^2\vert \p_uS^n\psi\vert^2+v^2V\vert S^n\psi\vert^2du+A\int_{\Sigma_{u=v_0-R}\cap\{v\geq v_0\}}v^2\vert \p_vS^n\psi\vert^2+u^2V\vert S^n\psi\vert^2du\\\nonumber
	&\qquad+A(l+1)^2\sum_{k=0}^{n-1}F_k.
	\end{align}
	
	As $F_0$ is bounded by \eqref{eq:Spacelike Infinity}, we can inductively obtain
	\begin{align}
	\sum_{k+m\leq n}\Bigg(\int_{\Sigma_{v_0}\cap{u\leq v_0-R}}u^2(l+1)^{2m}&\vert \p_u(S^k\psi)\vert^2+v^2V(l+1)^m\vert S^k\psi\vert^2du\\\nonumber
	&\qquad+\int_{\mathcal{I}^-\cap\{v\geq v_0\}}v^2(l+1)^{2m}\vert \p_v((v\p_v)^k\psi)\vert^2+u^2V(l+1)^{2m}\vert((v\p_v)^k\psi)\vert^2dv\Bigg)\\\nonumber
	&\leq A\sum_{k+m\leq n}\Bigg(\int_{\mathcal{I}^+\cap\{u\leq v_0-R\}}u^2(l+1)^{2m}\vert \p_u((u\p_u)^k\psi)\vert^2+v^2V(l+1)^{2m}\vert((u\p_u)^k\psi)\vert^2du\\\nonumber
	&\qquad+\int_{\Sigma_{u=v_0-R}\cap\{v\geq v_0\}}v^2(l+1)^{2m}\vert \p_v(S^k\psi)\vert^2+u^2V(l+1)^{2m}\vert (S^k\psi)\vert^2dv\Bigg),
	\end{align}
	along with the time reversed result
	\begin{align}\nonumber
	\sum_{k+m\leq n}\Bigg(\int_{\mathcal{I}^+\cap\{u\leq v_0-R\}}u^2(l+1)^{2m}&\vert \p_u((u\p_u)^k\psi)\vert^2+v^2V(l+1)^{2m}\vert((u\p_u)^k\psi)\vert^2du\\\nonumber
	&\qquad+\int_{\Sigma_{u=v_0-R}\cap\{v\geq v_0\}}v^2(l+1)^{2m}\vert \p_v(S^k\psi)\vert^2+u^2V(l+1)^{2m}\vert (S^k\psi)\vert^2dv\Bigg)\\
	&\leq A\sum_{k+m\leq n}\Bigg(\int_{\Sigma_{v_0}\cap{u\leq v_0-R}}u^2(l+1)^{2m}\vert \p_u(S^k\psi)\vert^2+v^2V(l+1)^{2m}\vert (S^k\psi)\vert^2du\\\nonumber
	&\qquad+\int_{\mathcal{I}^-\cap\{v\geq v_0\}}v^2(l+1)^{2m}\vert \p_v((v\p_v)^k\psi)\vert^2+u^2V(l+1)^{2m}\vert((v\p_v)^k\psi)\vert^2dv\Bigg).
	\end{align}
	
	All that is now left for the result is to bound 
	\begin{align}\nonumber
	\sum_{k+m\leq n}\int_{\Sigma_{u=v_0-R}\cap\{v\geq v_0\}}v^2(l+1)^{2m}\vert \p_v(S^k\psi)\vert^2+u^2V(l+1)^{2m}\vert (S^k\psi)\vert^2dv&\leq 	\sum_{k+m+j\leq n}\int_{\Sigma_{u=v_0-R}\cap\{v\geq v_0\}}v^{2+2k}(l+1)^{2m}\vert \p_v^{k+1}\p_t^j\psi\vert^2\\
	&\qquad\qquad\qquad+v^{2k}V(l+1)^{2m}\vert (\p_v^k\p_t^j\psi)\vert^2dv,
	\end{align}
	for fixed and arbitrarily large $R,v_0$. We have used \eqref{eq:RadWave} to remove any $\p_u\p_v$ derivatives, and have replaced any $\p_u$ derivatives with $\p_t+\p_v$ derivatives. As $\p_t$ and $\Omega$ are Killing fields, it is sufficient to bound
	\begin{equation}
	\int_{\Sigma_{u=v_0-R}\cap\{v\geq v_0\}}v^{2k+2}\vert \p_v^{k+1}\psi\vert^2dv\leq A\int_{\Sigma_{u=v_0-R}}\chi\left(\frac{r-R}{M}-1\right)r^{2(k+1)}\vert\p_v^{k+1}\psi\vert^2dv,
	\end{equation}
	for $k\geq 0$. 
	
	We can immediately apply Proposition \ref{Prop:PureReissner-Nordstrom} (with time reversed) to obtain the $k=0$ case
	\begin{equation}
	\int_{\Sigma_{u=v_0-R}}\chi r^2\vert\p_v\psi\vert^2du\leq A\int_{\mathcal{I}^+}(M^2+u^2)\vert\p_u\psi_+\vert^2+l(l+1)\vert\psi_+\vert^2du.
	\end{equation}
	Here the constant $A$ depends on choice of $v_0$ and $R$. We would now like to generalise this to the following result (closely based on Proposition $7.7$ in \cite{ERNScat}).
	\begin{equation}\label{eq:higher order r weighted Sigma_u}
	\int_{\Sigma_{u=v_0-R}}\chi\left(\frac{r-R}{M}-1\right)r^{2k}\vert\p_v^k\psi\vert^2dv\leq A\sum_{1\leq m+j\leq k}\int_{u=v_0-R}^{\infty}\left(M^{2m}+(u-u_R)^{2m}\right)(l+1)^{2j}\vert\p_u^m\psi_+\vert^2du,
	\end{equation}
	where $A$ depends on $M,n,R$. From here, we will denote $v_0-R=u_R$.
	
	We will prove this inductively. First, we consider commuting \eqref{eq:RadWave} with $\p_v$ to obtain
	\begin{equation}\label{eq:wave commute p_v}
	\p_u\p_v(\p_v^n\psi)+V\p_v^n\psi=-\p_v^n(V\psi)+V\p_v^n\psi=-\sum_{j=0}^{n-1}\binom{n}{j}\p_{r^*}^{n-j}V\p_v^j\psi\leq A\sum_{j=0}^{n-1}\frac{(l+1)^2}{r^{2+n-j}}\vert\p_v^j\psi\vert.
	\end{equation}
	
	We then look at applying this to the following generalisation of the right hand side of \eqref{eq:higher order r weighted Sigma_u}
	\begin{align}\nonumber
	\int_{\Sigma_{u=u_R}}\chi r^{p}\vert\p_v^k\psi\vert^2dv&=\int_{u\geq u_R}D(\chi'r^p+p\chi r^{p-1})\vert\p_v^k\psi\vert^2+2\R\left(\chi r^p\p_v^k\bar\psi\sum_{j=0}^{k-1}\binom{k}{j}\p_v^{k-j}V\p_v^j\psi\right)-D\p_{r^*}\left(\chi r^pV\right)\vert\p_v^{k-1}\psi\vert dudv\\\label{eq:rp higher order backwards}
	&\qquad+\int_{\mathcal{I}^+}r^{p-2}l(l+1)\vert\p_v^{k-1}\psi_+\vert^2du\\\nonumber
	&\leq A\int_{u\geq u_R, r^*\in[R,R+M]}\sum_{m+j\leq k-1}-dt((l+1)^{2j}J^{\p_t}[\p_t^m\psi])dudv+A\int_{u\geq u_R}\chi\sum_{j=0}^{k}\frac{(l+1)^{2j}}{r^{1+2j-p}}\vert\p_v^{k-j}\psi\vert^2dudv\\\nonumber
	&\leq A\int_{u=u_R}\sum_{m+j\leq k-1}(l+1)^{2j}\vert\p_u^{m+1}\psi_+\vert^2du+A\int_{u\geq u_R}\chi\sum_{j=0}^{k}\frac{(l+1)^{2j}}{r^{1+2j-p}}\vert\p_v^{k-j}\psi\vert^2dudv,
	\end{align}
	where we have used Proposition \ref{Prop:ILED}.
	
	For our induction argument, we will assume we have proved \eqref{eq:higher order r weighted Sigma_u} for $k\leq n$, where $n\geq 1$. We first consider \ref{eq:rp higher order backwards}, with $k=n+1$ and $p=1+2n$.
	\begin{align}\label{eq:rp-1 higher order backwards}
	\int_{\Sigma_{u=u_R}}\chi r^{1+2n}\vert\p_v^{n+1}\psi\vert^2dv&\leq\int_{u=u_R}\sum_{m+j\leq n}(l+1)^{2j}\vert\p_u^{m+1}\psi_+\vert^2du+\int_{u\geq u_R}\chi\sum_{j=0}^{n+1}\frac{(l+1)^{2j}}{r^{2(j-n)}}\vert\p_v^{1+n-j}\psi\vert^2dudv\\\nonumber
	&\leq A\int_{u=u_R}\sum_{m+j\leq n}(l+1)^{2j}\vert\p_u^{m+1}\psi_+\vert^2du\\\nonumber
	&\qquad+A\int_{u\geq u_R}\chi\sum_{j=0}^{n}\frac{(l+1)^{2j}}{r^{2(j-n)}}\vert\p_v^{1+n-j}\psi\vert^2+(l+1)^{2n}\chi\vert\p_v\psi\vert^2dudv\\\nonumber
	&\leq A\int_{u=u_R}\sum_{m+j\leq n}(l+1)^{2j}\vert\p_u^{m+1}\psi_+\vert^2du+A\int_{u\geq u_R}\chi\sum_{j=0}^{n}(l+1)^{2(n-j)}r^{2j}\vert\p_v^j(\p_t+\p_u)\psi\vert^2dudv\\\nonumber
	&\leq A\int_{u=u_R}\sum_{m+j\leq n}(l+1)^{2j}\vert\p_u^{m+1}\psi_+\vert^2du+A\int_{u\geq u_R}\chi\sum_{j=0}^{n}(l+1)^{2(n-j)}r^{2j}\vert\p_v^{j-1}(V\psi)\vert^2dudv\\\nonumber
	&\qquad+A\int_{u=u_R}\sum_{1\leq m+j\leq n}\int_{u'=u}^\infty\left(M^{2m}+(u-u_R)^{2m}\right)(l+1)^{2j}\vert\p_u^m\p_t\psi_+\vert^2du'du\\\nonumber
	&\leq A\sum_{0\leq m+j\leq n}\int_{u=u_R}^\infty\left(M^{2m}+(u-u_R)^{2m+1}\right)(l+1)^{2j}\vert\p_u^{m+1}\psi_+\vert^2du\\\nonumber
	&\qquad +A\int_{u\geq u_R}\chi\sum_{j=0}^{n}(l+1)^{2(n-j)+2}r^{2j}\vert\p_v^{j-1}(V\psi)\vert^2dudv\\\nonumber
	&\leq A\int_{u=u_R}\sum_{1\leq m+j\leq n+1}(l+1)^{2j}\vert\p_u^m\psi_+\vert^2du,
	\end{align}
	where we have used that $\p_t$ is a Killing field along with our induction hypothesis in the final three lines.
	
	We then proceed to prove \eqref{eq:higher order r weighted Sigma_u}:
	\begin{align}
	\int_{\Sigma_{u=u_R}}\chi r^{2+2n}\vert\p_v^{n+1}\psi\vert^2dv&\leq A\int_{u=u_R}\sum_{m+j\leq n}(l+1)^{2j}\vert\p_u^{m+1}\psi_+\vert^2du\\\nonumber
	&\qquad+A\int_{u\geq u_R}\chi\sum_{j=0}^{n}\frac{(l+1)^{2j}}{r^{2(j-n)-1}}\vert\p_v^{1+n-j}\psi\vert^2+(l+1)^{2n}r\chi\vert\p_v\psi\vert^2dudv\\\nonumber
	&\leq A\int_{u=u_R}\sum_{m+j\leq n}(M+(u-u_R))(l+1)^{2j}\vert\p_u^{m+1}\psi_+\vert^2du\\\nonumber
	&\qquad+A\int_{u\geq u_R}\chi\sum_{j=0}^{n}(l+1)^{2(n-j)}r^{2j+1}\vert\p_v^j(\p_t+\p_u)\psi\vert^2dudv\\\nonumber
	&\leq A\sum_{1\leq m+j\leq n+1}\int_{u=v_0-R}^{\infty}\left(M^{2m}+(u-u_R)^{2m}\right)(l+1)^{2j}\vert\p_u^m\psi_+\vert^2du,
	\end{align}
	applying \eqref{eq:rp-1 higher order backwards}, along with identical reasoning as used in \eqref{eq:rp-1 higher order backwards}.
\end{proof}

\begin{Proposition1}[Integrated Decay of Higher Order Energy]\label{Prop:IntegratedDecay}
	Let $\psi_+$ be a Schwartz function. Let $\psi$ be the solution of \eqref{eq:RadWave}, as given by Theorem \ref{Thm:RNExist}, on a sub-extremal \RNS background $\mathcal{M}_{RN}$, with radiation field on $\mathcal{I}^+$ equal to $\psi_+$, and which vanishes on $\mathcal{H}^+$. Let $R$ be a constant, and let $t_0$ be a fixed value of $t$. Then for each $n\in\mathbb{N}_0$, we have the following bounds:
	\begin{align}\label{eq:IEDHigherOrder}
	\int_{t_{2n+1}=-\infty}^{t_0}\int_{t_{2n}=-\infty}^{t_n}&...\int_{t_1=-\infty}^{t_2}\int_{t=-\infty}^{t_1}\left(\int_{\bar{\Sigma}_{t,R}}-dt(J^{\p_t}[\p_t^n\phi])\right)dtdt_1dt_2..dt_{2n+1}\\\nonumber
	&\qquad+\sum_{j+\vert\alpha\vert+m\leq n}\int_{v=t_0+R,r^*\geq R}\int_{v\leq t_0+R, r^*\geq R}r^{1+2j}\left(\vert\p_u^{1+j}\p_t^m\Omega^\alpha\psi\vert^2+jV\vert\p_u^j\p_t^m\Omega^\alpha\psi\vert^2\right) dudv\\\nonumber
	&\qquad+\sum_{j+\vert\alpha\vert+m\leq n}\int_{u=t_0+R,r^*\leq -R}(-r^*)^{1+2j}\left(\vert\p_v^{1+j}\p_{t}^m\Omega^\alpha\psi\vert^2+(-r^*)V\vert\p_v^j\p_{t}^m\Omega^\alpha\psi\vert^2\right)dudv\\\nonumber
	&\leq A_n\sum_{j+\vert\alpha\vert+m\leq n}\int_{v=t_0+R,r^*\geq R}r^{2+2j}\vert\p_u^{1+j}\p_{t}^m\Omega^\alpha\psi\vert^2du\\\nonumber
	&\qquad+A_n\sum_{j+\vert\alpha\vert+m\leq n}\int_{u=t_0+R,r^*\leq -R}(-r^*)^{2+2j}\vert\p_v^{1+j}\p_{t}^m\Omega^\alpha\psi\vert^2dv\\\nonumber
	&\qquad+A_n\sum_{j+\vert\alpha\vert\leq 2n+2}\int_{\Sigma_{t_0}}-dt(J^{\p_t}[\p_t^j\Omega^\alpha\phi]),
	\end{align}
	where $A_n=A_n(M,n,R)$.
\end{Proposition1}

\begin{proof}
	This proof again closely follows that of \cite{ERNScat}. We will consider $T$ energy through a null foliation, $\bar{\Sigma}_{t,R}$ (see \eqref{eq:SigmaBarDefinition}).
	
	We first look at how the wave operator commutes with both $\p_u$ and $\p_v$:
	\begin{align}\label{eq:wave commute p_u}
	\p_u\p_v(\p_u^n\psi)+V\p_u^n\psi=-\p_u^n(V\psi)+V\p_u^n\psi&=\sum_{j=0}^{n-1}\binom{n}{j}(-1)^{n-j+1}\p_{r^*}^{n-j}V\p_u^j\psi\leq A\sum_{j=0}^{n-1}\frac{V}{r^{n-j}}\vert\p_u^j\psi\vert\\\label{eq:wave commute p_v2}
	\p_u\p_v(\p_v^n\psi)+V\p_v^n\psi=-\p_v^n(V\psi)+V\p_v^n\psi&=-\sum_{j=0}^{n-1}\binom{n}{j}\p_{r^*}^{n-j}V\p_v^j\psi\leq A\sum_{j=0}^{n-1}V\kappa^{n-j}\vert\p_v^j\psi\vert
	\end{align}
	
	We apply the $r^p$ and $r^{*p}$ methods to the null segments of $\bar{\Sigma}_{t_0,R}$ to obtain:
	\begin{align}\label{eq:rp}
	\int_{v=t_0+R,r^*\geq R}r^p\chi\left(\frac{r^*-R}{M}\right)\vert\p_u^k\psi\vert^2du&=\int_{v\leq t_0+R,r^*\geq R}\left(pr^{p-1}D\chi+\frac{r^p}{M}\chi'\right)\vert\p_u^k\psi\vert^2-r^p\chi V\p_u\left(\vert\p_u^{k-1}\psi\vert^2\right)dvdu\\\nonumber
	&\qquad+\int_{v\leq t_0+R,r^*\geq R}2r^p\chi \R\left(\p_u^k\bar{\psi}\sum_{j=0}^{k-2}\binom{k-1}{j}(-1)^{k-j}\p_{r^*}^{k-1-j}V\p_u^j\psi \right)dudv\\\nonumber
	&\geq\int_{v\leq t_0+R,r^*\geq R}\left(pr^{p-1}D\chi+\frac{r^p}{M}\chi'\right)\vert\p^k_u\psi\vert^2-\p_{r^*}\left(r^p\chi V\right)\left(\vert\p_u^{k-1}\psi\vert^2\right)dvdu\\\nonumber
	&\qquad-A\int_{v\leq t_0+R,r^*\geq R}r^p\chi\vert\p_u^k\psi\vert\sum_{j=0}^{k-2}Vr^{1-k+j}\vert\p_u^j\psi \vert dudv+\int_{\mathcal{I}^-}r^pV\vert\p_u^{k-1}\psi\vert^2dv\\\nonumber
	&\geq a\int_{v\leq t_0+R,r^*\geq R}\chi r^{p-1}\left(p\vert\p_u^k\psi\vert^2+(p-2)V\vert\p_u^{k-1}\psi\vert^2\right)dudv\\\nonumber
	&\qquad-A\int_{R\leq r^*\leq M+R, t\leq t_0}\vert\p_u^k\psi\vert^2d+V\vert\p_u^{k-1}\psi\vert^2dr^*dt\\\nonumber
	&\qquad-A\sum_{j=0}^{k-2}\int_{v\leq t_0+R,r^*\geq R}\chi V^2r^{3-2k+2j+p}\vert\p_u^j\psi \vert^2 dudv.
	\end{align}
	
	\begin{align}\nonumber
	\int_{u=t_0+R}(-r^*)^p\chi\left(\frac{-r^*-R}{M}\right)\vert\p_v^k\psi\vert^2dv&=\int_{u\leq t_0+R}\left(p(-r^*)^{p-1}\chi+\frac{(-r^*)^p}{M}\chi'\right)\vert\p_v^k\psi\vert^2\\\nonumber
	&\qquad-(-r^*)^p\chi V\p_v\left(\vert\p_v^{k-1}\psi\vert^2\right)-2\chi(-r^*)^p \R\left(\p_v^k\bar{\psi}\sum_{j=0}^{k-2}\binom{k-1}{j}\p_{r^*}^{k-1-j}V\p_v^j\psi\right)dvdu\\\nonumber
	&\geq\int_{u\leq t_0+R}\left(p(-r^*)^{p-1}\chi+\frac{(-r^*)^p}{M}\chi'\right)\vert\p_v^k\psi\vert^2+\p_{r^*}\left((-r^*)^p\chi V\right)\left(\vert\p_v^{k-1}\psi\vert^2\right)\\\label{eq:r^*p}
	&\qquad-A\chi (-r^*)^p\vert\p_v^k\psi\vert\sum_{j=0}^{k-2}V\kappa^{k-1-j}\vert\p_v^j\psi\vert dvdu\\\nonumber
	&\geq a\int_{u\leq t_0+R}\chi(-r^*)^{p-1}\left(p\vert\p_v^k\psi\vert^2+(-r^*\kappa-p)V\vert\p_v^{k-1}\psi\vert^2\right)dudv\\\nonumber
	&\qquad-A\int_{-r^*\geq M+R, t\leq t_0}\vert\p_v^k\psi\vert^2+V\vert\p_v^{k-1}\psi\vert^2dr^*dt\\\nonumber
	&\qquad-A\sum_{j=0}^{k-2}\int_{u\leq t_0+R}\chi(-r^*)^{p+1}\kappa^{2k-2j-2}V^2\vert\p_v^j\psi\vert^2dudv.
	\end{align}
	
	By summing \eqref{eq:rp} and \eqref{eq:r^*p} when $p=1$, $k=1$ (as then the two summations vanish), we obtain:
	\begin{align}\nonumber
	\int_{v=t_0+R,r^*\geq R}r\chi\left(\frac{r^*-R}{M}\right)\vert\p_u\psi\vert^2du&+\int_{u=t_0+R}(-r^*)\chi\left(\frac{-r^*-R}{M}\right)\vert\p_v\psi\vert^2dv+\int_{\Sigma_{t_0}}\sum_{j=0}^1-(l+1)^{2-2j}dt(J^{\p_t}[\p_t^j\phi])\\
	&\geq a\int_{t=-\infty}^{t_0}T\text{-energy}(\bar{\Sigma}_{t,R})dt
	\end{align}
	Here we have used Proposition \ref{Prop:ILED}.
	
	We then consider the $p=2$, $k=1$ case to obtain
	\begin{align}\nonumber
	\int_{v=t_0+R,r^*\geq R}r^2\chi\left(\frac{r^*-R}{M}\right)\vert\p_u\psi\vert^2du&+\int_{u=t_0+R}(-r^*)^2\chi\left(\frac{-r^*-R}{M}\right)\vert\p_v\psi\vert^2dv+\int_{\Sigma_{t_0}}\sum_{j=0}^2-(l+1)^{4-2j}dt(J^{\p_t}[\p_t^j\Omega^\alpha\phi])\\\label{eq:p=2}
	&\geq a	\int_{t=-\infty}^{t_0}\Bigg(\int_{v=t+R}r\chi\left(\frac{r^*-R}{M}\right)\vert\p_u\psi\vert^2du\\\nonumber
	&\qquad+\int_{u=t+R}(-r^*)\chi\left(\frac{-r^*-R}{M}\right)\left(\vert\p_v\psi\vert^2+((-r^*)\kappa-2)V\vert\psi\vert^2\right)dv\\\nonumber
	&\qquad+\int_{\Sigma_{t}}\sum_{j=0}^1-(l+1)^{2-2j}dt(J^{\p_t}[\p_t^j\phi])\Bigg)\\\nonumber
	&\geq a\int_{t=-\infty}^{t_0}\int_{t'=-\infty}^{t}T\text{-energy}(\bar{\Sigma}_{t',R})dt'dt.
	\end{align}
	By using mean value theorem and $T$-energy boundedness (see \cite{NewPhysSpace} for an example of this), one can thus obtain
	\begin{equation}
	\int_{\bar{\Sigma}_{t,R}}T\text{-energy}\leq A (-t)^{-2}\int_{\mathcal{I}^+}(M^2+u^2)\vert\p_u\psi_+\vert^2+l(l+1)\vert\psi_+\vert^2du.
	\end{equation}
	
	By considering $T$-energy boundedness between $\bar{\Sigma}_{t,R}$ and $\mathcal{H}^-\cup\mathcal{I}_-$, we can also obtain:
	\begin{align}\label{eq:ForwardWeightedHigherOrdern=0}
	\int_{t=-\infty}^{t_0}\int_{t'=-\infty}^{t}&\left(\int_{u=-\infty}^{t'+R}\vert\p_u\psi_{\mathcal{H}^-}\vert^2du+\int_{v=-\infty}^{t'+R}\vert\p_v\psi_-\vert^2dv\right)\\\nonumber
	&=\int_{u=-\infty}^{t_0+R}(u-t_0-R)^2\vert\p_u\psi_{\mathcal{H}^-}\vert^2du+\int_{v=-\infty}^{t_0+R}(v-t_0-R)^2\vert\p_v\psi_{RN}\vert^2dv\\\nonumber
	&\leq A\int_{\mathcal{I}^+}(M^2+u^2)\vert\p_u\psi_+\vert^2+l(l+1)\vert\psi_+\vert^2du.
	\end{align}
	
	We now proceed to prove the result inductively, given the case $n=0$ is \eqref{eq:p=2} (Provided $R>3\kappa$). We first look to bound the $r$ and $r^*$ weighted summations. We take $p=2+2n, k=n+1$ in \eqref{eq:rp}
	\begin{align}\label{eq:rpInduct}
	\int_{v\leq t_0+R}\chi r^{1+2n}\left(\vert\p_u^{1+n}\psi\vert^2+V\vert\p_u^n\psi\vert^2\right)dudv&\leq A\int_{v=t_0+R}\chi r^{2+2n}\vert\p_u^{1+n}\psi\vert^2du\\\nonumber
	&\qquad+A\int_{R\leq r^* \leq M+R, t\geq t_0}\vert\p_u^{n+1}\psi\vert^2+V\vert\p_u^n\psi\vert^2dr^*dt\\\nonumber
	&\qquad+A\sum_{m=1}^{n-1}\int_{v\leq t_0+R}\chi(l+1)^2r^{1+2m}V\vert\p_u^m\psi\vert^2dudv\\\nonumber
	&\qquad+\int_{v\leq t_0+R}2\chi r^{2+2n}\R\left(\p_u^{n+1}\bar{\psi}(-1)^n\p_{r^*}^nV\psi \right)dudv\\\nonumber
	&\leq A\int_{v=t_0+R}\chi r^{2+2n}\vert\p_u^{1+n}\psi\vert^2du+A\sum_{m+\vert\alpha\vert\leq n+1}(l+1)^2\int_{\Sigma_{t_0}}-dt(J^{\p_t}[\p_t^m\psi])\\\nonumber
	&\qquad+A\sum_{j+k+m\leq n}\int_{v=t_0+R,r^*\geq R}r^{2+2j}(l+1)^{2k}\vert\p_u^{1+j}\p_{t}^m\psi\vert^2du\\\nonumber
	&\qquad+A\sum_{j+k+m\leq n}\int_{u=t_0+R,r^*\leq -R}(-r^*)^{2+2j}(l+1)^{2k}\vert\p_v^{1+j}\p_{t}^m\psi\vert^2dv\\\nonumber
	&\qquad+A\sum_{j+m\leq 2n+2}\int_{\Sigma_{t_0}}-dt(J^{\p_t}[\p_t^j\Omega^\alpha\phi])+\int_{v\geq t_0+R,r^*\geq R}\chi (l+1)^2rV\vert\psi\vert^2 dudv\\\nonumber
	&\qquad+\int_{v\leq t_0+R}2\chi r^{2+2n}\R\left(\p_u^{n+1}\bar{\psi}(-1)^n\p_{r^*}^nV\psi \right)dudv\\\nonumber
	&\leq A\sum_{j+k+m\leq n}\int_{v=t_0+R,r^*\geq R}r^{2+2j}(l+1)^{2k}\vert\p_u^{1+j}\p_{t}^m\psi\vert^2du\\\nonumber
	&\qquad+A\sum_{j+k+m\leq n}\int_{u=t_0+R,r^*\leq -R}(-r^*)^{2+2j}(l+1)^{2k}\vert\p_v^{1+j}\p_{t}^m\psi\vert^2dv\\\nonumber
	&\qquad+A\sum_{j+m\leq 2n+2}\int_{\Sigma_{t_0}}-dt(J^{\p_t}[\p_t^j\Omega^\alpha\phi])\\\nonumber
	&\qquad+\int_{v\leq t_0+R}2\chi r^{2+2n}\R\left(\p_u^{n+1}\bar{\psi}(-1)^n\p_{r^*}^nV\psi \right)dudv
	\end{align}

	In order to bound the final term in \eqref{eq:rpInduct}, we first note that the usual method of separating does not work:
	\begin{equation}
	\int_{v=t+R}\chi r^{2+2n}\R\left(\p_u^{n+1}\bar{\psi}(-1)^n\p_{r^*}^nV\psi \right)du\leq	A\int_{v=t+R}\chi r^{2n+1}\vert\p_u^{n+1}\psi\vert^2+ r(l+1)^2V\vert\psi\vert^2 du.
	\end{equation}
	
	Unfortunately, we have no way to bound $rV\vert\psi\vert^2$. If we consider lower order terms in $r$, we can use Hardy's inequality.
	\begin{equation}\label{eq:Hardyrp}
	\int_{v=t+R}V\vert\psi\vert^2\leq A\int_{v=t+R}(l+1)^2\vert\p_u\psi\vert^2+A\int_{v=t+R, \vert r^*\vert\leq R}V\vert\psi\vert^2du.
	\end{equation}
	
	Thus, the only term we need to be concerned about in \eqref{eq:rpInduct} is the leading order in $r$  behaviour of the final term. This behaves as follows:
	\begin{align}\label{eq:rpInductn=1}
	\int_{v=t+R}2l(l+1)\chi r^{n}\R\left(\p_u^{n+1}\bar{\psi}\psi \right)du&=\int_{v=t+R}2l(l+1)Re\left(\p_u\bar{\psi}\sum_{j=0}^n\binom{n}{j}\p_{r^*}^{n-j}(-1)^{j}(\chi r^n)\p_u^j\psi\right) du\\\nonumber
	&\leq A\int_{v=t+R}l(l+1)r\vert\p_u\psi\vert^2+l(l+1)\sum_{j=1}^{n-1}r^{2j-1}\vert\p_u^j\psi\vert^2du\\\nonumber
	&\qquad +\int_{v=t+R}l(l+1)\p_{r^*}^n(\chi r^n)\p_u\left(\vert\psi\vert^2\right)du\\\nonumber
	&\leq A\int_{v=t+R}l(l+1)r\vert\p_u\psi\vert^2+l(l+1)\sum_{j=1}^{n-1}r^{2j-1}\vert\p_u^j\psi\vert^2du\\\nonumber
	&\qquad +\int_{v=t+R}l(l+1)\p_{r^*}^{n+1}(\chi r^n)\vert\psi\vert^2du-l(l+1)n!(-1)^n\vert\psi\vert_{\mathcal{I}^-}^2.
	\end{align}
	As $\p_{r^*}^{n+1}(r^n)\leq Ar^{-2}$, we can use \eqref{eq:Hardyrp} to bound this. Combining \eqref{eq:rpInduct}, \eqref{eq:Hardyrp} and \eqref{eq:rpInductn=1}, we obtain
	
	\begin{align}
	\int_{v\leq t_0+R}\chi r^{1+2n}\left(\vert\p_u^{1+n}\psi\vert^2+V\vert\p_u^n\psi\vert^2\right)dudv&\leq A\sum_{j+k+m\leq n}\int_{v=t_0+R,r^*\geq R}r^{2+2j}(l+1)^{2k}\vert\p_u^{1+j}\p_{t}^m\psi\vert^2du\\\nonumber
	&\qquad+A\sum_{j+k+m\leq n}\int_{u=t_0+R,r^*\leq -R}(-r^*)^{2+2j}(l+1)^{2k}\vert\p_v^{1+j}\p_{t}^m\psi\vert^2dv\\\nonumber
	&\qquad+A\sum_{j+m\leq 2n+2}\int_{\Sigma_{t_0}}-dt(J^{\p_t}[\p_t^j\Omega^\alpha\phi]),
	\end{align}
	as required.
	
	The $(-r^*)^p$ section is made much more easy by the exponential behaviour of the potential. We take $p=2+2n, k=n+1$ in \eqref{eq:r^*p}:
	\begin{align}
	\int_{u\leq t_0+R}\chi (-r^*)^{1+2n}\left(\vert\p_v^{n+1}\psi\vert^2+(-r^*)V\vert\p_v^n\psi\vert^2\right)dudv&\leq A\int_{u=t_0+R}\chi(-r^*)^{2+2n}\vert\p_v^{n+1}\psi\vert^2dv\\\nonumber
	&\qquad+A\int_{u\leq t_0+R, -r^*\leq 2(2+2n)/\kappa}\chi (-r^*)^{2+2n} V\vert\p_v^n\psi\vert^2dudv\\\nonumber
	&\qquad+A\int_{R\leq-r^*\leq R+M,t\leq t_0}\vert\p_v^{n+1}\psi\vert^2+V\vert\p_v^k\psi\vert^2dr^*dt\\\nonumber
	&\qquad+A\sum_{j=0}^{n-1}\int_{u\leq t_0+R}\chi(-r^*)^{3+2n}\kappa^{2n-2j}V^2\vert\p_v^j\psi\vert^2dudv\\\nonumber
	&\leq A\int_{u=t_0+R}\chi(-r^*)^{2+2n}\vert\p_v^{n+1}\psi\vert^2dv\\\nonumber
	&\qquad+A\int_{R\leq-r^*\leq \max\{R+M, 2(2+2n)/\kappa\},t\leq t_0}\vert\p_v^{n+1}\psi\vert^2+V\vert\p_v^n\psi\vert^2dr^*dt\\\nonumber
	&\qquad+A\frac{(l+1)^2}{M^2}\sum_{j=0}^{n-1}\int_{u\leq t_0+R}\chi(-r^*)^{1+2j}\kappa^{2n-2j}V\vert\p_v^j\psi\vert^2dudv\\\nonumber
	&\leq A\sum_{j+\vert\alpha\vert+m\leq n}\int_{v=t_0+R,r^*\geq R}r^{2+2j}\vert\p_u^{1+j}\p_{t}^m\Omega^\alpha\psi\vert^2du\\\nonumber
	&\qquad+A\sum_{j+\vert\alpha\vert+m\leq n}\int_{u=t_0+R,r^*\leq -R}(-r^*)^{2+2j}\vert\p_v^{1+j}\p_{t}^m\Omega^\alpha\psi\vert^2dv\\\nonumber
	&\qquad+A\sum_{j+\vert\alpha\vert\leq 2n+2}\int_{\Sigma_{t_0}}-dt(J^{\p_t}[\p_t^j\Omega^\alpha\phi]),
	\end{align}
	as required. In the final inequality, we have used the induction hypothesis.
	
	By repeating the above argument, we can also show that
	\begin{align}
	\int_{u\leq t_0+R}\chi (-r^*)^{2n}\left(\vert\p_v^{n+1}\psi\vert^2+(-r^*)V\vert\p_v^n\psi\vert^2\right)dudv&+\int_{v\leq t_0+R}\chi r^{2n}\left(\vert\p_u^{1+n}\psi\vert^2+V\vert\p_u^n\psi\vert^2\right)dudv\\\nonumber
	&\leq A\sum_{j+\vert\alpha\vert+m\leq n}\int_{v=t_0+R,r^*\geq R}r^{1+2j}\vert\p_u^{1+j}\p_{t}^m\Omega^\alpha\psi\vert^2du\\\nonumber
	&\qquad+A\sum_{j+\vert\alpha\vert+m\leq n}\int_{u=t_0+R,r^*\leq -R}(-r^*)^{1+2j}\vert\p_u^{1+j}\p_{t}^m\Omega^\alpha\psi\vert^2dv\\\nonumber
	&\qquad+A\sum_{j+\vert\alpha\vert\leq 2n+1}\int_{\Sigma_{t_0}}-dt(J^{\p_t}[\p_t^j\Omega^\alpha\phi])
	\end{align} 
	
	We now look to prove the final part. Assuming the result is true in the $n$ case, apply \eqref{eq:IEDHigherOrder} to $\p_t\psi$. We then integrate  twice with respect to $t_0$ to obtain
	\begin{align}
	\int_{t_{2n+3}=-\infty}^{t_0}\int_{t_{2n+2}=-\infty}^{t_{2n+3}}...&\int_{t_1=-\infty}^{t_2}\int_{t=-\infty}^{t_1}\left(\int_{\bar{\Sigma}_{t,R}}-dt(J^{\p_t}[\p_t^{n+1}\phi])\right)dtdt_1dt_2..dt_{2n+3}\\\nonumber
	&\leq A\sum_{j+k+m\leq n}\int_{-\infty}^{t_0}\int_{-\infty}^{t_{2n+3}}\int_{v=t+R,r^*\geq R}r^{2+2k}(l+1)^{2k}\vert\p_u^{1+k}\p_{t}^{m+1}\psi\vert^2dudt_{2n+2}dt_{2n+3}\\\nonumber
	&\qquad+A\sum_{j+k+m\leq n}\int_{-\infty}^{t_0}\int_{-\infty}^{t_{2n+3}}\int_{u=t+R,r^*\leq -R}(-r^*)^{2+2k}(l+1)^{2k}\vert\p_v^{1+k}\p_{t}^{m+1}\vert^2dvdt_{2n+2}dt_{2n+3}\\\nonumber
	&\qquad+A\sum_{j+\vert\alpha\vert\leq 2n+2}\int_{-\infty}^{t_0}\int_{-\infty}^{t_{2n+3}}\left(\int_{\Sigma_{t}}-dt(J^{\p_t}[\p_t^{j+1}\Omega^\alpha\phi])\right)dt_{2n+2}dt_{2n+3}.
	\end{align}
	
	The final term here can be immediately bounded using Proposition \ref{Prop:ILED}. To bound the earlier terms, we note that $\p_u+\p_v=\p_t$, and we can use \eqref{eq:RadWave} to remove any mixed $u,v$ derivatives.
	
	\begin{align}\label{eq:GrossBigr}
	\sum_{j+k+m\leq n}\int_{-\infty}^{t_0}\int_{-\infty}^{t_{2n+3}}&\int_{v=t+R,r^*\geq R}r^{2+2j}(l+1)^{2k}\vert\p_u^{1+j}\p_{t}^{m+1}\psi\vert^2dudt_{2n+2}dt_{2n+3}\\\nonumber
	&\leq A\sum_{j+k+m\leq n}\int_{-\infty}^{t_0}\int_{-\infty}^{t_{2n+3}}\int_{v=t+R,r^*\geq R}r^{2+2j}(l+1)^{2k}\vert\p_u^{2+j}\p_{t}^m\psi\vert^2dudt_{2n+2}dt_{2n+3}\\\nonumber
	&\qquad+A\sum_{j+k+m\leq n}\int_{-\infty}^{t_0}\int_{-\infty}^{t_{2n+3}}\int_{v=t+R,r^*\geq R}r^{2+2j}(l+1)^{2k}\vert\p_u^j\p_{t}^m (V\psi)\vert^2dudt_{2n+2}dt_{2n+3}\\\nonumber
	&\leq A\sum_{j+k+m\leq n}\int_{-\infty}^{t_0}\int_{-\infty}^{t_{2n+3}}\int_{v=t+R,r^*\geq R}r^{2+2j}(l+1)^{2k}\vert\p_u^{2+j}\p_{t}^m\psi\vert^2dudt_{2n+2}dt_{2n+3}\\\nonumber
	&\qquad+A\sum_{j+k+m\leq n}\int_{-\infty}^{t_0}\int_{-\infty}^{t_{2n+3}}\int_{v=t+R,r^*\geq R}r^{2j}(l+1)^{2k+2}V\vert\p_u^j\p_{t}^m\psi\vert^2dudt_{2n+2}dt_{2n+3}\\\nonumber
	&\leq A\sum_{j+k+m\leq n+1}\int_{-\infty}^{t_0}\int_{v=t+R,r^*\geq R}r^{1+2j}(l+1)^{2k}\vert\p_u^{1+j}\p_{t}^m\psi\vert^2dudt_{2n+3}\\\nonumber
	&\qquad+ A\sum_{j+k+m\leq n+1}\int_{-\infty}^{t_0}\int_{u=t+R,r^*\leq R}(-r^*)^{1+2j}(l+1)^{2k}\vert\p_v^{1+j}\p_{t}^m\psi\vert^2dudt_{2n+3}\\\nonumber
	&\qquad+A\sum_{j+\vert\alpha\vert\leq 2n+3}\int_{-\infty}^{t_0}\left(\int_{\Sigma_{t_0}}-dt(J^{\p_t}[\p_t^j\Omega^\alpha\phi])\right)dt_{2n+3}\\\nonumber
	&\leq A\sum_{j+k+m\leq n+1}\int_{v=t+R,r^*\geq R}r^{2+2j}(l+1)^{2k}\vert\p_u^{1+j}\p_{t}^m\psi\vert^2du\\\nonumber
	&\qquad+ A\sum_{j+k+m\leq n+1}\int_{u=t+R,r^*\leq R}(-r^*)^{2+2j}(l+1)^{2k}\vert\p_v^{1+j}\p_{t}^m\psi\vert^2dv\\\nonumber
	&\qquad+A\sum_{j+\vert\alpha\vert\leq 2n+4}\int_{\Sigma_{t_0}}-dt(J^{\p_t}[\p_t^j\Omega^\alpha\phi]),
	\end{align}
	as required. An identical argument follows for the $-r^*\geq R$ region.
\end{proof}

\begin{Theorem1}[Boundedness of the $u$ and $v$ Weighted Energy]\label{Thm:RNScatBound}
	Let $\psi_+$ be a Schwartz function on the cylinder. Let $\phi$ be a solution to \eqref{eq:wave} on a sub-extremal \RNS background $\mathcal{M}_{RN}$. Further, let $\phi$ vanish on $\mathcal{H}^+$ and have future radiation field equal to $\psi_+$. Then there exists a constant $A_n=A_n(M,n)$ (which also depends on the choice of origin of $u,v$) such that
	\begin{align}
	\sum_{k=0}^2\sum_{2j+m+2\vert\alpha\vert\leq 2n}\int_{\mathcal{H}^-\cap\{u\leq 0\}}(M^{2(j+1)-k}+u^{2(j+1)-k})\vert\p_u^{j+m+k+1}\Omega^\alpha\psi_{\mathcal{H}^-}\vert^2du&\\\nonumber
	+\sum_{k=0}^2\sum_{1\leq j+\vert\alpha\vert, 2j+2\vert\alpha\vert+m\leq 2n+2}\int_{\mathcal{I}^-\cap\{v\leq 0\}}(M^{2j-k}+v^{2j-k})\vert\p_v^{j+k+m}\Omega^\alpha\psi_-\vert^2dv&\\\nonumber
	\leq A\sum_{k=0}^2\sum_{1\leq j+\vert\alpha\vert, 2j+2\vert\alpha\vert+m\leq 2n+2}&\int_{\mathcal{I}^-}(M^{2j-k}+u^{2j-k})\vert\p_u^{j+k+m}\Omega^\alpha\psi_+\vert^2du
	\end{align}
\end{Theorem1}

\begin{proof}
	This result again follows closely that of \cite{ERNScat}. It is an easy combination of Propositions \ref{Prop:r^* Weighted Bound} and \ref{Prop:IntegratedDecay}, applied to $T^m\Omega^\alpha$, for $\alpha\leq n-j$ and $m\leq 2n-2k-2\alpha$. All that remains is to note
	\begin{align}
	\int_{t_{2n+1}=-\infty}^{t_0}&\int_{t_{2n}=-\infty}^{t_n}...\int_{t_1=-\infty}^{t_2}\int_{t=-\infty}^{t_1}\left(\int_{\bar{\Sigma}_{t,R}}-dt(J^{\p_t}[\p_t^n\phi])\right)dtdt_1dt_2..dt_{2n+1}\\\nonumber
	&=\int_{t_{2n+1}=-\infty}^{t_0}\int_{t_{2n}=-\infty}^{t_n}...\int_{t_1=-\infty}^{t_2}\int_{t=-\infty}^{t_1}\left(\int_{-\infty}^{t+R}\vert\p_u\psi_{\mathcal{H}^-}\vert^2\sin\theta d\theta d\varphi du+\int_{-\infty}^{t+R}\vert\p_v\psi_-\vert^2\sin\theta d\theta d\varphi dv\right)dtdt_1dt_2..dt_{2n+1}\\\nonumber
	&=\frac{1}{(2n+2)!}\left(\int_{-\infty}^{t+R}(u-t_0-R)^{2n+2}\vert\p_u\psi_{\mathcal{H}^-}\vert^2\sin\theta d\theta d\varphi du+\int_{-\infty}^{t+R}(v-t_0-R)^{2n+2}\vert\p_v\psi_-\vert^2\sin\theta d\theta d\varphi dv\right),
	\end{align}
	by repeated integration by parts.
\end{proof}

\begin{Theorem1}[Arbitrary polynomial decay of I.E.~Terms]\label{Thm:Decay of IE}
	Let $\psi_+$ be a Schwartz function on the cylinder, with $\hat{\psi}_+$ supported on $\omega\geq 0$. Then for each $n$, there exists an $A_n(M,\psi_+)$ such that
	\begin{equation}
	I.E.[\psi_+,v_c,u_1,u_0]\leq A_n\left((u_0-u_1)^{-n}+(u_0-v_c)^{-n}\right).
	\end{equation}
	Here $I.E.$ is as defined in Theorem \ref{Thm:Hawking}.
\end{Theorem1}

\begin{proof}
This proof is identical to that of Theorem \ref{Thm:ExtremalI.E.decay}.
\end{proof}

\section{Proof of the Main Result}\label{Sec:Proof}
In this section we will prove the main result of this paper, Theorem \ref{Thm:HawkingVague} as stated in the introduction.

\begin{proof}[Proof of Theorem \ref{Thm:HawkingVague}]
	By Lemma \ref{Lem:Reduction} and Theorem \ref{Thm:Hawking}, we can bound
	\begin{equation}
		2\left\vert\int_{\omega=-\infty}^0\vert\omega\vert\vert\hat{\psi}_-\vert^2 -\int_{\omega=-\infty}^\infty\alpha\vert\omega\vert\vert\hat{\psi}_{\mathcal{H}^-}\vert^2\right\vert=\left\vert\int_{\omega=-\infty}^\infty\vert\omega\vert\vert\hat{\psi}_-\vert^2 -\int_{\omega=-\infty}^\infty\vert\omega\vert\left(\coth\left(\frac{\pi}{\kappa}\vert\omega\vert\right)\vert\hat{\psi}_{\mathcal{H}^-}\vert^2+\vert\hat{\psi}_{RN}\vert^2\right)\right\vert.
	\end{equation}
	
	By applying Theorem \ref{Thm:Decay of IE} to the $I.E.$ terms, and choosing $u_1$ such that $e^{-\kappa u_1}=u_0^{-n}$ in the extremal case, we obtain the required result.
\end{proof}

As discussed in the introduction, Theorem \ref{Thm:HawkingVague} is the calculation of radiation of frequency $\hat{\psi}_+$ given off by the RNOS model of a collapsing black hole, see \cite{hawking1975} for a full discussion of this. We will however, comment that the quantity of particles emitted by extremal RNOS models is integrable. This means that the total number of particles given off by the forming extremal black hole is finite, and thus the black hole itself may never evaporate.

\section*{Acknowledgements}
We would like to thank Mihalis Dafermos for many insightful comments and for proof reading the manuscript. We would like to thank Owain Salter Fitz-Gibbon for many insightful discussions. We would also like to thank Dejan Gajic for useful discussions of the extremal case. Last but by no means least, we would like to thank Claude Warnick and Bernard Kay for their comments and suggestions.

This work was part funded by EPSRC DTP, 
$[1936235]$. This work was supported by the Additional Funding Programme for
Mathematical Sciences, delivered by EPSRC (EP/V521917/1) and the
Heilbronn Institute for Mathematical Research.

\vspace{10mm}
\textbf{Declarations:} The author has no relevant financial or non-financial interests to disclose.

\textbf{Data availability statement:} Data sharing not applicable to this article as no datasets were generated or analysed.

\bibliographystyle{unsrt}
\bibliography{SES-O.bib}

\end{document}